\documentclass[
  journal=largetwo,
  manuscript=article-type,
  year=2020,
  volume=37,
]{cup-journal}

\usepackage{amsmath}
\usepackage[nopatch]{microtype}
\usepackage{booktabs}
\usepackage{caption}

\title{Nucleation regions in the Large-Scale Structure I.  A catalogue of \emph{cores} in nearby rich superclusters}

\author{J. M. Z\'u\~niga}
\affiliation{Departamento de Astronom\'ia, DCNE-CGT, Universidad de Guanajuato, Callej\'on de Jalisco s/n, CP 36023, Guanajuato, Gto., Mexico}
\email[J. M. Z\'u\~niga]{jm.zuniga@ugto.mx}

\author{C. A. Caretta}
\affiliation{Departamento de Astronom\'ia, DCNE-CGT, Universidad de Guanajuato, Callej\'on de Jalisco s/n, CP 36023, Guanajuato, Gto., Mexico}

\author{H. Andernach}
\affiliation{Departamento de Astronom\'ia, DCNE-CGT, Universidad de Guanajuato, Callej\'on de Jalisco s/n, CP 36023, Guanajuato, Gto., Mexico}
\alsoaffiliation{Currently on leave at Th\"uringer Landessternwarte, Sternwarte 5, D-07778 Tautenburg, Germany}


\begin{document}

\begin{abstract}
We applied a Density-Based Clustering algorithm on samples of galaxies and galaxy systems belonging to 53 rich superclusters from the \textit{Main SuperCluster Catalogue} (MSCC) to identify the presence of ``central regions'', or \emph{cores}, in these large-scale structures. \emph{Cores} are defined here as large gravitationally bound galaxy structures, comprised of two or more clusters and groups, with sufficient matter density to survive cosmic expansion and virialize in the future. We identified a total of 105 galaxy structures classified as \emph{cores}, which exhibit a high density contrast of mass and galaxies. The Density-based \textit{Core} Catalogue (DCC), presented here, includes \emph{cores} that were previously reported in well-known superclusters of the Local Universe, and also several newly identified ones. We found that 83\% of the rich superclusters in our sample have at least one \emph{core}. While more than three \emph{cores} with different dynamical state are possible, the presence of a single \emph{core} in the superclusters is more common. Our work confirms the existence of nucleation regions in the internal structure of most rich superclusters and points to the fact that these \emph{cores} are the densest and most massive features that can be identified in the cosmic web with high probability for future virialization. 
\end{abstract}

\section{Introduction}
\label{intro}
Superclusters are the largest coherent, relatively isolated, conglomerates of galaxies that can be detected by averaging the observed --or simulated-- galaxy distribution over scales of $\sim 150$ $h_{70}^{-1}$ Mpc \citep[\textit{e.g.},][]{ba1996,ei2001,lii2012}, where $h_{70}$ is the Hubble constant in units of 70 km s$^{-1}$ Mpc$^{-1}$. 
These structures can be described as a hierarchical assembly of galaxies and systems (\textit{i.e.}, groups and clusters), which are frequently connected to their neighbors by galaxy bridges \citep[\textit{e.g.},][]{ei1984}, forming a complex and intricate network \citep[the `cosmic web', \textit{e.g.},][]{lib2018} of knots, filaments, walls and cosmic voids, called the Large-Scale Structure \citep[LSS, \textit{e.g.},][]{pe1980,ei2010} of the Universe.
{They can also be defined as convergent galaxy velocity fields \citep[see, \textit{e.g.},][for more details on supercluster definitions]{san2023}.} \\

{On the other hand,} there is no theoretical or observational consensus to define the boundaries of a supercluster within the LSS, and it is even discussed whether these structures are only large transitory density enhancements in the current galaxy distribution or are really physically {bound} systems with established dynamics \citep[\textit{e.g.},][]{tl2014,boh2021}. In practice, the identification of superclusters is carried out using algorithms based mainly on techniques such as cluster analysis and percolation theory \citep[\textit{e.g.},][]{sh1983,ei1984}, extended percolation \citep[\textit{e.g.},][]{ei2018}, analysis of smoothed density fields or luminosity density maps \citep[\textit{e.g.},][]{cd2011,lu2011}, and {velocity field analysis} \citep[\textit{e.g.},][]{tl2014,pr2021} among others, all {combined} with different empirical selection criteria (threshold density contrasts, characteristic lengths, etc). \\

Several catalogues of superclusters have been compiled using some of the above identification techniques: on observational samples of galaxies in large redshift surveys \citep[\textit{e.g.},][]{ei2007a,lii2012}, on catalogues of systems \citep[\textit{e.g.},][]{ei2001,ch2013,chow2014}{, taking them} as second-order galaxy agglomerations \citep[or `clusters of clusters', \textit{e.g.},][]{ab1961,ba1984}, or on cosmological simulations \citep[\textit{e.g.},][]{cd2011,ei2019}. 
In any case, superclusters appear predominantly as filamentary, planar or lumpy galaxy sets, surrounding large cosmic voids, very low-density galaxy regions of comparable size \citep[\textit{e.g.},][]{ba1996,sh2004}.\\

Some authors claim that most of the identified superclusters are not gravitationally bound structures as a whole \citep[with internal dynamics generally dominated by the Hubble flow, \textit{e.g.},][]{pea2014}, with a rather irregular and varied morphology and rarely {have well-defined} ``central regions'' \citep[\textit{e.g.},][]{ei2007b,ch2013}. 
{Even being bound,} they are assumed to be dynamically unrelaxed structures because their crossing times are of the order or higher than the age of the Universe and, thus, they still maintain the memory of their formation history \citep[\textit{e.g.},][]{oo1983,ch2013,ei2019}. 
{Thus}, it is still debated which of the current superclusters will {survive} in the future and become virialized structures \citep[\textit{e.g.},][]{lu2011,ch2015}; it may be that only the densest and most massive clumps inside superclusters will survive the {effect} of the cosmic expansion and begin a process of collapse --or some may already be collapsing-- as indicated by various simulations \citep[\textit{e.g.},][]{gra2002,du2006}. 
{Anyway,} the study of the internal structure of observed superclusters can help to understand their current evolutionary states and their future within the LSS, as well as the effect of the different cosmological density environments on the evolution of their member galaxies and systems.\\

The filaments, long chains of galaxies and smaller galaxy systems, are the most notable features within superclusters \citep[\textit{e.g.},][]{ei2007b,ir2020}. 
However, in rich superclusters it is also possible to identify other prominent agglomerations of galaxies and massive systems, ``central parts'' {\citep[\textit{e.g.},][]{ei2008,san2023}} generally, but not necessarily, located at intersections of filaments.
These regions, sometimes referred to in the literature as \emph{cores} \citep[\textit{e.g.},][]{bar1994,sm1997, ko1998,ei2008,ei2015,ei2016}, could be substructures in a relatively more advanced dynamical state than the rest of the supercluster and with enough mass density to gravitationally bind and collapse. 
The \emph{cores} are themselves a kind of `compact superclusters' with a higher probability of future virialization, the marginal `island universes' \citep[\textit{e.g.},][and references therein]{am2009}. \\

Based on the hierarchical structure formation models \citep[\textit{e.g.},][]{pe1980}, one could expect that \emph{cores} are structures with an intermediate dynamical state between the richest galaxy clusters and the ordinary superclusters. 
{In other words, \emph{cores} are for the present state of the LSS what the proto-clusters were in the past, some kind of ``nucleation regions'' that already have a chance to grow, collapse and virialize in our accelerated expanding Universe. In this context of continuous and unfinished structure formation, we will use indistinctly \emph{cores} and \emph{nucleation regions} to refer to these large, massive and relatively dense ``central parts'' of superclusters.} \\

{The currently massive rich clusters are found preferably at nodes
of the LSS, where the network of filaments converges \citep[e.g.,][]{ei2024}. It is natural to expect that these nodes, or at least some
of these, are the mere \textit{cores}.}
Thus, massive systems inside superclusters could serve as markers of \emph{cores}. The results of a preliminary study \citep[\textit{e.g.},][]{cw2019t} based on percolation and using the Most Massive Clusters (MMCs) of rich superclusters as ``seeds'' of \emph{cores} show that about 86\% of them --in an all-sky sample-- may contain at least one \emph{core}. 
In addition, several works {have already pointed} \emph{cores} 
{in well-studied superclusters} \citep[such as \textit{Shapley} and \textit{Corona-Borealis} superclusters, \textit{e.g.},][and references therein]{bre1994,bar1994,bar2000,mar2004,pea2014}, which give us a {preliminary} idea of their general properties. \\

In this work we intend to broaden the knowledge panorama about the internal structure of superclusters in relation to {the presence of} \emph{cores} and their evolutionary states. For this, we have developed a percolation-based algorithm to identify accumulations of galaxy systems within rich superclusters. 
The algorithm makes use of a set of search criteria (see Section \ref{cores_id}) imposed to identify the most probable \emph{core} candidates within each supercluster of the sample and is able to provide feedback to the user (if necessary). 
The study of the internal structure of superclusters provides valuable information to understand the formation and evolution of virialized structures on the largest scales. 
The \emph{cores}, {embedded} in rich superclusters, are precisely the large-scale regions with the highest probability of future virialization, which is why they represent interesting laboratories to study the hierarchical aggregation of matter, its influence on the evolution of smaller structures within them {and to test cosmological models \citep[\textit{e.g.},][]{ei2021}}.\\

{This paper is organized as follows. 
Section \ref{s:data} describes the samples of superclusters, galaxy systems and galaxies selected for {our study}, as well as the process for constructing the supercluster boxes to which the identification algorithms are applied. Section \ref{s:tools} {presents} the {basic} analysis tools used throughout this work, {while} the clustering algorithms (applied to galaxies and systems) for {systems and} structures identification, {as well as} for sample refinement, are discussed in Section \ref{s:clust}. The characterization of the systems and structures is also presented in this section.
Section \ref{cores_id} describes the method used for \emph{core} selection from the structures identified previously. 
Section \ref{flucc} presents the catalogue of \emph{cores} obtained in the present study and describes some of their observational properties, comparing some of them with those reported previously in the literature. 
Finally, the discussion and conclusions of this work are presented in Section \ref{s:conc}.}\\
 
Through this paper a flat $\Lambda$CDM cosmology is used with the following
parameters:  Hubble constant {$H_0=70$ $h_{70}^{-1}$ km s$^{-1}$ Mpc$^{-1}$}, matter density $\Omega_m=0.3$, curvature and radiation density $\Omega_k=\Omega_r=0$, and dark energy density $\Omega_{\Lambda}=0.7$.

\section{Data}\label{s:data}
\subsection{Supercluster sample}
We selected a sample of 53 rich superclusters from the \textit{Main SuperCluster Catalogue} \citep[MSCC,][]{chow2014}, an all-sky catalogue containing 601 superclusters, identified by a tunable Friends-of-Friends algorithm based on a 2012 version of the spectroscopic redshift compilation of Abell/ACO-clusters \citep[see][for a description]{and2005}, with redshifts in the interval $0.02\leq z\leq 0.15$. 
The multiplicity $m_\mathrm{sc}$ of each MSCC-supercluster was defined as the number of its member {clusters in the Abell/ACO catalogue \citep[\textit{e.g.},][]{ab1958,ab1989}.}\\

The criteria to select the sample were: (i) choose only rich superclusters with $m_\mathrm{sc}\geq 5$; (ii) select all the well-sampled superclusters inside the Sloan Digital Sky Survey \citep[SDSS-DR13,][]{al2017}, {as a relatively complete subsample}; 
and (iii) include known and well-studied superclusters of the Local Universe in the 
southern sky with available redshifts in the 2dF Galaxy Redshift Survey \citep[2dFGRS,][]{col2001} and the 6dF Galaxy Survey \citep[6dFGS,][]{jo2009}. 
The first criterion allows better identification of internal structures in superclusters from samples of member galaxies and systems. 
The second criterion provides better statistics due to the 
relative homogeneity of SDSS, 
the currently largest, densest and most complete {available} galaxy redshift survey, containing 116 MSCC-superclusters of all $m_\mathrm{sc}$. 
Finally, the third criterion offers the possibility of having information from independent sources and including the most studied superclusters in order to be able to compare our results with those of previous studies, and to obtain validation support to our analysis.\\

Thus, our sample contains 45 superclusters within the SDSS region, which correspond to those selected by \citet{ir2020}, along with another 8 in the southern sky with a relatively good number of galaxies in the 2dFGRS or 6dFGS regions. The selected sample includes several well-known superclusters such as \textit{Shapley}, \textit{Pisces-Cetus}, \textit{Ursa Major}, \textit{Coma-Leo}, \textit{Corona-Borealis}, \textit{Hercules}, \textit{B\"ootes}, among others. 
The complete sample of 53 superclusters is presented in Table \ref{tab:SCsample}: column 1 shows the IDs of the superclusters in the MSCC; column 2 shows their proper names when they exist; columns 3 and 4 present, respectively, the mean RA and Dec (Equinox=J2000) of the supercluster positions taken from the MSSC; column 5 shows their mean redshifts $\bar{z}$, and column 6 presents their multiplicities $m_\mathrm{sc}$. The other columns of Table \ref{tab:SCsample} will be described below.

\begin{table*}
\resizebox{13cm}{!}{
\begin{threeparttable}
\caption{Sample of MSCC-superclusters}
\label{tab:SCsample}
\begin{tabular}{rcrrrrrrcrr}
\toprule \toprule

  \multicolumn{1}{c}{ID} &
  \multicolumn{1}{c}{Name$^{1}$} &
  \multicolumn{1}{c}{RA} &
  \multicolumn{1}{c}{Dec} &
  \multicolumn{1}{c}{$\bar{z}$} &
  \multicolumn{1}{c}{$m_\mathrm{sc}$} &
  \multicolumn{1}{c}{$N_{g_\text{box}}$} &
  \multicolumn{1}{c}{$N_\text{sys}$} &
  \multicolumn{1}{c}{Survey} &
  \multicolumn{1}{c}{$\varepsilon_\text{sc}$} &
  \multicolumn{1}{c}{$N_{g_\text{sc}}$} \\

  \multicolumn{1}{c}{MSCC} &
  \multicolumn{1}{c}{} &
  \multicolumn{1}{c}{[deg]} &
  \multicolumn{1}{c}{[deg]} &
  \multicolumn{1}{c}{} &
  \multicolumn{1}{c}{} &
  \multicolumn{1}{c}{} &
  \multicolumn{1}{c}{} &
  \multicolumn{1}{c}{} &
  \multicolumn{1}{c}{[$h_{70}^{-1}$ Mpc]} &
  \multicolumn{1}{c}{} \\
  
  \multicolumn{1}{c}{(1)} &
  \multicolumn{1}{c}{(2)} &
  \multicolumn{1}{c}{(3)} &
  \multicolumn{1}{c}{(4)} &
  \multicolumn{1}{c}{(5)} &
  \multicolumn{1}{c}{(6)} &
  \multicolumn{1}{c}{(7)} &
  \multicolumn{1}{c}{(8)} &
  \multicolumn{1}{c}{(9)} &
  \multicolumn{1}{c}{(10)} &
  \multicolumn{1}{c}{(11)}\\

\midrule \midrule

1 & Pisces-Cetus (Southern)     &   0.77 & -26.72 & 0.064 &  9 &  6567 & 114 & 2dF  &  13.7 & 5891 \\
 27 & Pisces-Cetus (Central)     &  11.32 & -22.87 & 0.061 &  9 &  2602 &  34 & 2dF  &  13.7 & 2306 \\
 33 & Sculptor             &  13.02 & -26.68 & 0.112 & 24 &  8893 & 137 & 2dF  &  12.5 & 6346 \\
 39 & Pisces-Cetus (Northern)     &  14.65 & -11.34 & 0.054 & 11 &  1241 &  15 & 6dF  &  18.2 & 736  \\
 55 &                      &  17.73 &  15.43 & 0.061 &  5 &   812 &  16 & SDSS &  13.9 & 519  \\
 72 &                      &  25.19 &   0.63 & 0.080 &  5 &  1941 &  50 & SDSS &  16.9 & 1670 \\
 75 &                      &  28.09 &  -5.15 & 0.094 &  7 &  1607 &  16 & SDSS &  26.0 & 1182 \\
 76 &                      &  28.36 &  -2.61 & 0.130 & 16 &  2617 &  25 & SDSS &  28.5 & 1982 \\
117 & Horologium-Reticulum &  49.89 & -48.46 & 0.066 & 26 &  2848 &  25 & 6dF  &  18.7 & 1448 \\
175 &                      & 125.30 &  17.07 & 0.094 &  6 &  2504 &  33 & SDSS &  17.1 & 1884 \\
184 &                      & 130.10 &  30.23 & 0.106 &  6 &  2101 &  19 & SDSS &  23.0 & 1457 \\
211 &                      & 147.87 &  64.87 & 0.119 &  8 &  1484 &   5 & SDSS &  20.9 & 619  \\
219 &                      & 153.99 &  19.13 & 0.116 &  5 &  1913 &  21 & SDSS &  22.7 & 1635 \\
222 &                      & 155.15 &  49.20 & 0.138 & 10 &  1865 &  16 & SDSS &  19.5 & 1127 \\
223 &                      & 155.24 &  62.94 & 0.140 &  5 &   776 &   4 & SDSS &  19.3 & 222  \\
229 &                      & 156.15 &  33.03 & 0.142 &  7 &  1855 &   8 & SDSS &  36.3 & 1291 \\
236 &                      & 156.77 &  10.37 & 0.033 &  6 &  8636 & 162 & SDSS &  9.9 & 7299  \\
238 &                      & 156.98 &  39.54 & 0.107 & 21 &  8328 & 106 & SDSS &  25.9 & 6815 \\
248 &                      & 159.49 &  44.26 & 0.125 &  5 &  1263 &   7 & SDSS &  27.0 & 823  \\
264 &                      & 165.30 &  12.20 & 0.116 &  8 &  1704 &  23 & SDSS &  24.8 & 1322 \\
266 &                      & 165.91 &  11.84 & 0.127 &  8 &   958 &   9 & SDSS &  17.3 & 502  \\
272 &                      & 167.85 &  41.33 & 0.076 &  6 &  1379 &  15 & SDSS &  16.6 & 1126 \\
277 &                      & 169.40 &  49.66 & 0.110 &  7 &  2748 &  24 & SDSS &  18.4 & 1756 \\
278 & Leo                  & 169.40 &  28.46 & 0.033 &  6 &  7920 & 113 & SDSS &  14.9 & 7387 \\
283 &                      & 170.79 &  20.33 & 0.138 & 12 &  2320 &  24 & SDSS &  24.8 & 1490 \\
295 & Coma-Leo             & 173.60 &  23.12 & 0.022 &  5 & 14308 & 153 & SDSS &  10.0 & 12182\\
310 & Ursa Major           & 175.88 &  55.24 & 0.064 & 21 & 12286 & 244 & SDSS &  12.4 & 10228\\
311 & Virgo-Coma A         & 176.15 &   9.90 & 0.083 &  8 &  5270 &  81 & SDSS &  19.1 & 4551 \\
314 &                      & 177.14 &  -2.00 & 0.079 &  6 &   558 &  13 & SDSS &  8.9 & 405   \\
317 &                      & 177.47 &  -1.58 & 0.128 & 13 &   840 &  14 & SDSS &  18.6 & 658  \\
323 &                      & 179.67 &  27.25 & 0.140 & 12 &  3330 &  35 & SDSS &  17.7 & 1722 \\
333 &                      & 181.43 &  29.32 & 0.081 &  9 &  1968 &  40 & SDSS &  8.1 & 1066  \\
335 &                      & 182.41 &  29.48 & 0.073 &  6 &  3099 &  65 & SDSS &  12.1 & 2138 \\
343 & Virgo-Coma B         & 183.89 &  14.30 & 0.081 &  5 &  2679 &  44 & SDSS &  9.1 & 1706  \\
360 & {Draco}                & 190.93 &  64.40 & 0.106 & 11 &  2199 &  28 & SDSS &  22.4 & 1944 \\
386 &                      & 199.53 &  38.31 & 0.072 &  5 &  3256 &  68 & SDSS &  16.2 & 2671 \\
389 & Shapley              & 201.05 & -32.89 & 0.049 & 24 &  4649 &  79 & 6dF  &  16.9 & 3790 \\
407 &                      & 208.56 &  26.70 & 0.136 &  6 &  1126 &  10 & SDSS &  23.3 & 635  \\
414 & Boötes               & 211.30 &  27.33 & 0.071 & 24 & 10902 & 293 & SDSS &  10.4 & 9012 \\
419 &                      & 212.34 &   7.17 & 0.112 &  5 &  1723 &  44 & SDSS &  15.4 & 1340 \\
422 &                      & 213.21 &  28.95 & 0.143 &  9 &  1065 &   8 & SDSS &  35.2 & 821  \\
430 &                      & 216.72 &  25.64 & 0.098 &  6 &  1603 &  27 & SDSS &  16.5 & 1303 \\
440 & Boötes A             & 223.17 &  22.27 & 0.117 &  9 &  3442 &  38 & SDSS &  20.7 & 2617 \\
441 &                      & 223.22 &  28.39 & 0.125 &  5 &  1058 &   5 & SDSS &  28.5 & 646  \\
454 &                      & 228.26 &   7.18 & 0.046 &  6 &  5704 & 185 & SDSS &  15.2 & 5590 \\
457 &                      & 228.61 &   6.97 & 0.079 &  6 &  4072 &  97 & SDSS &  18.6 & 3819 \\
460 &                      & 229.69 &  31.16 & 0.114 &  9 &  3499 &  50 & SDSS &  15.6 & 2598 \\
463 & Corona Borealis      & 232.20 &  30.42 & 0.074 & 14 &  8466 & 223 & SDSS &  10.8 & 7126 \\
474 & Hercules             & 241.50 &  16.22 & 0.036 &  5 &  7424 &  92 & SDSS &  9.9 & 6623  \\
484 &                      & 245.58 &  42.39 & 0.136 &  7 &  1319 &   9 & SDSS &  9.1 & 242   \\
509 & Pavo-Indus           & 311.70 & -39.98 & 0.023 &  6 &  3008 &  51 & 6dF  &  17.3 & 2584 \\
574 & Aquarius B           & 348.10 & -21.69 & 0.122 & 42 &  5982 &  65 & 2dF  &  19.3 & 4020 \\
579 &                      & 351.75 &  14.79 & 0.043 &  5 &  1477 &  40 & SDSS &  15.5 & 1190 \\

\bottomrule
\end{tabular}
\begin{tablenotes}[hang]
\item[]Table note
\item[1]Taken from \citet{ei2001}, \citet{chow2014}, \citet{ir2020}, and references therein.
\end{tablenotes}
\end{threeparttable}
}
\end{table*}

\subsection{Samples of galaxies and systems}

From now on we reserve the term `systems' to refer to first-order galaxy agglomerations such as groups and clusters, and the term `structures' to refer to {larger and} second-order galaxy agglomerations (clusters of systems) such as filaments, \emph{cores} and superclusters. 
Catalogues of higher density of objects make it possible to better define the systems, favoring the detection of the poorest ones and increasing the number of systems available for the detection of possible larger structures formed by them. These catalogues also improve the possibility of studying the correlations between the properties of galaxy populations and their surroundings, as well as analyzing the connectivity between member systems and substructures within superclusters. 
So, in order to detect \emph{cores} in MSCC-superclusters and study their properties, it is more convenient to apply the identification algorithms on a denser and more homogeneous galaxy sample to include those systems poorer and smaller than the Abell/ACO-clusters used by \citet{chow2014} to compile the MSCC. \\

\subsubsection{The SDSS galaxy sample}
We use the GalCat and SysCat catalogues, compiled by \citet{ir2020}, to study the internal structure of superclusters within the SDSS region and the galaxy properties in these environments. The first catalogue contains all the SDSS-DR13 \citep[\textit{e.g.},][]{al2017} galaxies, {belonging to the SpecObj sample of extragalactic objects (\textit{i.e.}, galaxies and low-$z$-quasars) with spectroscopic redshifts,} inside a rectangular box that encloses each supercluster, \textit{i.e.}, the `supercluster box', as well as photometric ($u,g,r,i,z$ magnitudes) and spectroscopic data from the survey. The second catalogue contains all the galaxy systems, {with at least 5 members,} identified by hierarchical clustering \citep[HC, \textit{e.g.},][]{se1996,the2009} {applied to} GalCat, as well as the estimated virial properties of these systems in each supercluster box.\\

The SDSS-DR13 covers an area of the sky of $\sim$14,555 deg$^2$ ($\sim35$\% of the sky) and provides homogeneus photometric and spectroscopic data with precision of 0.1 arcsec rms and uncertainty in radial velocities of $\sim$30 km s$^{-1}$ \citep[\textit{e.g.},][]{bol2012} of more than 2.6 million galaxies and quasars, being {almost complete to the magnitude limit of the main galaxy sample ($r_\mathrm{pet}=17.77$) corresponding to a median redshift of $z\sim 0.1$.} 
Thus, the SDSS-DR13 is a suitable database for the identification of galaxy systems and structures in the LSS and the study of galaxy properties in different cosmological environments. The GalCat and SysCat catalogues were used successfully by \citet{ir2020} to identify filaments in MSCC-superclusters. 

\subsubsection{Southern galaxy sample}
For superclusters located outside the SDSS region we compiled catalogues analogous to GalCat and SysCat (see the procedure below), but taking galaxies from the 2dFGRS or 6dFGS surveys (or just 2dF and 6dF for short), depending on where each of these superclusters had higher number density and greater coverage and homogeneity of galaxies. Both redshift surveys have also been quite useful and historically important for the study of the distribution of galaxies in the LSS.\\ 

The 2dF survey provides reliable redshifts for 221,414 galaxies brighter than an extinction-corrected nominal magnitude limit of $b_J = 19.45$ {\citep[$r_F \sim 18.3$ for early-type or $r_F\sim 18.6$ for late-type galaxies, \textit{e.g.},][]{cole2005}}, covering an area of $\sim$1,500 deg$^2$ (only $\sim 4$\% of the sky) in regions of high Galactic latitude in both the northern and southern Galactic hemispheres and with median redshift of $z = 0.11$. 
The four superclusters of our sample inside this database are all in the southern hemisphere. \\

The 6dF survey offers a catalogue of 125,071 galaxies, making near-complete samples with magnitude limits $K \leq 12.65$, $H \leq 12.95$, $J \leq 13.75$, $r_F \leq 15.60$, and $b_J \leq 16.75$ \citep[e.g.,][]{jo2009}, for almost half of the sky ($\sim$17,000 deg$^2$ on the southern sky, $|b| > 10$ deg), with a median redshift of $z=0.053$. 
Although 6dF covers a larger area of the sky ($\sim 41$\%), it is shallower than 2dF and SDSS, allowing to cover only four MSCC-superclusters with $z\leq 0.08$ in our sample. \\

To build the supercluster boxes we first transform the redshift-angular coordinates of 2dF/6dF galaxies and Abell/ACO-clusters to rectangular coordinates. Thus, if $(\alpha,\delta)$ are the equatorial coordinates (RA and Dec) of a galaxy or system, its rectangular coordinates can be estimated in the form
\begin{equation}\label{xyz}
\begin{split}
& X=D_\text{c}\cos{\delta}\cos{\alpha},\\
& Y=D_\text{c}\cos{\delta}\sin{\alpha},\\
& Z=D_\text{c}\sin{\delta},
\end{split}
\end{equation}
where,
\begin{equation}\label{D_c}
D_\text{c}(z)=\frac{c}{H_0}\int_{0}^{z}\frac{dz'}{E(z')},
\end{equation}
is the line-of-sight comoving distance \citep[\textit{e.g.},][]{hog2000} of the object defined by its redshift $z$, $c$ is the speed of light, and 
\begin{equation}\label{E}
E(z)\equiv\sqrt{\Omega_r(1+z)^4+\Omega_m(1+z)^3+\Omega_k(1+z)^2+\Omega_{\Lambda}}.
\end{equation}

Then, in a similar way as in \citet{ir2020}, we take all the 2dF/6dF galaxies located within the corresponding rectangular volume box that encloses the member Abell/ACO-clusters of each supercluster, including those found up to a distance of 20 $h_{70}^{-1}$ Mpc beyond the center of the farthest member cluster in each direction. 
For each supercluster box, we compile the southern GalCat catalogue containing the equatorial coordinates (FK5, Equinox=J2000.0), the redshift and the magnitudes of each galaxy. 
For 2dF galaxies we select the final $b_J$ magnitude corrected from extinction and the SuperCosmos R magnitude \citep[\textit{e.g.},][]{ham2001}, while for 6dF galaxies we select the recalibrated $b_J$ and $r_F$ magnitudes \citep[\textit{e.g.},][]{cole2005}. 
Photometric and spectroscopic data were taken directly from the 2dF and 6dF catalogues, both available in VizieR\footnote{see \url{https://cdsarc.cds.unistra.fr/viz-bin/cat/VII/250} and \url{https://cdsarc.cds.unistra.fr/viz-bin/cat/J/MNRAS/399/683}.}. In each supercluster box, a system detection algorithm was applied to build the corresponding southern SysCat. 
The details of this algorithm are described in Section \ref{s:clust}. The left panel of Figure \ref{f:sha_i} below shows, for example, the raw supercluster box for the \textit{Shapley Supercluster} (MSCC 389 and MSCC 401) extracted from 6dF.\\

Columns 7 and 8 of Table \ref{tab:SCsample} show, respectively, the number $N_{g_\text{box}}$ of galaxies (GalCat) and the number $N_\text{sys}$ of detected galaxy systems (SysCat) within each supercluster box either from \citet{ir2020} or the ones we detected from 2dF or 6dF; column 9 shows the survey from which the galaxy sample for each supercluster was drawn. In most cases, the number of systems identified within each supercluster box (in any of the SDSS, 2dF or 6dF regions) significantly exceeds the number of its member Abell/ACO-clusters (compare columns 6 and 8 of Table \ref{tab:SCsample}). \\

\subsubsection{Luminosity limits of the sub-samples}
\begin{figure}[t] 
\centering
 \includegraphics[trim={3cm 0cm 3cm 0.5cm},clip, width=\columnwidth]{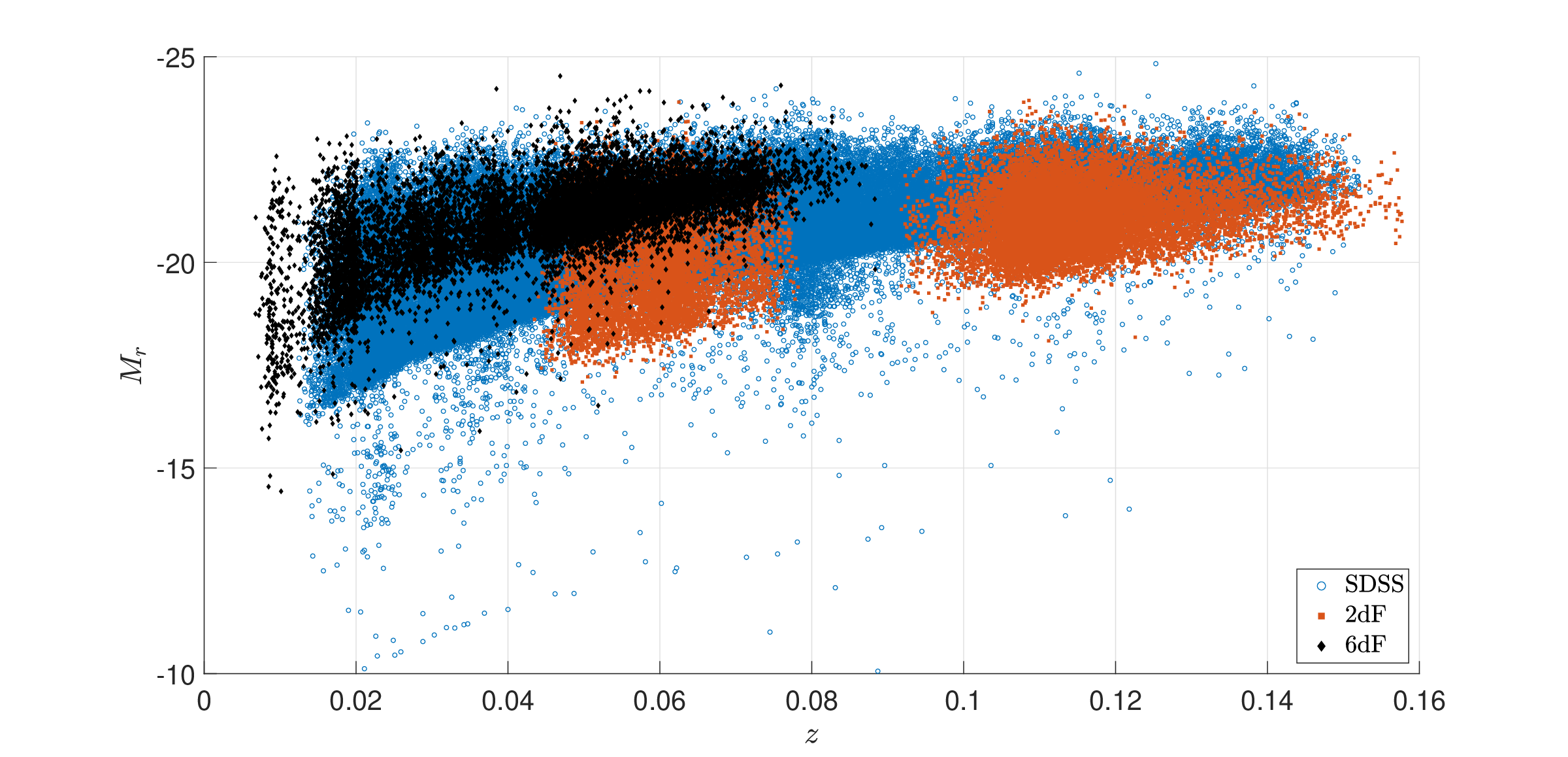} 
\caption[]{Distribution of $M_r$ galaxy absolute magnitudes as a function of redshift. Included here are the three galaxy samples taken from the SDSS (blue circles), 2dF (red boxes) and 6dF (black diamonds) Surveys.}
\label{f:M_z}
\end{figure}

{Although the three sub-samples are not homogeneous in depth, they were selected to identify galaxy systems with enough membership completeness for estimating their dynamical parameters necessary for our study. Figure \ref{f:M_z} shows the absolute magnitudes $M_r$ of our total sample (SDSS + 2dF + 6dF sub-samples) of galaxies as a function of redshift.} \\

{For SDSS, the faint limit for $M_r$ is mainly due to $r_\mathrm{pet}=17.77$, which corresponds to SDSS DR7 data, while we used SDSS DR13 which contains spectra for some fainter objects below this limit. 
This apparent magnitude limit corresponds to $M_r - 5 log(h) = -20.7$, or $M_r = -21.5$ for our assumed cosmology, at the redshift limit of our sample. This corresponds to about $M_r^*$ for early-type galaxies \citep[\textit{e.g.,}][]{na2003}.
For 2dF, the limit in apparent magnitudes is fainter and less sharp ($r_F\sim 18.3$, corresponding to $M_r = -21.0$, taking color correction from $r_F$ to $r$ to be very small), a limit slightly fainter than the one for SDSS as be seen for the two red patches along the $z$ axis ($0.045\leq z \leq 0.075$ and $0.095\leq z \leq 0.150$) that cover our used data.
The case of 6dF (black points) is particular: since this survey is much shallower than the others, having a limiting apparent magnitude of only $r_F \sim 15.6$, we used these data only to study nearby superclusters ($z\leq 0.08$, which correspond to a luminosity limit of $M_r = -22.2$).} 
{Thus, in the worst case of the redshift limits of our sample we have at least galaxies brighter than $M_r^*$ as potential members of the galaxy systems we detect. This is important to guarantee that the calculated dynamical parameters are reliable for our analysis. 
We will come back to the discussion of uncertainties in dynamical parameters below.} \\

\section{Analysis tools}\label{s:tools}
{The methodology used in the present work was divided in 3 sections: the basic analysis tools, the clustering algorithms and characterization of systems and structures, and the strategy to identify the \emph{cores}. Here, we start with the basic tools. }

\subsection{Distances and projection angle}
For two galaxy systems observed with angular separation $\theta_{jk}$ (in radians) and redshifts $z_j$ and $z_k$, the physical distance $r_{jk}$ between them and their projected separation $R_{jk}$ in the sky-plane (both defined in units of $h_{70}^{-1}$ Mpc) are given, respectively, by
\begin{equation}\label{dij}
r_{jk}=\sqrt{D^2_\text{c}(z_j)+D^2_\text{c}(z_k)-2D_\text{c}(z_j)D_\text{c}(z_k)\cos{\theta_{jk}}},
\end{equation}
and,
\begin{equation}\label{Rij}
R_{jk}\simeq\theta_{jk}D_A(\bar{z})=\theta_{jk}\frac{D_\text{c}(\bar{z})}{1+\bar{z}},
\end{equation}
where $D_A(z)$ is the angular diameter distance \citep[\textit{e.g.},][]{hog2000}, and $\bar{z}$ the average redshift of the systems. \\

Furthermore, if $z_j\leq z_k$, the tangent of the projection angle $\chi$ between the separation vector along $r_{jk}$ and the sky-plane, at the midpoint between the pair of systems, is  
\begin{equation}\label{chi}
\tan{\chi}=\frac{z_k-z_j}{2z_j\tan{(\theta_{jk}/2)}},
\end{equation}
with $0\leq\chi\leq\pi/2$ \citep[\textit{e.g.},][]{st1977}. As can be seen below, this angle can be used to test the dynamical state of the pair of systems. 

\subsection{Binding Tests} \label{g_lig}
\subsubsection{Pairwise gravitational binding}
{Consider} two galaxy systems of masses $m_j$ and $m_k$, with redshifts $z_j\leq z_k$ and separated by a distance $r_{jk}$. Assuming linear orbits for the systems, \textit{\textit{i.e.}}, with no rotations or discontinuities around the center of mass, one can use the Newtonian energy criterion, $\upsilon_{jk}^2\leq 2G\mathcal{M}/r_{jk}$, where $\upsilon_{jk}$ is the relative velocity between systems, $G$ the universal gravitational constant and $\mathcal{M}=m_j+m_k$ the pair total mass, to check the state of gravitational binding of the pair \citep[\textit{e.g.},][]{br1982}. Thus, since $R_{jk}=r_{jk}\cos{\chi}$ is the projected separation (in the sky-plane) between systems, and $\upsilon_{r_{jk}}=\upsilon_{jk}\cos{\chi}\approx H_0\,\ r_{jk}$ their relative radial velocity, then the pair binding criterion can be evaluated observationally in the form:
\begin{equation}\label{lig1}
\upsilon_{r_{jk}}^2R_{jk}\leq 2G\mathcal{M}\sin^2{\chi}\cos{\chi}.
\end{equation}  
 
\subsubsection{Gravitationally bound structures} 
An alternative method can be tested to check the gravitational binding state of a `cluster of systems' as a whole, if the individual masses the member systems are known. A structure composed of $N$ objects of masses $m_j$, peculiar velocities $\upsilon_j$ {with} respect to its center of mass, and relative separations $r_{jk}$ between them, will be gravitationally bound if $K\leq |W|$, where $K$ and $W$ are the internal kinetic and potential energies, respectively, so that 
\begin{equation}\label{K<W}
\frac{1}{2}\sum_{j=1}^N\upsilon_j^2m_j\leq \frac{1}{2}\sum_{j\neq k}^N \frac{Gm_jm_k}{r_{jk}}.
\end{equation}
The above inequality can be expressed in the observationally more practical form \citep[\textit{e.g.},][]{sc2015}:
\begin{equation}\label{ligp}
\frac{1}{2}\sigma_\text{sys}^2\leq \frac{G\mathcal{M}}{r_G},
\end{equation}
where $\sigma_\text{sys}^2:=\left\langle \upsilon_i^2 \right\rangle$ is the squared velocity dispersion of systems inside the structure, $\mathcal{M}:=\sum m_k$, and 
\begin{equation}\label{r_G}
r_G := 2\mathcal{M}^2\left( \sum_{i\neq j} \frac{m_j m_k}{r_{jk}} \right)^{-1},
\end{equation}
is the gravitational radius of the structure. \\

Since only the line-of-sight (radial) component of the velocity, $\upsilon_{r_j}$, can be estimated for each galaxy system, one can, to a first approximation, assume that $\sigma_\text{sys}^2= \beta\left\langle \upsilon_{r_j}^2 \right\rangle$, with $\beta\approx 3$ assuming a quasi-Maxwellian distribution of peculiar velocities of the systems inside the structures \citep[see][]{ba1996} or $\beta\approx 2.5$ asuming a weak anisotropy in the system velocity distributions \citep[similar to that used for galaxy velocity distributions, \textit{e.g.},][]{tl2015}. Thus, the binding condition \eqref{ligp} can be evaluated in the form
\begin{equation}\label{lig2}
\beta\sigma_{\upsilon_\text{sys}}^2 r_G \leq 2G\mathcal{M},
\end{equation}
where, $\sigma_{\upsilon_\text{sys}}^2=\left\langle \upsilon_{r_j}^2 \right\rangle$ is the line-of-sight squared velocity dispersion of galaxy systems inside the structure.

\subsection{Future virialized structures} \label{f_vir}
Following the strategy adopted by \citet{du2006}, \citet{ch2015} and references therein, and thinking of each galaxy structure as {an} overdense region of mean mass density $\rho_\text{ov}$, one can define the ratio 
\begin{equation}\label{chon1}
\mathcal{R}\equiv\frac{\rho_\text{ov}}{\rho_\text{b}},
\end{equation}
between the overdensity and the local background\footnote{see Section \ref{core_sel} for more details on the estimation of $\rho_\mathrm{b}$ here.} mean mass density $\rho_\text{b}$, as well as the the density contrast
\begin{equation}\label{chon2}
\Delta_\text{cr}\equiv\frac{\rho_\text{ov}}{\rho_\text{cr}}-1,
\end{equation}
of the overdensity  with respect to the critical density of the Universe, $\rho_\text{cr}=3H^2(z)/8\pi G$, where $H(z)=H_0E(z)$ is the Hubble parameter, \textit{i.e.}, the Hubble constant at redshift $z$.\\

It is possible to use physically motivated density criteria to decide which structures --{on scales} of superclusters or \emph{cores}-- will be able to survive the cosmic expansion and become virialized systems in the future. Assuming spherically symmetric overdense regions, the density criteria are based on theoretical estimates of the mean density that must be enclosed by its last --or `critical'-- shell, at a given cosmological epoch, so that it remains gravitationally bound to the overdensity in a future dominated by dark energy \citep[\textit{e.g.},][]{ch2002,lh2002,du2006}. Using the spherical collapse model in an $\Lambda$CDM scenario, \citet{ch2015} estimated, for various cosmologies, threshold values for $\mathcal{R}$ and $\Delta_\text{cr}$ that characterize structures that are currently {at} turn-around {\citep[\textit{i.e.}, structures that have already decoupled from the Hubble flow and are at rest in the Eulerian frame of reference, so they are now beginning to collapse,][]{ch2015}} compared to those that will collapse marginally. Likewise, \citet{du2006} estimated very consistent values at the current time for $\mathcal{R}$ and $\Delta_\text{cr}$ within the last layer that will eventually stop its expansion. The threshold values of these two parameters are shown in Table \ref{tab:R_d}. \\ 

Although the threshold values for $\mathcal{R}$ and $\Delta_\text{cr}$ parameters are valid only for spherical overdensities, they are approximately reasonable {for other} realistic structures: these density criteria have been successfully used to define \emph{bound structures} in future extended $N$-body cosmological simulations \citep{du2006}; to build catalogues of zones ---from the SDSS-DR7 region--- that will become \emph{future virialized structures} \citep[FVSs,][]{lu2011}; to study whether currently known superclusters will survive cosmic expansion \citep[the \emph{superstes-clusters},][]{ch2015}; and to identify and study the properties of \emph{quasi-spherical superclusters} also in SDSS-DR7 region \citep[\textit{e.g.,}][]{he2022}. {As stated in \citet{san2023}, a definition of ``superclusters'' based on large gravitationally bound structures most often leads to the identification of the central regions of the superclusters identified by other criteria like overdensities in galaxy or light distribution and converging peculiar velocity fields regions. Thus, the \emph{cores} we define here may not be confused with ordinary superclusters, but are a generalization of the bound or future virialized structures described above}.

\begin{table}
\caption{Current-epoch threshold values for density ratio ($\mathcal{R}$) and density contrast ($\Delta_\text{cr}$) characterizing structures at turn-around and those that are marginally collapsing, assuming a flat cosmology with $\Omega_{m,0}+\Omega_{\Lambda,0}=1$, where $\Omega_{m,0}=0.3$, $\Omega_{\Lambda,0}=0.7$ and $h_{70}^{-1}=0.7$.}
\label{tab:R_d}
\begin{tabular}{lcccc}
\hline 
\hline
   \multicolumn{1}{c}{Reference} &
   \multicolumn{2}{c}{at turn-around} & 
   \multicolumn{2}{c}{future collapse} \\
 & $\mathcal{R}$ & $\Delta_\text{cr}$ & $\mathcal{R}$ & $\Delta_\text{cr}$\\

\hline
  Dünner et al. 2006 & \,\ - & - & 7.88 & 1.36\\
  Chon et al. 2015 & 12.15 & 2.65 & 7.86 & 1.36 \\
\hline
\end{tabular} \end{table}

\subsection{Percolation and FoF algorithm}
The galaxy clustering, present in a hierarchical way at different scales, can be studied by percolation theory \citep[\textit{e.g.},][]{st1979,sh1983}, introduced in cosmological studies by \citet{zel1982}, \citet{me1983} and \citet{ei1984}. In this context, the set of coordinates of galaxies and/or systems constitutes the point distribution space where the clustering will be analyzed.\\

The well-known Friends-of-Friends (FoF) algorithm is a percolation technique that uses a single input parameter, the \textit{linking length} $\varepsilon$, as a distance criterion to link points and detect agglomerates in a data space: two points $p$ and $q$ are linked to each other (`friends') if the distance between them is $d_{pq}\leq\varepsilon$; also a third point $s$, such that $d_{ps}>\varepsilon$, is linked {to both $p$ and $q$} if $d_{qs}\leq\varepsilon$ (`friend of friend'). The set of {all points that are mutually friend constitute a `cluster'}. We will use the term `cluster', in quotes, to refer to any agglomeration of points in a generic data space and avoid confusion with conventional {clusters} of galaxies. If $\varepsilon$ is very small, the number of `clusters' will be reduced to those with the highest density of points. As $\varepsilon$ gets large, the dense `clusters' will begin to link to each other through their boundary points and sparse points in the data volume, so the number of detected `clusters' will begin to decrease until, on some large scale, the entire sample percolates \citep[\textit{e.g.},][]{ei1984}. The value $\varepsilon=\varepsilon_c$ for which the number of detected `clusters' in the percolation process is maximized is called critical linking length \citep[or critical percolation radius, \textit{e.g.},][]{sh1983,chow2014}.\\

The FoF technique is attractive, among other things, because it produces a single `cluster' catalogue for each linking volume and does not assume any particular shape or geometry for the `clusters' \citep[\textit{e.g.},][]{ber2006}. In astronomy, the FoF algorithm has often been used to detect galaxy clusters in redshift surveys \citep[\textit{e.g.},][]{hu1982,ber2006}, superclusters or filamentary structures in samples of galaxies and clusters \citep[\textit{e.g.},][]{ei1984,ei2001,ca2002,chow2014}, and identify dark matter halos in $N$-body simulations \citep[\textit{e.g.},][]{da1985}.

\subsection{The DBSCAN algorithm}
Density-Based Clustering is a family of unsupervised learning methods capable of identifying distinctive `clusters' in a data space. The method is based on the idea that a `cluster' is a region of high density of points, separated from other `clusters' by neighboring regions of low density of points, typically considered noise/outliers \citep[\textit{e.g.},][]{sa2011,kr2011}. In this work we use the Density-Based Spatial Clustering of Applications with Noise \citep[DBSCAN,][]{es1996}, one of the most popular and cited clustering algorithms in the scientific literature. This algorithm is advantageous because it does not require knowing, as an input parameter, the number $k$ of `clusters' to be detected (unlike $k$-\textit{means} or $k$-\textit{medoid} algorithms). Rather, it determines this parameter automatically from the data set ($\mathcal{D}$), the radius of the neighborhood around each point ($\varepsilon$), and the minimum number of points in each `cluster' ($N_{\text{min}}$). Once the input parameters $(\mathcal{D},\varepsilon,N_{\text{min}})$ are established, DBSCAN searches for `clusters' according to the following definitions:\\ 

Definition 1: ($\varepsilon$-neighborhood) The $\varepsilon$-\textit{neighborhood} of a point $p$, denoted by $N_{\varepsilon}(p)$, is defined by $N_{\varepsilon}(p)=\left\lbrace q\in \mathcal{D}| d_{pq}\leq \varepsilon \right\rbrace$, where $d_{pq}$ is the distance between $p$ and $q$. \\

Definition 2: (directly density-reachable) A point $p$ is \textit{directly
density-reachable} from a point $q$ with respect to $\varepsilon$ and $N_\text{min}$ if: (i) $p\in N_{\varepsilon}(q)$ and, (ii) $|N_{\varepsilon}(q)|\geq N_\text{min}$.\\

Definition 3: (density-reachable) A point $p$ is \textit{density-reachable} from a point $q$ with respect to $\varepsilon$ and $N_\text{min}$ if there is a chain of points $p_1,..., p_n$, $p_1=q$, $p_n=p$ such that $p_{i+1}$ is directly
density-reachable from $p_i$.\\

Definition 4: (density-connected) A point $p$ is \textit{density-connected} to a point $q$ with respect to $\varepsilon$ and $N_\text{min}$ if there is a point $s$ such that both, $p$ and $q$ are density-reachable from $s$ with respect to $\varepsilon$ and $N_\text{min}$.\\

Definition 5: (`cluster') Let $\mathcal{D}$ be a database of points. A \textit{`cluster'} $C$ with respect to $\varepsilon$ and $N_\text{min}$ is a non-empty subset of $\mathcal{D}$ satisfying the following conditions: (i) $\forall p,q$: if $p\in C$ and $q$ is density-reachable from p wrt. $\varepsilon$ and $N_\text{min}$, then $q\in C$; (ii) $\forall p,q \in C$: $p$ is density-{connected} to $q$ with respect to $\varepsilon$ and $N_\text{min}$.\\

Definition 6: (noise) Let $C_1,...,C_k$ be the `clusters' of the database $\mathcal{D}$ with respect to parameters $\varepsilon_i$ and $N_{\text{min}_i}$, $i=1,...,k$. Then the \textit{noise} is defined as the set of points in the database $\mathcal{D}$ not belonging to any cluster $C_i$, \textit{i.e.} $\text{noise}=\left\lbrace p\in \mathcal{D}|\forall i: p\notin C_i\right\rbrace$.\\

As a result, DBSCAN allows to discover `clusters' with arbitrary shapes (spherical, linear, elongated, etc., depending on the chosen $d_{pq}$ metric function), using few input parameters and with good efficiency in large databases \citep[see][]{es1996}, so that it can be successfully used for astronomical analysis. The DBSCAN technique is basically {a} percolation-based algorithm for linking points, supplemented with sufficient density criteria to define agglomerates and noise. A similar extended percolation method using the density field instead of points was used by \citet{ei2018} to study the connectivity of over- and under-dense regions in the cosmic web.\\

\section{Clustering analysis} \label{s:clust}
\subsection{Detection of galaxy systems in the southern sample}
The detection of systems in superclusters (from the 2dF or 6dF regions) to compile the {southern} SysCat catalogue was performed by an automated algorithm based on those presented by \citet{bi2006} and \citet{ir2020}, executing the following steps in each box:

\begin{itemize}

\item[(1)] The galaxies in the supercluster box were represented by the set of points with coordinates $\mathcal{D}=\lbrace(\alpha_k, \delta_k, 1000z_k), \,\ k=1,...,N_{g_\text{box}}\rbrace$, compiled in its southern GalCat catalogue. The factor of 1000 {was applied to redshift $z$ values} to be comparable to the sky coordinate values $\alpha$ and $\delta$. The set of points in this form represents a pseudo-three-dimensional space where the --dimensionless- distances can be estimated with the metric
\begin{equation}
d_{jk}=\sqrt{(\Delta\alpha_{jk}\cos{\bar{\delta}})^2+(\Delta\delta_{jk})^2+(1000\Delta z_{jk})^2}
\end{equation}
where $\Delta\alpha_{jk}=\alpha_j-\alpha_k$, $\Delta\delta_{jk}=\delta_j-\delta_k$, $\Delta z_{jk}=z_j-z_k$ and $\bar{\delta}$ is the mean declination between galaxies $j$ and $k$.\\

\begin{figure}[t] 
\centering
 \includegraphics[trim={2.5cm 0cm 3.3cm 0.7cm},clip, width=\columnwidth]{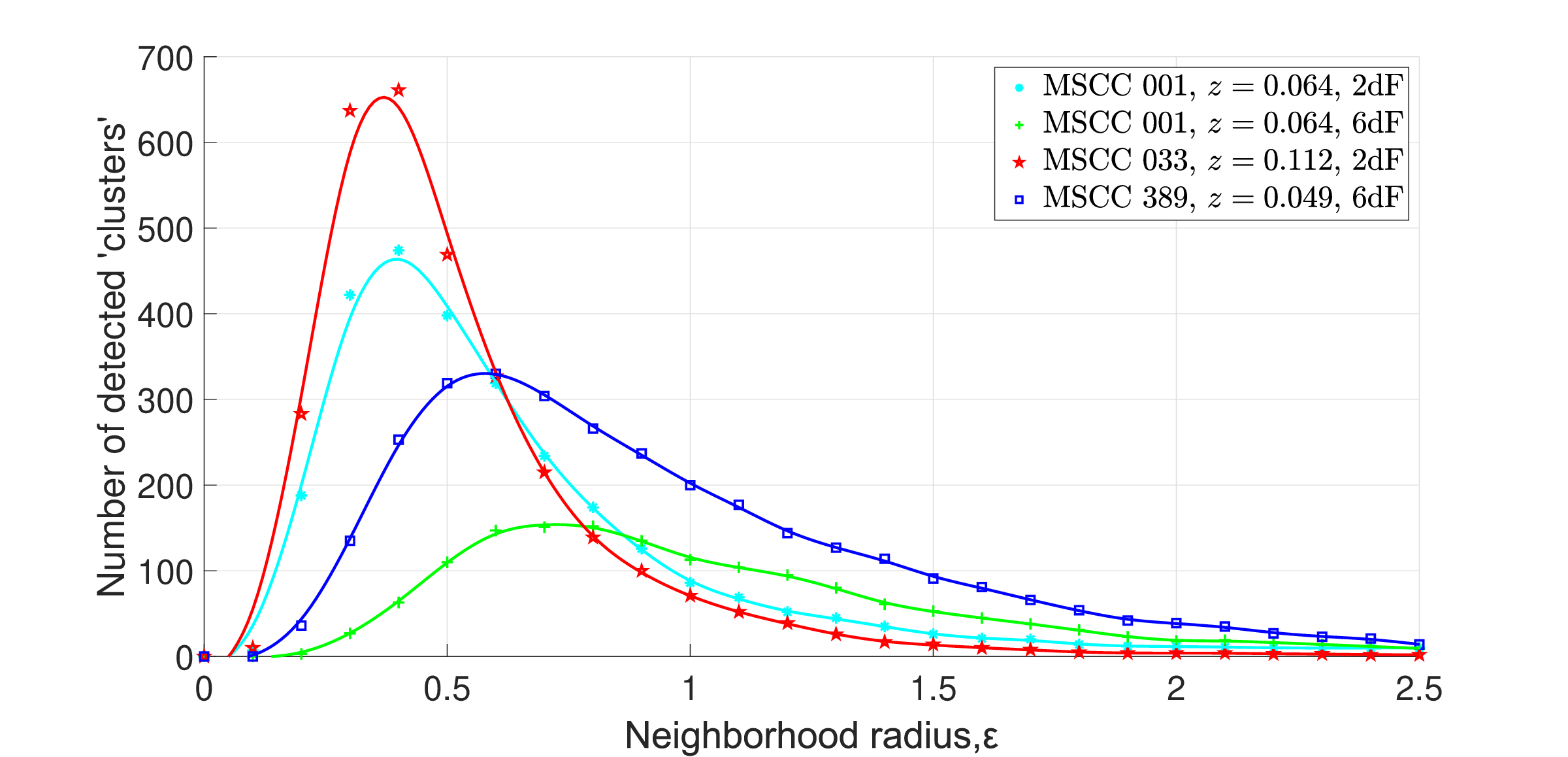} 
\caption[]{Percolation curves (PCs) obtained for three superclusters sampled in the 2dF or 6dF region, using galaxies as input data. The PCs show the variation in the number of first-order galaxy `clusters', with $N_\text{min}=3$ galaxies, detected by the DBSCAN algorithm as the neighborhood radius varies. The lines represent the smoothing-spline of data and their maxima are located at the critical neighborhood radius, $\varepsilon_c$.}
\label{f:eps}
\end{figure}

\item[(2)] A DBSCAN-based algorithm, with input parameters $(\mathcal{D},\varepsilon,3)$, was applied iteratively for a wide range of neighborhood radii, $0\leq \varepsilon\leq 2.5$, to analyze the percolation properties of galaxies and determine the --dimensionless-- critical neighborhood radius $\varepsilon_c$. \\

Figure \ref{f:eps} shows an example of four percolation curves (PCs, \textit{i.e.}, number of detected `clusters' \textit{vs.} neighborhood radius) obtained for three superclusters. The PCs were obtained with the data from the best box (in 2dF or 6dF) for each supercluster, except for one of them where the PCs were obtained with the data from both boxes (in 2dF and 6dF) for comparison. It can be seen that, in addition to its dependence on redshift, the radius $\varepsilon_c$ depends on factors such as the density and homogeneity of the survey in the region of a given supercluster. The critical neighborhood radii obtained for the 8 superclusters in the 2dF and 6dF regions have mean and median values of $0.6$ and $0.5$, respectively, similar to the dimensionless neighborhood radius value obtained by \citet{ei1984} to detect galaxy clusters (and groups) by FoF.\\

\item[(3)] The radius $\varepsilon_c$ was established as the most appropriate to find `clusters' with $N_\text{min}=3$ in the box. The projected centroid, $c_i=(\alpha_i, \delta_i)$, and line-of-sight (radial) velocitiy, $\upsilon_{r_i}$, of each detected $i$-`cluster' was initially estimated as median values of its member galaxies. Not all the `clusters' identified in this step are necessarily physical galaxy systems, but around their positions it is more likely to find them.\\

\item[(4)] All galaxies contained within an initial projected aperture $R_{a_i}=1$ $h_{70}^{-1}$ Mpc from each centroid $c_i$ and whose radial velocities were in the interval $\upsilon_{r_i}\pm 3S_a$, with $S_{a_i}=1000$ km s$^{-1}$, were taken. That is, galaxy cylinders oriented along the line-of-sight in redshift space with radius $R_{a_i}$ and depth $6S_{a_i}$ were taken centered on $(c_i,\upsilon_{r_i})$.\\

\item[(5)] The cylinders with 5 or more galaxies were accepted as candidates for galaxy systems, while the others were rejected. For these candidates, line-of-sight velocities $V_{\text{LOS}_i}$ and velocity dispersions $\sigma_{\upsilon_i}$ were estimated using Tukey's biweight method \citep[see][]{br1990}, and centroids were recalculated. \\

\item[(6)] The virial mass of each candidate within the cylinder was determined as \citep[\textit{e.g.},][]{bi2006, tl2015}:
\begin{equation}\label{M_vi}
\mathcal{M}_{\text{vir}_i}=\frac{\beta \pi}{2G}\sigma_{\upsilon_i}^2R_{\text{vp}_i}, 
\end{equation} 
with $\beta$ being, as above, an anisotropy parameter for the galaxy velocity distributions, and $R_{\text{vp}_i}$, the projected mean radius, calculated in the form
\begin{equation}
R_{\text{vp}_i}=\frac{N_i(N_i-1)}{\sum_{j<k} 1/R^{(i)}_{jk}},
\end{equation}
where $R^{(i)}_{jk}$ is the projected distance (in the sky-plane) between pairs of galaxies and $N_i$ the number of them within the $i$-th cylinder. Furthermore, assuming a spherical model for nonlinear collapse with virialization density $\rho_{\text{vir}}=18\pi^2[3H^2(z)/8\pi G]$ \citep[\textit{e.g.},][]{bry1998}, the virial radius is then 
\begin{equation}\label{R_v}
r_{\text{vir}_i}^3=\frac{3\mathcal{M}_{\text{vir}_i}}{4\pi\rho_{\text{vir}_i}}=\frac{\beta\sigma_{\upsilon_i}^2 R_{\text{vp}_i}}{18\pi H^2(z_i)}.
\end{equation}

\item[(7)] For each candidate, the aperture $R_{a_i}$ was updated to the corresponding calculated $r_{\text{vir}_i}$ value, the median radial velocity $v_{r_i}$ to $V_{\text{LOS}_i}$, and $S_a$ to $\sigma_{\upsilon_i}$, defining a new cylinder (including or excluding galaxies as the case may be). The process, from step (4) to (7), was repeated iteratively unti finding the $r_{\text{vir}_i}\sim R_{a_i}$ convergence.\\

Here, we assume an $r_{\text{vir}_i}\sim R_{a_i}$ convergence when
\begin{equation}
\frac{|r_{\text{vir}_i}- R_{a_i}|}{R_{a_i}}\leq 0.05,
\end{equation}
that is, if the relative difference between $r_{\text{vir}_i}$ and $R_{a_i}$ was less than or equal to 5\%.\\

\item[(8)] The candidates for which $r_{\text{vir}_i}$ converged before 20 iterations were accepted as real galaxy systems and their dynamical properties correspond to those estimated at the end of the last iteration, while those that did not converge were rejected. All the systems for which convergence $r_{\text{vir}_i}\sim R_{a_i}$ occurred achieved it well before 20 iterations.
\end{itemize} 

Accepted systems become part of the corresponding southern SysCat catalogue of the respective supercluster, indicating for each member system its centroid $(\text{RA},\text{Dec},\bar{z})$, position of its Brightest Cluster Galaxy (BCG), the number of member galaxies ($N_g$), its line-of-sight velocity ($V_\text{LOS}$), its radial galaxy velocity dispersion ($\sigma_\upsilon$), its virial mass ($\mathcal{M}_\text{vir}$) and its projected ($R_{\text{vp}}$) and virial ($r_\text{vir}$) radii.\\

\begin{figure}[t] 
\centering
 \includegraphics[trim={8cm 0cm 8.7cm 1cm},clip, width=\columnwidth]{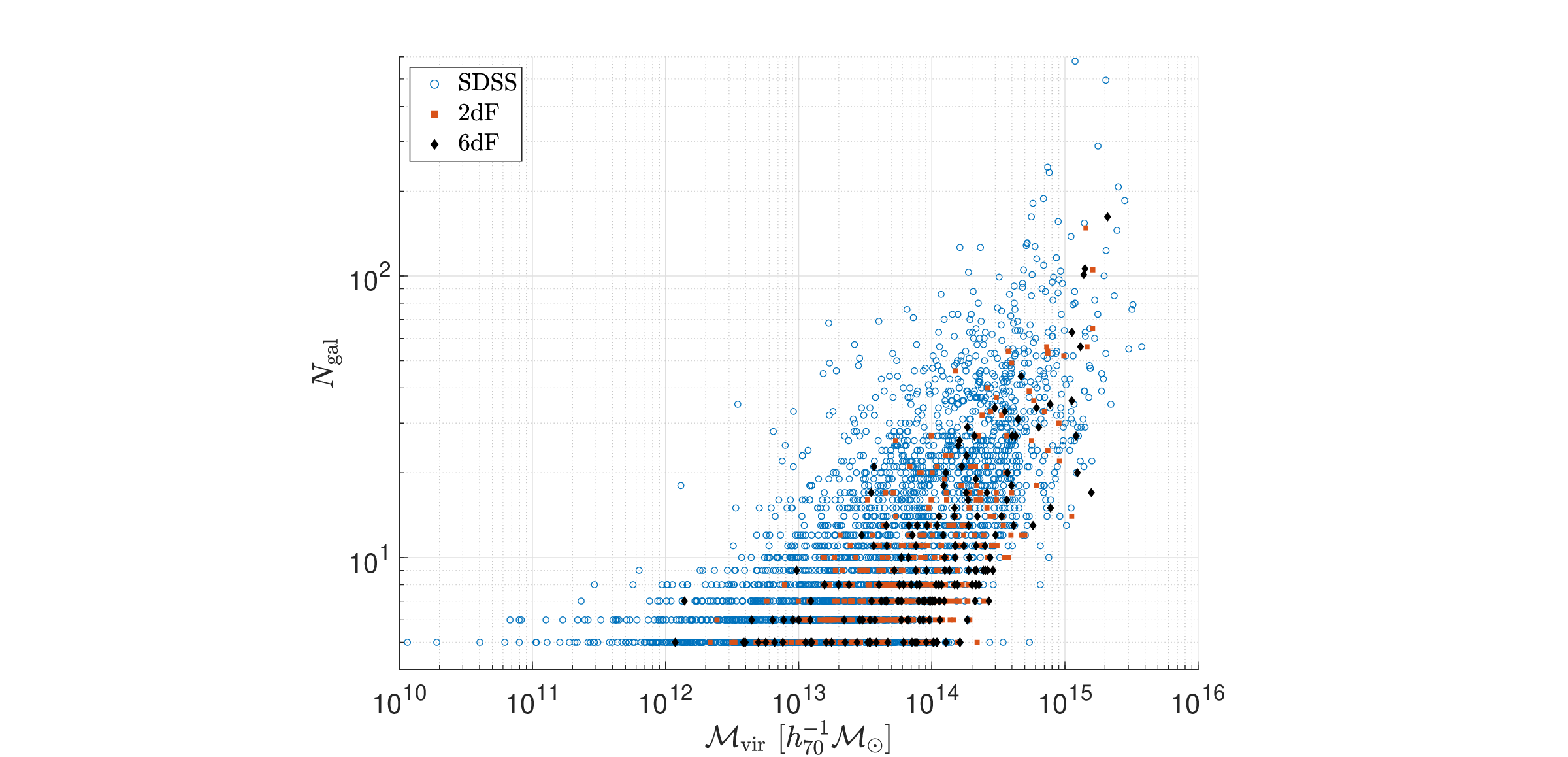} 
\caption[]{Richness as a function of mass for the complete sample of 3,337 SysCat systems identified in the SDSS, 2dF, and 6dF regions.}
\label{f:N_M}
\end{figure}

{Figure \ref{f:N_M} shows the distribution of richness versus mass for our total sample of 3,337 SysCat systems (\textit{i.e.}, combining those identified in the SDSS, 2dF and 6dF regions). This distribution is very similar to the one obtained by \citet{tm2014} for systems with $N_\mathrm{gal} \geq 5$.} 

\subsubsection{FoG-effect correction}
For each accepted system, a simple correction for the Finger-of-God effect \citep[FoG, \textit{e.g.},][]{co2012} was performed by adjusting the position of its member galaxies in the final cylinder, so that their comoving distances were rescaled to stay within the calculated virial radius. Thus, if $D_{\text{c}_k}=D_\text{c}(z_k)$ is the initial comoving distance of the $k$-th member galaxy, at redshift $z_k$, in the $i$-th system, its rescaled comoving distance is
\begin{equation}\label{FoG}
D_{\text{c}_k}'=\frac{2r_{\text{vir}_i}}{\epsilon} \left( D_{\text{c}_k}-D_\text{c}^{(i)} \right)+D_\text{c}^{(i)},
\end{equation}

where $D_\text{c}^{(i)}=D_\text{c}(V_{\text{LOS}_i}/c)$ is the comoving distance to the centroid of the system, and $\epsilon=\max_k{\left\lbrace D_{\text{c}_k} \right\rbrace}-\min_k{\left\lbrace D_{\text{c}_k} \right\rbrace}$, with $k=1,...,N_i$, is the distance (in the line of sight) between the nearest and the most distant galaxy in the system. The surface pressure term correction based on the concentration parameter was not applied here, however {a virial} approximation is sufficient for the objective of this work \citep[\textit{e.g.},][]{ir2020}.\\

\begin{figure*}
\includegraphics[trim={2.7cm 1cm 2.7cm 3cm},clip,width=\textwidth]{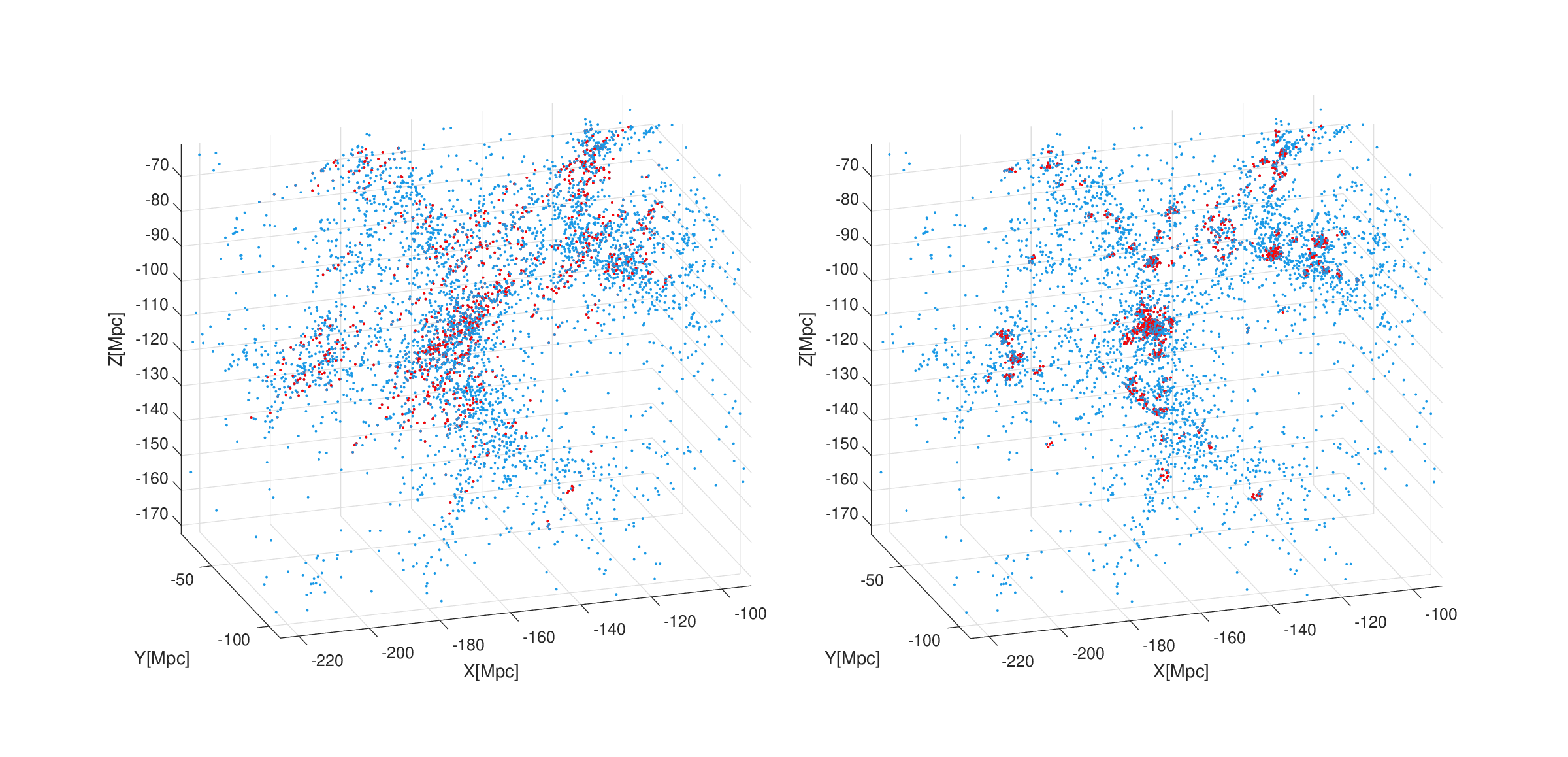}  
 \caption{The initial supercluster box for the \textit{Shapley Supercluster} (MSCC 389 and MSCC 401): each dot (4,649 in total) represents an observed 6dF galaxy in 3D-rectangular coordinates. On both panels the red dots represent the members of galaxy systems. \textit{Left}: galaxy positions before the FoG-effect correction. \textit{Right}: galaxy positions after FoG-effect correction.}
 \label{f:sha_i}
\end{figure*}

The FoG-effect can be seen in the three-dimensional distribution of galaxies obtained by transforming their coordinates from raw redshift-angular (provided by the source survey) to rectangular directly through {eq.} \eqref{xyz}, as shown in the left panel of Figure \ref{f:sha_i} for the \textit{Shapley Supercluster}. The {regions of red points} are galaxy systems which appear elongated in the line-of-sight due to the effect of the peculiar velocities of member galaxies \citep[\textit{e.g.},][]{co2012}. After the FoG-correction, re-estimating the rectangular coordinates only for member galaxies of systems using the rescaled comoving distances {from eq. \eqref{FoG} in eq. \eqref{xyz}}, the systems and their spatial distribution within the supercluster can be clearly distinguished as shown in the right panel of Figure \ref{f:sha_i}.\\

\subsubsection{System mass uncertainties}
{Although \citet{ir2020} did not provide uncertainties for the dynamical masses of their systems, we estimate such uncertainties for all SysCat systems, including both those identified by them from the SDSS galaxy sample and those identified by us from the 2dF and 6dF galaxy samples. The mass uncertainties were estimated here through simple error propagation (\textit{i.e.}, $\Delta y = [\sum (\Delta x_i\partial y/\partial x_i)^2]^{1/2}$) such that, for each $i$-system 
\begin{equation}
\Delta \mathcal{M}_{\mathrm{vir}_i}= \frac{\beta\pi}{2G}\sigma_{\upsilon_i}\sqrt{(2 R_{\mathrm{vp}_i}\Delta\sigma_{\upsilon_i})^2+(\sigma_{\upsilon_i} \Delta R_{\mathrm{vp}_i})^2}, 
\end{equation}
where $\Delta\sigma_{\upsilon_i}$ represents the uncertainty in the system velocity dispersion, taken as $\sigma_{\upsilon_i}/\sqrt{N_i}$ \citep[\textit{e.g.},][]{br1990}, and
\begin{equation}
\Delta R_{\mathrm{vp}_i}\approx \frac{2R_{\mathrm{vp}_i}^2}{N_i(N_i-1)}\times \frac{\Delta\theta (\bar{z}+1)}{D_\mathrm{c}(\bar{z})}\sqrt{\sum_{j<k} \left[\theta^{(i)}_{jk}\right]^{-6}},
\end{equation}
is the propagated uncertainty in the projected mean radius of the system (at average redshift $\bar{z}$) due to the astrometric precision $\Delta\theta$ (in radians) of the galaxy angular position on the sky-plane (\textit{e.g.}, $\Delta\theta\sim 0.1$ arsec for SDSS and $\Delta\theta\sim 0.2$ arcsec for 2dF and 6dF). Here, $\theta^{(i)}_{jk}$ represents the angular separation between pairs of observed member galaxies in the $i$-system.}\\

\begin{figure}[t] 
\centering
 \includegraphics[trim={5.5cm 0cm 2.3cm 0cm},clip, width=\columnwidth]{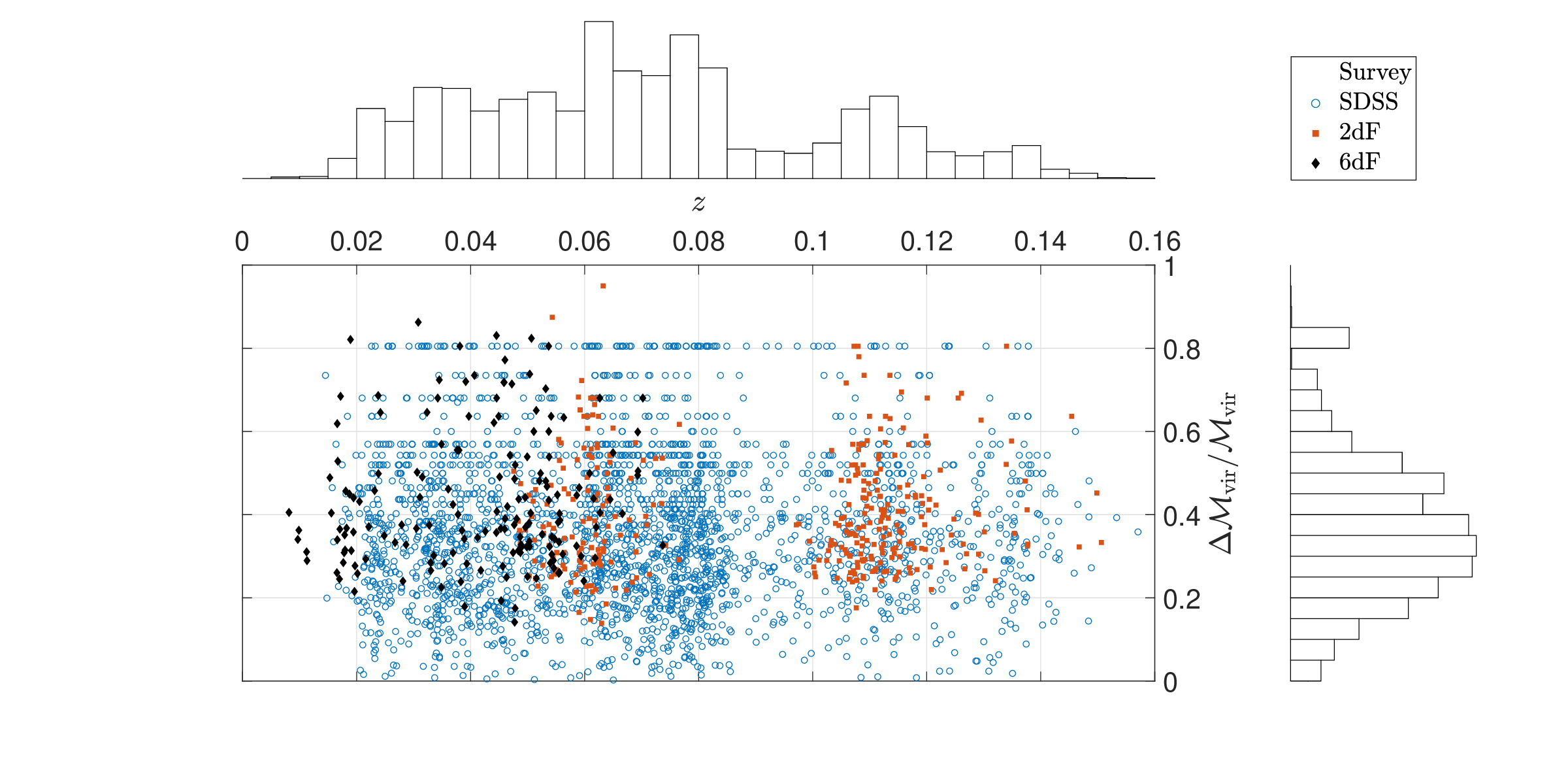} 
\caption[]{Distribution of relative mass uncertainties $\Delta \mathcal{M}_{\mathrm{vir}}/ \mathcal{M}_{\mathrm{vir}}$ as a function of redshift for SysCat systems. The marginal histograms in the horizontal and vertical axes show the total distribution of systems (SDSS+2dF+6dF) with respect to redshift and relative mass uncertainties, respectively.}
\label{f:errM}
\end{figure}

{The 
 mass uncertainties with respect to redshift for the total sample of SysCat systems is shown in Figure \ref{f:errM}. Note that there is no significant trend of the relative uncertainties (to increase or decrease) with respect to redshift. It can be seen that the uncertainties obtained are typical for this type of systems when compared with other catalogues: our relative uncertainties have mean and median values of 0.367 and 0.351, respectively (see the marginal histogram in the vertical axis of Fig. \ref{f:errM}), similar to those obtained, for example, for the GalWCat19 \citep[][]{ad2020}, which have mean and median values of 0.372 and 0.343, respectively.}\\

\begin{figure}[t] 
\centering
 \includegraphics[trim={10cm 0cm 10cm 1cm},clip, width=\columnwidth]{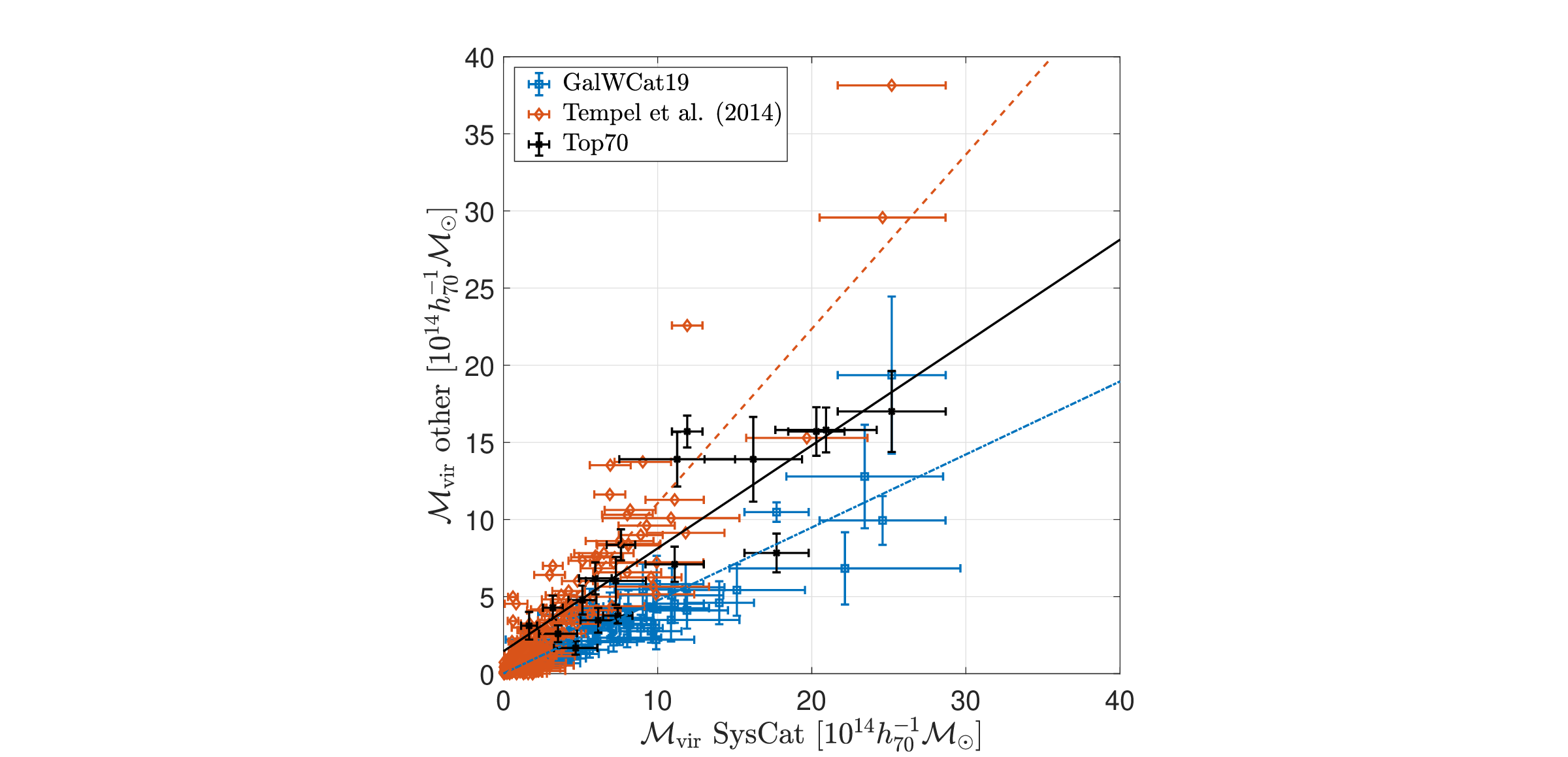} 
\caption[]{Comparison of SysCat masses with system masses from other similar catalogues: GalWCat19 \citep{ad2020} masses are $\mathcal{M}_{100}$ calculated from a NFW profile, the ones closest to $\mathcal{M}_\mathrm{vir}$; \citet{tm2014} are total masses estimated assuming a Herniquist density profile; and Top70 \citep{ca2023,zu2024} are also virial masses, estimated in a similar way but with a different (and more complete) database. SysCat and Tempel groups tend to slightly overestimate the masses with respect to Top70, while GalWCat19 slightly underestimates them}
\label{f:mass_match}
\end{figure}

{We have also compared the masses obtained by our method with those from other sources in the literature such as Top70 \citep[][]{ca2023,zu2024}, which comprise 70 nearby well sampled galaxy clusters; GalWCat19 \citep[][]{ad2020}, a catalogue with 1,800 groups up to $z=0.2$ from SDSS-DR13; and \citet{tm2014}, a flux-limited catalogue of groups from SDSS-DR10, all considering only galaxies with spectroscopic redshifts. The comparison (Fig. \ref{f:mass_match}) shows consistent mass estimates between the four samples and methods.}

\subsection{Detection of galaxy structures}\label{gsd}
The detection of structures inside superclusters of any {redshift survey} region (SDSS, 2dF or 6dF) was performed by an automated algorithm, similar to that of the detection of systems, but using the SysCat catalogues as input data. The procedure for each supercluster was:

\begin{itemize}
\item[(1)]  Using transformations {from eq.} \eqref{xyz}, the member systems in the corresponding box were represented by the set $\mathcal{D}=\left\lbrace (X_k,Y_k,Z_k),\,\ k=1,...,N_\text{sys} \right\rbrace$ of points, in rectangular coordinates, of their centroids compiled in SysCat. In this case, the set of points represents a three-dimensional space where the distances (in units of $h_{70}^{-1}$ Mpc) can be calculated with the standard Euclidean metric.\\

\begin{figure}[t] 
\centering
 \includegraphics[trim={2.5cm 0cm 3.3cm 0.7cm},clip, width=\columnwidth]{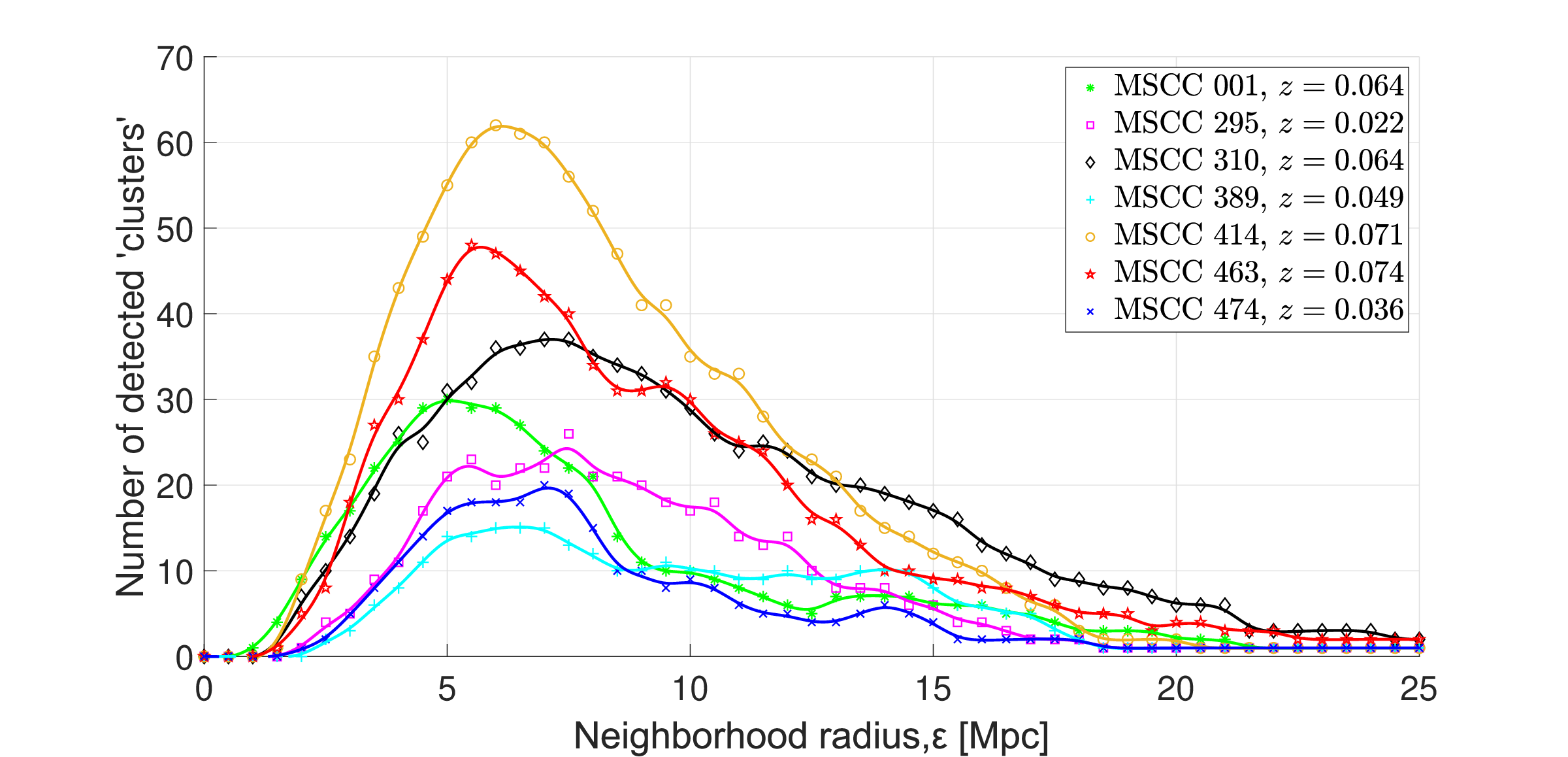} 
\caption[Percolation curves (PCs) obtained for seven of the superclusters in the sample, using galaxy systems as input data.]{Percolation curves (PCs) obtained for seven of the superclusters in the sample, using galaxy systems as input data. The PCs show the variation in the number of second-order galaxy `clusters', \textit{i.e.}, `clusters' of galaxy systems with $N_\text{min}=2$, detected by the DBSCAN algorithm as the neighborhood radius varies. The lines represent the smoothing-spline of data and their maxima are located at the critical neighborhood radius, $\varepsilon_c$.}
\label{f:eps2}
\end{figure}

\begin{figure}[t] 
\centering
 \includegraphics[trim={2.2cm 0cm 3.3cm 0.7cm},clip, width=\columnwidth]{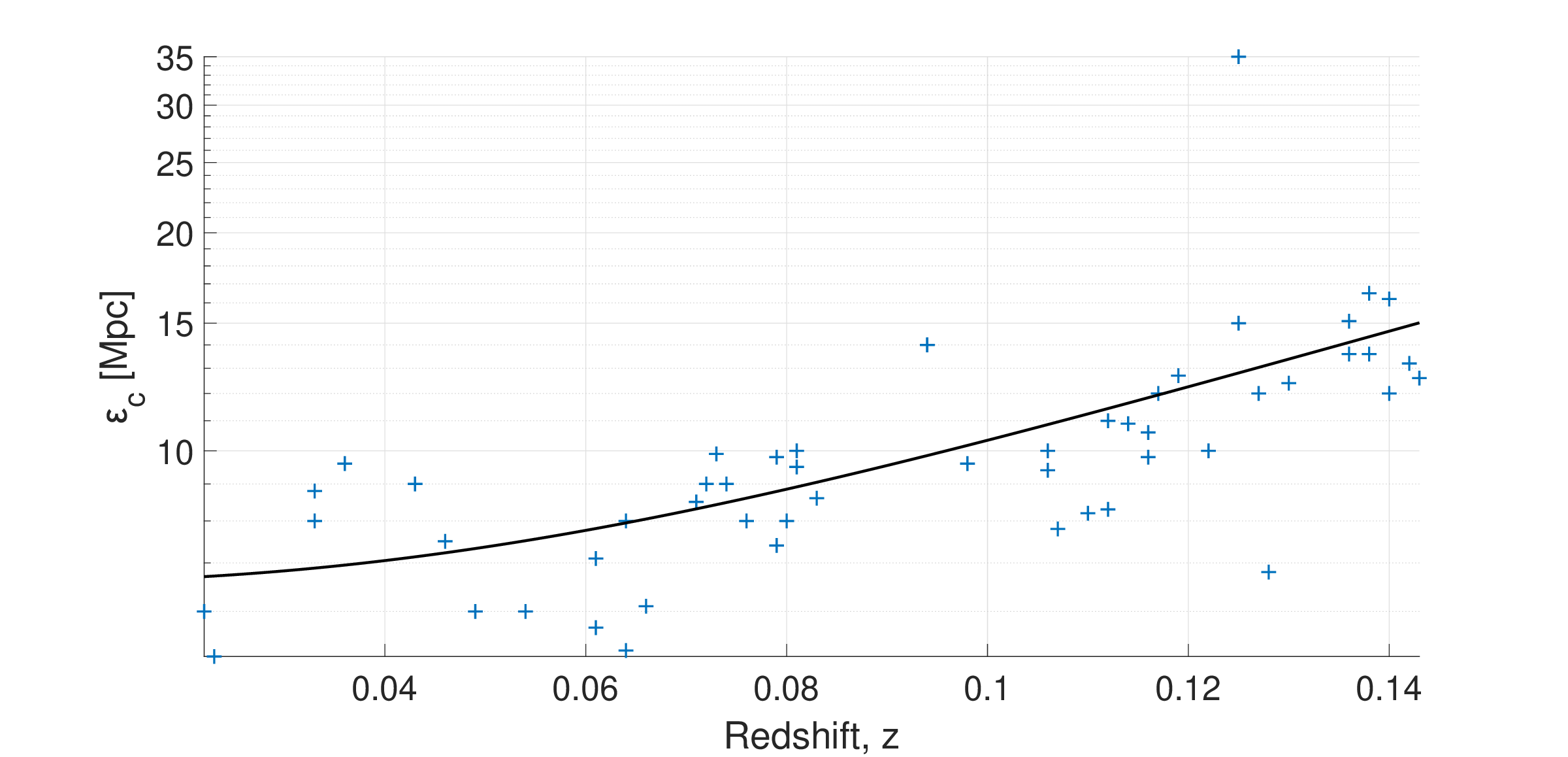} 
\caption{Percolation function (PF) obtained for the supercluster sample: critical neighborhood radius for detecting galaxy structures increases as a function of supercluster redshift. The solid black line represents the best fit (goodness, $\mathcal{R}^2_\text{det}=0.34$) to the data: a power law $az^{b}+c$ with coefficients $a=685.20$, $b=2.26$ and $c=6.58$. In general, the critical radius used to detect galaxy structures increases with mean redshift due to incompleteness and thus the loss of galaxy density.}
\label{f:PF}
\end{figure}

\item[(2)] A DBSCAN-based algorithm, with input parameters $(\mathcal{D},\varepsilon,2)$, was applied for a range of neighborhood radii, $0\leq\varepsilon\leq 25$ $h_{70}^{-1}$ Mpc, to analyze the percolation properties of galaxy systems and determine the critical radius $\varepsilon_c$.\\

Figure \ref{f:eps2} shows an example of PCs obtained for seven of the superclusters of Table \ref{tab:SCsample} sampled in some of the surveys (SDSS, 2dF or 6dF). Again, the curves reach their maximum, \textit{i.e.}, {greatest} number of detected `clusters', when $\varepsilon=\varepsilon_c$. The value of $\varepsilon_c$ obtained for each supercluster is shown in column 9 of Table \ref{tab:SCprop}. The variation of $\varepsilon_c$ with respect to the mean redshift $\bar{z}$ of the supercluster, \textit{i.e.}, the so-called percolation function \citep[PF, \textit{e.g.},][]{chow2014}, can be seen in Figure \ref{f:PF}. The critical radius that maximizes the detected amount of `clusters' of galaxy systems in superclusters is, as expected, an increasing function of their mean redshift. For the total sample, $\varepsilon_c$ has mean and median values of $10.4$ $h_{70}^{-1}$ Mpc and $9.6$ $h_{70}^{-1}$ Mpc, respectively. \\

\item[(3)] The radius $\varepsilon_c$ was established as the most appropriate to find galaxy structures with $N_\text{min}=2$ in the supercluster box. \\

The number $N_\text{str}$ of structures detected by the algorithm within each supecluster box is shown in column 10 of Table \ref{tab:SCprop}.\\
\end{itemize}
A total of 155 structures were identified in the full sample of superclusters. While in principle each SysCat-object represents a virialized --physical-- galaxy system, the structures detected at this stage are only distance-linked `clusters' of systems, but there is no guarantee that they are gravitationally bound structures or that they have enough density to survive the cosmic expansion. Several of the detected structures correspond to filaments, chains of galaxy systems that generally connected with each other if the neighborhood radius $\varepsilon$ is allowed to grow as reported by \citet{ei1984}. In what follows we will focus on selecting only those structures that meet the necessary criteria to be considered \emph{cores} of superclusters.

\begin{table*}
\resizebox{14cm}{!}{
\begin{threeparttable}
\caption{General properties of the MSCC-superclusters in the sample}
\label{tab:SCprop}
\begin{tabular}{rrrr |r@{\hspace{6pt}}c@{\hspace{6pt}}r| r |rc@{\hspace{6pt}}r| crrc}
\toprule \toprule

  \multicolumn{1}{c}{ID} &
  \multicolumn{1}{c}{RA$_{\mathrm{CM}}$} &
  \multicolumn{1}{c}{Dec$_{\mathrm{CM}}$} &
  \multicolumn{1}{c}{$z_{\mathrm{CM}}$} &
  \multicolumn{3}{c}{$\mathcal{M}^\text{sc}_\text{ext}$} &
  \multicolumn{1}{c}{$V_\text{sc}$} &
  \multicolumn{3}{c}{$\rho_\text{sc}=\mathcal{M}^\text{sc}_\text{ext}/V_\text{sc}$} &
  \multicolumn{1}{c}{$n_\text{sc}=N_{g_\text{sc}}/V_\text{sc}$}  &
  \multicolumn{1}{c}{$\varepsilon_\text{c}$} &
  \multicolumn{1}{c}{$N_\text{str}$} &
  \multicolumn{1}{c}{$N_\text{crs}$} \\

  \multicolumn{1}{c}{MSCC} &
  \multicolumn{1}{c}{[deg]} &
  \multicolumn{1}{c}{[deg]} &
  \multicolumn{1}{c}{[deg]} &
  \multicolumn{3}{c}{[$10^{14} h_{70}^{-1} \mathcal{M}_\odot$]} &
  \multicolumn{1}{c}{[$10^3$ $h_{70}^{-3
  }$ Mpc$^{3}$]} &
  \multicolumn{3}{c}{[$10^{10}h_{70}^{2}\mathcal{M}_\odot$Mpc$^{-3}$]} &
  \multicolumn{1}{c}{[$h_{70}^{3}$ Mpc$^{-3}$]} &
  \multicolumn{1}{c}{[$h_{70}^{-1}$ Mpc]} &
  \multicolumn{1}{c}{} &
  \multicolumn{1}{c}{} \\

  \multicolumn{1}{c}{(1)}  &
  \multicolumn{1}{c}{(2)}  &
  \multicolumn{1}{c}{(3)}  &
  \multicolumn{1}{c}{(4)}  &
  \multicolumn{3}{c}{(5)}  &
  \multicolumn{1}{c}{(6)}  &
  \multicolumn{3}{c}{(7)}  &
  \multicolumn{1}{c}{(8)}  &
  \multicolumn{1}{c}{(9)}  &
  \multicolumn{1}{c}{(10)}  &
  \multicolumn{1}{c}{(11)} \\

\midrule \midrule

  1 &    0.32 & -29.24 &  0.061 & 143.24 & $\pm$ &  7.38 &  150.4 &  9.5 & $\pm$ &  0.4 &  0.0391 &  5.3 &  4  & 3 \\
 27 &    9.61 & -26.97 &  0.062 &  19.59 & $\pm$ &  1.67 &   73.8 &  2.6 & $\pm$ &  0.2 &  0.0312 &  5.7 &  1  & 1 \\
 33 &   10.68 & -29.85 &  0.110 & 201.42 & $\pm$ &  9.14 &  268.3 &  7.5 & $\pm$ &  0.3 &  0.0236 &  8.3 &  5  & 5 \\
 39 &   13.18 & -11.12 &  0.054 &  41.84 & $\pm$ &  5.45 &   57.0 &  7.3 & $\pm$ &  0.9 &  0.0129 &  6.0 &  2  & 2 \\
 55 &   18.03 &  15.56 &  0.060 &  32.53 & $\pm$ &  5.95 &   24.7 & 13.1 & $\pm$ &  2.4 &  0.0210 &  7.1 &  2  & 1 \\
 72 &   25.13 &   0.00 &  0.080 &  66.25 & $\pm$ &  6.13 &   60.5 & 10.9 & $\pm$ &  1.0 &  0.0276 &  8.0 &  4  & 2 \\
 75 &   27.93 &  -1.04 &  0.086 &  72.06 & $\pm$ &  8.94 &  100.0 &  7.2 & $\pm$ &  0.8 &  0.0118 & 14.0 &  2  & 0 \\
 76 &   29.17 &  -1.64 &  0.123 &  59.68 & $\pm$ &  7.95 &  450.6 &  1.3 & $\pm$ &  0.1 &  0.0043 & 12.4 &  2  & 0 \\
117 &   51.57 & -47.59 &  0.061 &  91.11 & $\pm$ & 11.16 &  129.3 &  7.0 & $\pm$ &  0.8 &  0.0111 &  6.1 &  1  & 1 \\
175 &  125.13 &  18.78 &  0.094 &  65.23 & $\pm$ &  5.59 &  122.5 &  5.3 & $\pm$ &  0.4 &  0.0153 & 14.0 &  3  & 1 \\
184 &  130.06 &  29.95 &  0.103 &  37.19 & $\pm$ &  4.40 &  204.1 &  1.8 & $\pm$ &  0.2 &  0.0071 &  9.4 &  2  & 1 \\
211 &  149.10 &  64.08 &  0.118 &   7.44 & $\pm$ &  3.01 &   27.6 &  2.6 & $\pm$ &  1.0 &  0.0223 & 12.7 &  1  & 0 \\
219 &  153.45 &  19.20 &  0.114 &  77.87 & $\pm$ &  8.45 &  257.6 &  3.0 & $\pm$ &  0.3 &  0.0063 & 10.6 &  1  & 1 \\
222 &  156.66 &  50.04 &  0.138 &  47.46 & $\pm$ &  6.98 &  107.8 &  4.4 & $\pm$ &  0.6 &  0.0104 & 13.6 &  2  & 2 \\
223 &  153.74 &  62.78 &  0.138 &   7.23 & $\pm$ &  4.42 &   12.9 &  5.6 & $\pm$ &  3.4 &  0.0172 & 12.0 &  1  & 0 \\
229 &  157.19 &  34.24 &  0.136 &  30.21 & $\pm$ &  6.21 &  368.0 &  0.8 & $\pm$ &  0.1 &  0.0035 & 13.2 &  1  & 0 \\
236 &  160.88 &  16.12 &  0.032 & 155.02 & $\pm$ & 10.78 &  158.6 &  9.7 & $\pm$ &  0.6 &  0.0459 &  8.8 &  5  & 4 \\
238 &  158.64 &  38.40 &  0.105 & 177.01 & $\pm$ & 14.26 & 1412.2 &  1.2 & $\pm$ &  0.1 &  0.0048 &  7.8 & 10  & 4 \\
248 &  159.07 &  44.66 &  0.125 &  16.37 & $\pm$ &  2.70 &  197.8 &  0.8 & $\pm$ &  0.1 &  0.0041 & 35.0 &  1  & 0 \\
264 &  167.30 &  12.29 &  0.117 &  16.77 & $\pm$ &  7.14 &  202.4 &  0.8 & $\pm$ &  0.3 &  0.0065 &  9.8 &  1  & 0 \\
266 &  166.28 &  12.28 &  0.126 &  43.29 & $\pm$ &  7.83 &   29.5 & 14.6 & $\pm$ &  2.6 &  0.0169 & 12.0 &  1  & 1 \\
272 &  167.54 &  40.96 &  0.075 &  23.93 & $\pm$ &  2.22 &   36.3 &  6.5 & $\pm$ &  0.6 &  0.0309 &  8.0 &  1  & 1 \\
277 &  169.47 &  50.11 &  0.109 & 103.33 & $\pm$ & 10.20 &  103.8 &  9.9 & $\pm$ &  0.9 &  0.0169 &  8.2 &  2  & 3 \\
278 &  171.27 &  25.17 &  0.031 & 107.07 & $\pm$ &  4.62 &  268.7 &  3.9 & $\pm$ &  0.1 &  0.0274 &  8.0 &  4  & 3 \\
283 &  171.76 &  20.82 &  0.134 &  80.42 & $\pm$ &  9.08 &  290.3 &  2.7 & $\pm$ &  0.3 &  0.0051 & 16.5 &  4  & 3 \\
295 &  183.39 &  22.27 &  0.022 & 159.78 & $\pm$ &  6.83 &  153.8 & 10.3 & $\pm$ &  0.4 &  0.0791 &  6.0 &  2  & 2 \\
310 &  173.20 &  54.32 &  0.060 & 251.39 & $\pm$ &  9.50 &  351.5 &  7.1 & $\pm$ &  0.2 &  0.0290 &  8.0 &  6  & 5 \\
311 &  174.51 &  11.89 &  0.083 & 142.23 & $\pm$ &  7.78 &  294.4 &  4.8 & $\pm$ &  0.2 &  0.0154 &  8.6 &  3  & 3 \\
314 &  177.19 &  -1.73 &  0.079 &  26.85 & $\pm$ &  4.32 &    3.4 & 78.5 & $\pm$ & 12.6 &  0.1184 &  7.4 &  1  & 1 \\
317 &  177.01 &  -1.59 &  0.124 &  48.89 & $\pm$ &  7.14 &   77.2 &  6.3 & $\pm$ &  0.9 &  0.0085 &  6.8 &  2  & 2 \\
323 &  179.79 &  26.05 &  0.138 & 106.49 & $\pm$ & 11.73 &  174.0 &  6.1 & $\pm$ &  0.6 &  0.0098 & 16.2 &  3  & 3 \\
333 &  181.55 &  30.27 &  0.079 &  33.33 & $\pm$ &  5.05 &   12.8 & 25.9 & $\pm$ &  3.9 &  0.0829 &  9.5 &  2  & 2 \\
335 &  181.74 &  29.43 &  0.073 &  55.74 & $\pm$ &  7.83 &   57.8 &  9.6 & $\pm$ &  1.3 &  0.0369 &  9.9 &  5  & 1 \\
343 &  182.05 &  12.66 &  0.082 &  66.17 & $\pm$ &  5.16 &   21.8 & 30.3 & $\pm$ &  2.3 &  0.0781 & 10.0 &  5  & 4 \\
360 &  190.45 &  63.67 &  0.105 &  39.88 & $\pm$ &  4.65 &  299.3 &  1.3 & $\pm$ &  0.1 &  0.0064 & 10.0 &  2  & 1 \\
386 &  198.67 &  39.60 &  0.069 &  77.39 & $\pm$ &  5.58 &  146.9 &  5.2 & $\pm$ &  0.3 &  0.0181 &  9.0 &  3  & 2 \\
389 &  201.48 & -31.99 &  0.046 & 171.33 & $\pm$ &  9.01 &  191.0 &  8.9 & $\pm$ &  0.4 &  0.0198 &  6.0 &  6  & 3 \\
407 &  208.70 &  25.63 &  0.136 &  12.81 & $\pm$ &  2.19 &   92.1 &  1.3 & $\pm$ &  0.2 &  0.0068 & 15.1 &  1  & 0 \\
414 &  212.27 &  26.76 &  0.067 & 298.68 & $\pm$ & 11.43 &  242.0 & 12.3 & $\pm$ &  0.4 &  0.0372 &  8.5 &  7  & 5 \\
419 &  212.98 &   7.18 &  0.112 &  96.06 & $\pm$ &  9.39 &   70.5 & 13.6 & $\pm$ &  1.3 &  0.0189 & 11.0 &  3  & 3 \\
422 &  210.19 &  26.78 &  0.139 &   7.35 & $\pm$ &  1.79 &  199.9 &  0.3 & $\pm$ &  0.0 &  0.0041 & 12.6 &  1  & 0 \\
430 &  218.56 &  24.33 &  0.095 &  48.33 & $\pm$ &  4.67 &   95.3 &  5.0 & $\pm$ &  0.4 &  0.0136 &  9.6 &  4  & 2 \\
440 &  223.15 &  21.55 &  0.115 &  83.57 & $\pm$ &  7.21 &  244.3 &  3.4 & $\pm$ &  0.2 &  0.0107 & 12.0 &  2  & 1 \\
441 &  223.96 &  28.02 &  0.125 &  15.35 & $\pm$ &  3.73 &  130.7 &  1.1 & $\pm$ &  0.2 &  0.0049 & 15.0 &  1  & 1 \\
454 &  228.78 &   6.96 &  0.044 & 172.79 & $\pm$ &  8.04 &  249.3 &  6.9 & $\pm$ &  0.3 &  0.0224 &  7.5 &  5  & 4 \\
457 &  229.57 &   5.64 &  0.079 & 173.07 & $\pm$ &  9.61 &  177.1 &  9.7 & $\pm$ &  0.5 &  0.0215 &  9.8 &  4  & 4 \\
460 &  229.89 &  31.06 &  0.114 & 169.83 & $\pm$ & 11.18 &  163.9 & 10.3 & $\pm$ &  0.6 &  0.0158 & 10.9 &  4  & 3 \\
463 &  232.79 &  29.37 &  0.073 & 312.36 & $\pm$ & 13.20 &  326.1 &  9.5 & $\pm$ &  0.4 &  0.0218 &  9.0 &  6  & 5 \\
474 &  239.46 &  15.69 &  0.036 & 120.34 & $\pm$ &  4.98 &   64.2 & 18.7 & $\pm$ &  0.7 &  0.1030 &  9.6 &  5  & 3 \\
484 &  246.08 &  41.86 &  0.134 &  24.74 & $\pm$ &  4.35 &    4.4 & 56.3 & $\pm$ &  9.9 &  0.0550 & 13.6 &  1  & 1 \\
509 &  310.21 & -40.02 &  0.020 &  39.05 & $\pm$ &  3.24 &  176.9 &  2.2 & $\pm$ &  0.1 &  0.0146 &  5.2 &  3  & 2 \\
574 &  349.03 & -26.76 &  0.119 & 104.17 & $\pm$ &  7.94 &  412.6 &  2.5 & $\pm$ &  0.1 &  0.0097 & 10.0 &  3  & 2 \\
579 &  353.89 &   9.97 &  0.040 &  24.52 & $\pm$ &  2.77 &  121.2 &  2.0 & $\pm$ &  0.2 &  0.0098 &  9.0 &  2  & 1 \\

\bottomrule
\end{tabular}
\end{threeparttable}}
\end{table*}


\subsubsection{Extensive mass of structures} 
The multiplicity $m$ of each detected structure was defined as the number of systems (linked by DBSCAN) that compose it, and its \textit{extensive mass}, \textit{i.e.}, the sum\footnote{{Note that this sum does not necessarily have to correspond to the value obtained, for example, by applying the virial mass equation (\ref{M_vi}) to the sample of member galaxies of the structure. The calculation of virial mass can be under or over estimated due to the presence of substructures \citep[\textit{e.g.},][]{bi2006}, such as the member systems of a structure, so it is not an extensive parameter. The term `extensive' here is used because of the additive property of mass.}} of the dynamical masses of its constituent parts, can be estimated in the form
\begin{equation}\label{M_ex}
\mathcal{M}_\text{ext}=\sum_{i=1}^{m} \mathcal{M}_{\text{vir}_k},
\end{equation}
where $M_{\text{vir}_k}$ is the virial mass of each system in the structure. Assuming that all systems {inside the structure} are viralized, one might expect the total mass of a structure to be only slightly greater than its extensive mass, that is $\mathcal{M}_\text{tot}\geq \mathcal{M}_\text{ext}$. The difference in mass would be given by the matter (probably gas, galaxies or dark matter) that resides between systems or as a dispersed component, \textit{i.e.}, not contained in galaxy systems. \\

In particular, the mass $\mathcal{M}^\text{sc}_\text{ext}$ of the superclusters was estimated from the above equation, but using $m=N_\text{sys}$. The value of $\mathcal{M}^\text{sc}_\text{ext}$ {(and its respective propagated uncertainty)} for each supercluster in the sample is shown in column 5 of Table \ref{tab:SCprop}. 
{For comparison, the masses of some well-known supercluster are available in the literature: the total mass of the \textit{Coma-Leo} (MSCC 295) and \textit{Hercules} (MSCC 474) superclusters are estimated in the ranges of $(168-189)\times 10^{14}h_{70}^{-1}\mathcal{M}_\odot$ and $(67-260)\times 10^{14}h_{70}^{-1}\mathcal{M}_\odot$, respectively \citep[\textit{e.g.},][]{bc2021}; the mass of the \textit{Shapley Supercluster} (MSCC 389 and MSCC 401) is estimated in the range of $(142-285)\times 10^{14}h_{70}^{-1}\mathcal{M}_\odot$ \citep[\textit{e.g.},][]{ra2006,sd2011}; and the mass of the \textit{Corona-Borealis Supercluster} (MSCC 463) lie in the range of $(67-420)\times 10^{14}h_{70}^{-1}\mathcal{M}_\odot$ \citep[\textit{e.g.},][]{sm1998,ei2021}. As can be seen, our estimates using $\mathcal{M}^{\text{sc}}_\mathrm{ext}$ are consistent with the mass ranges determined by other authors.} \\ 

\begin{figure}[t]
\includegraphics[trim={19cm 4cm 0.7cm 4.5cm},clip,width=\columnwidth]{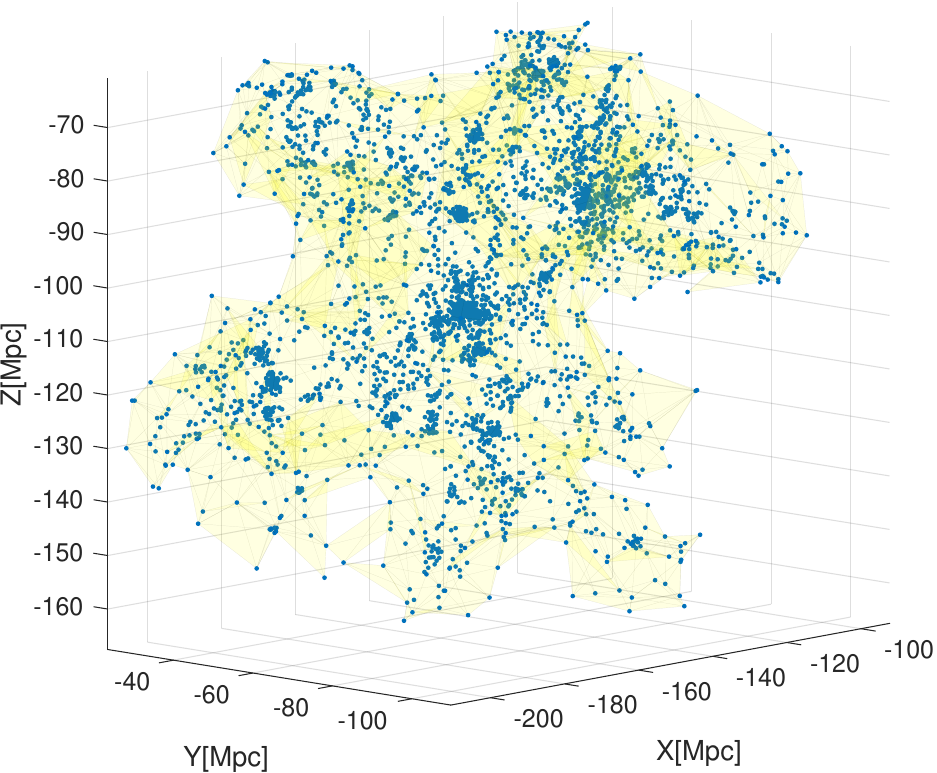}  
 \caption[]{Polyhedral surface (compact boundary, with shrink factor $s_\text{f}=1$) fitted to the ---purgued and FoG-corrected--- sample of member galaxies of \textit{Shapley Supercluster} (MSCC 389 and MSCC 401). The volume enclosed by the compact boundary is $V_\text{sc}=191.1$ $h_{70}^{-3}$ Mpc$^3$, smaller than that which would enclosed by the convex hull (\textit{e.g.}, with $s_\text{f}=0$) since the latter include voids of considerable size.}
 \label{f:alph}
\end{figure}

\subsubsection{Volume estimations} \label{vol}
The volume of any structure can be approximated by the volume enclosed by the surface enclosing its member galaxies (represented as points in FoG-corrected rectangular coordinates {from eq.} \eqref{xyz}). Such enclosing surfaces can be built by triangulating boundary points using alpha-shape based algorithms \citep[\textit{e.g.},][]{em1994} that allow polyhedral surfaces to be built around sets of points in three-dimensional space.\\

In particular, in this work a MATLAB alpha-shape based algorithm \citep[see][]{mw2022} was used, which allows the generation of a single polyhedral surface around a set of points $(x,y,z)$ in three-dimensional space and returns an estimate of the volume enclosed by it. In such algorithm the fit of polyhedral surfaces can be adjusted {(tightened/loosened)} by means of a `shrink factor' $s_\text{f}$ ranging from 0, generating the convex hull, to 1, generating the compact boundary. Figure \ref{f:alph} shows the polyhedral surface fit, with $s_\text{f}=1$, made for the \textit{Shapley Supercluster} from the galaxies contained in its --FoG-corrected and purified-- supercluster box. The volumes $V_\text{sc}$ of the sampled superclusters are shown in column 6 of Table \ref{tab:SCprop}. Such values were estimated by fitting polyhedral surfaces, with $s_\text{f}=1$, to the set of member galaxies in each refined supercluster box.

\section{Identification of supercluster \emph{cores}} \label{cores_id}
\subsection{{Purifying} the galaxy sample in superclusters}
To identify \textit{cores}, we first refined the samples of supercluster `member' galaxies from both southern and SDSS samples as follows:
\begin{itemize}
\item[(1)] Inside each supercluster box, a FoF algorithm was applied to the set of $N_\text{sys}$ systems identified there to determine the smallest radius $\varepsilon_\text{sc}$ linking them all. For this, each (SysCat) system was visualized as a single point, its centroid ($c_i$) in rectangular coordinates. \\

\item[(2)] For each galaxy ($\text{g}_*$) in the corresponding initial FoG-corrected supercluster box ($\mathcal{B}_\text{sc}$), its Euclidean distance $d(\text{g}_*,c_i)$ to each galaxy system was calculated. Galaxies were provisionally labeled as `linked' to the nearest galaxy system.\\

\item[(3)] All galaxies within a radius $\varepsilon_\text{sc}$ around the centroid of the system to which they were `linked' were taken as a `member' of the respective supercluster. The final{, purified,} supercluster box ($\mathcal{G}_\text{sc}$) only contains the galaxies selected as `members'. This set of `member' galaxies can be represented, for each supercluster, in the form
\begin{equation}
\mathcal{G}_\text{sc}=\left\lbrace \text{g}_* \in \mathcal{B}_\text{sc} | \min_i\left\lbrace d(\text{g}_*,c_i)\right\rbrace \leq \varepsilon_\text{sc}, i=1,...,N_\text{sys}  \right\rbrace.
\end{equation}
\end{itemize}
Galaxies not selected by this criterion were considered as `field' galaxies, not belonging to superclusters. There is no univocal criterion to define {the membership} of a galaxy to a supercluster, so here the criterion was to include all galaxies that are members of systems as well as those that are between them, forming galaxy bridges or being part of the disperse component of the supercluster up to a distance that we consider reasonable, \textit{e.g.}, $\varepsilon_\text{sc}$ from each identified galaxy system.\\

The radius $\varepsilon_\text{sc}$ and the number $N_{g_\text{sc}}$ of `member' galaxies found for each supercluster are shown respectively in columns 10 and 11 of Table \ref{tab:SCsample}. 

\subsection{Definition of \emph{cores}} \label{def}
Rich superclusters have central regions of high density of galaxies (and systems) that are absent in poor superclusters \citep[\textit{e.g.},][]{ei2007c,ei2008}. Such regions, commonly called \emph{cores}, may be observationally defined as collections of rich (and poor) clusters, groups and individual galaxies, being identified by their high galaxy density contrasts \citep[$\delta>10$, according to][]{ei2007c,ei2021}. Given these characteristics, it could be said that \emph{cores} are ``internal structures of superclusters that have begun to materialize as recognizable entities'' \citep[\textit{e.g.},][]{am2009}, like `compact superclusters' \citep[\textit{e.g.},][]{pea2014}. The most important characteristic of these entities is probably that they are the largest structures in the Universe that are expected to survive cosmic expansion, reaching, at least marginally, dynamical equilibrium. \\

In this work only rich MSCC-superclusters (with $m\geq 5$ member Abell/ACO-clusters) have been selected, so the presence of \emph{cores} is expected in them. To identify the \emph{cores} we think of them as \textit{massive and gravitationally bound galaxy structures, within superclusters, comprised of two or more galaxy systems (groups or clusters), with high probability of future collapse and virialization}. The \emph{cores} of superclusters involve high local densities in regions of small volume \citep[\textit{i.e.},][]{mar2004}. Based on the above, we choose as \emph{core} candidates those structures that, at the present-epoch, meet the following conditions: 

\begin{itemize}
\item[i.] Have extensive masses $\mathcal{M}_\mathrm{ext}\geq 5\times 10^{14}h_{70}^{-1}\mathcal{M}_{\odot}$.

\item[ii.] Are composed by gravitationally bound galaxy systems.

\item[iii.] Have {a} density ratio $\mathcal{R}\geq 7.86$ and {an} overdensity $\Delta_\mathrm{cr} \geq 1.36$.
\end{itemize}

The first condition allows the massive systems to be used as markers \citep[or ``seed'' points, \textit{e.g.},][]{cw2019t} of \emph{cores}; rich galaxy clusters, for example, are powerful indicators of local density in supercluster environments \citep[][]{ei2008}. The {minimum} mass $5\times 10^{14}h_{70}^{-1}\mathcal{M}_{\odot}$ (\textit{e.g.}, $\sim 10^{15}h_{70}^{-1}\mathcal{M}_{\odot}$ chosen by \citet{am2009} to select bound superclusters in cosmological simulations) is of the order of magnitude of the richest clusters, so any mass above that represents a considerable galaxy agglomeration. Thus, searching for systems with large virial masses increases the probability of finding nucleation regions, \textit{i.e.}, zones of higher concentration of matter in the form of clusters, groups, individual galaxies and hot gas. \\

The spherical density criterion described in Section \ref{f_vir} forms a key ingredient in our definition of \emph{cores} and for their selection process from the identified structures. Conditions (ii) and (iii) above, together, guarantee, at least theoretically, that galaxy structures once gravitationally bound will begin to collapse to virialize in the future according to the threshold values presented in Table \ref{tab:R_d}. In fact, condition (iii) could guarantee, by itself, that the structures under these density conditions will remain bound despite the expansion. However, an alternative method is used here to analyze the current gravitational binding state of the structures, which is why the second condition is taken independently.\\

\subsection{\textit{Core} selection} \label{core_sel}
Using the three conditions stated above to define \emph{cores}, these were selected from among the structures identified within each supercluster in the sample. The selection process was developed through a semi-automatic analysis as follows:

\begin{itemize}
\item[(1)] The extensive mass of each structure is calculated by {eq.} \eqref{M_ex} and used as an approximation for its total mass. Only structures with $\mathcal{M}_\mathrm{ext}\geq 5\times 10^{14}$ $h_{70}^{-1}\mathcal{M}_{\odot}$ are accepted as \emph{core} candidates.\\

\item[(2)] For each structure, its Most Massive Cluster (MMC), \textit{i.e.}, the member galaxy system with the largest virial mass, is identified and the distance $d_f$ from this to the farthest constituent system of the initial structure is determined. Then, all the systems in the surroundings of the structure (if any in the supercluster box), within the spherical shell between $d_f$ and $d_f+1.5\varepsilon_{c}$, centered on the MMC centroid, are provisionally included as members of the --enlarged-- structure (see Figure \ref{f:str_env}). Here,  $\varepsilon_{c}$ is the corresponding critical neighborhood radius used to identify the initial structure (see column 7 of Table \ref{tab:SCprop}). Here, a new $\mathcal{M}_\mathrm{ext}$ for the enlarged structure is calculated if it is the case. \\

\begin{figure}[t] 
\centering
 \includegraphics[trim={0cm 0cm 0cm 0cm},clip, width=\columnwidth]{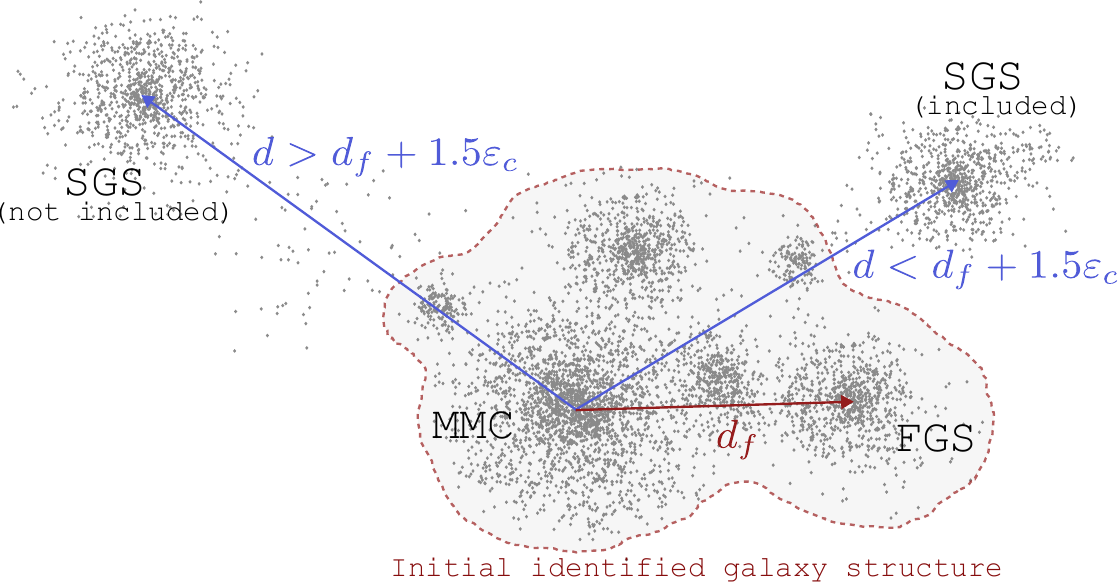} 
\caption[Schematic of a galaxy structure identified by DBSCAN with a critical neighborhood radius $\varepsilon_c$.]{Schematic of a galaxy structure identified by DBSCAN with a critical neighborhood radius $\varepsilon_c$. Its initial member systems are enclosed by the red dashed line, and $d_f$ is the distance between the Most Massive Cluster (MMC) and the Farthest Galaxy System (FGS) from it inside the structure. The systems outside the dashed line (if they exist) are called Surrounding Galaxy Systems (SGS). If any SGS is found in the spherical shell of radii $d_f$ and $d_f+1.5\varepsilon_{c}$, centered on the MCC, it is provisionally taken as a member of the structure. Whether or not a member system remains in the structure will depend on the gravitational binding criteria.}
\label{f:str_env}
\end{figure}

\item[(3)] The state of gravitational binding between pairs of member systems inside the enlarged structure is analyzed using the criterion {established in eq.} \eqref{lig1}: a member is classified as \textit{directly-bound} ($\mathcal{D}_b$-member) if it meets the criterion paired with the MMC or as \textit{indirectly-bound} ($\mathcal{I}_b$-member) if it does not meet the criterion with the MMC, but is linked to a third member that is directly-bound to both separately (like a `gravitational FoF'). These two sets of members form, together with the MMC, the \textit{pairwise bound region} of the structure.  Members of the enlarged structure that are neither \textit{directly}- nor \textit{indirectly-bound} to the MMC are classified as \textit{possibly-bound} ($\mathcal{P}_b$-member).  This can happen because these $\mathcal{P}_b$-members can still be bound to the larger mass of the whole structure, even if not bound to a specific neighbor.\\

\item[(4)] In a complementary way, the gravitational binding state of the structure as a whole is analyzed using the criterion established in eq. \eqref{lig2}. For this, the line-of-sight velocity dispersion $\sigma_{\upsilon_\mathrm{sys}}$ of systems inside the structure is estimated from their mean velocities using Tukey's biweight robust method \citep[\textit{e.g.},][]{br1990}, which is best suited for the small amount of member radial velocity data, and $\beta=2.5$, taken assuming a weak anisotropy in the velocity distribution of member systems. Furthermore, in order to estimate the gravitational radius $r_G$, the virial mass $\mathcal{M}_{\mathrm{vir}_k}$ of each member system of the structure is used in {eq.} \eqref{r_G}. \\

\item[(5)] If the structure contains a pairwise bound region and $\mathcal{P}_b$-members, then one of the following two cases could occur: 

\begin{itemize}
\item[(a)] If the structure is bound as a whole (or globally bound), then the $\mathcal{P}_b$-members are considered as bound members and the entire enlarged structure is accepted. \\

\item[(b)] Instead, if the structure is not globally bound, the farthest $\mathcal{P}_b$-member from the MMC is removed and step (4) is performed again. The process is repeated until a globally bound structure is obtained, stripped of unbound members, or until only the pairwise bound region remains.
\end{itemize}
On the other hand, cases may occur in which there are no pairwise bound systems within a structure (\textit{i.e.}, there is no a pairwise bound region), but it is globally bound. These type of structures are also accepted.\\ 

In any case, the structures accepted after this step continue to be considered \emph{core}-candidates of the corresponding supercluster, while the others are discarded.\\

\item[(6)] The member galaxies of each \emph{core}-candidate are defined as all galaxies up to a distance of $3.5R_{\mathrm{vir}_k}$ from the centroid of each $k$-th member system in the corresponding FoG-corrected supercluster box. At the chosen distance we expect to select the galaxies of member systems up to about their turn-around zone, those present in bridges between them, as well as galaxies that can be considered as the disperse component of the \emph{core}-candidate. The photometric and spectroscopic data of \emph{core} galaxies are extracted from the respective GalCat catalogue. \\

\item[(7)] The mean densities of each \emph{core}-candidate and its corresponding host supercluster are estimated, respectively, in the form 
\begin{equation}
\rho_\text{c}=\frac{\mathcal{M}_\text{ext}^c}{V_\text{c}}, \,\ \text{and} \,\ \rho_\text{sc}=\frac{\mathcal{M}^\text{sc}_\text{ext}}{V_\text{sc}},
\end{equation}
where $\mathcal{M}_\text{ext}^c$ and $\mathcal{M}^\text{sc}_\text{ext}$ are the extensive masses of the \emph{core}-candidate and the supercluster, and $V_\text{c}$ and $V_\text{sc}$ are the respective volumes estimated from the spatial distribution of their member galaxies using the alpha-shape based method described in Section \ref{vol}, with $s_\text{f}=0.5$ for \emph{core}-candidates and $s_\text{f}=1$ for superclusters. The reason for this choice of shrink factors is explained below.\\

\noindent The masses $\mathcal{M}^\text{sc}_\text{ext}$ and volumes $V_\text{sc}$ used here for superclusters are those compiled in columns 5 and 6 of Table \ref{tab:SCprop}, while the respective (mass and galaxy) densities calculated from these values are shown in columns 7 and 8 of the same table. The estimated gravitational radius, extensive mass, volume and density values for the \emph{core}-candidates are presented in Table \ref{tab:fluccP} only for those that were finally selected as \emph{cores} (see Section \ref{flucc}).\\

\noindent Now, assuming $\rho_\text{sc}$ to be the local background density of each \emph{core}-candidate in its respective supercluster, its density ratio is calculated from eq. \eqref{chon1} in the form
\begin{equation}\label{chon1b}
\mathcal{R}=\frac{\rho_\text{c}}{\rho_\text{sc}},
\end{equation}
and its density contrast with respect to the critical density {from eq.} \eqref{chon2} as 
\begin{equation}\label{chon2b}
\Delta_\text{cr}=\frac{8\pi G\rho_\text{c}}{3H^2(z)}-1
\end{equation}
where $z$ is the mean redshift of the host supercluster using the cosmology parameters described in Section \ref{intro}. Finally, only candidates that meet criteria $\mathcal{R}\geq 7.86$ and $\Delta_\text{cr} \geq 1.36$ (see Table \ref{tab:R_d}) are considered \emph{cores}, and the other structures were rejected. 
\end{itemize}

\begin{figure}[!t] 
\centering
\includegraphics[trim={0cm 0cm 0cm 0cm},clip,width=\columnwidth]{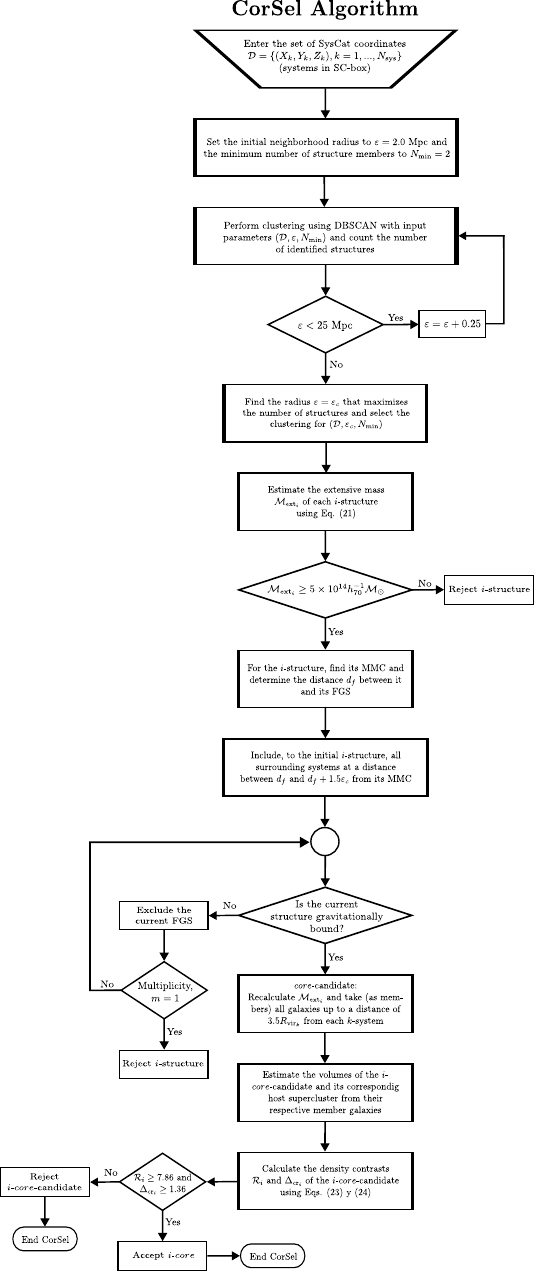} \\ 
 \caption{Flow chart of the semi-automatic CorSel-Algorithm.} 
 \label{f:fc}
 \end{figure}

We took the values $s_\text{f}=0.5$ and $s_\text{f}=1$, respectively, for the polyhedral fit shrink factors of the \emph{core}-candidates and superclusters, because with these values we have the worst case scenario for the density ratio (contrast) of a \emph{core}-candidate, possibly underestimating its density and overestimating the density of its local environment (the host supercluster). If a candidate meets the criteria under these conditions, they will meet the criteria under any other volume setting. {A summary of the \textit{core} selection process (or CorSel Algorithm) is shown in Figure \ref{f:fc}.}\\

\begin{figure}[!t] 
\centering
\includegraphics[trim={1cm 5cm 0.8cm 6.5cm},clip,width=\columnwidth]{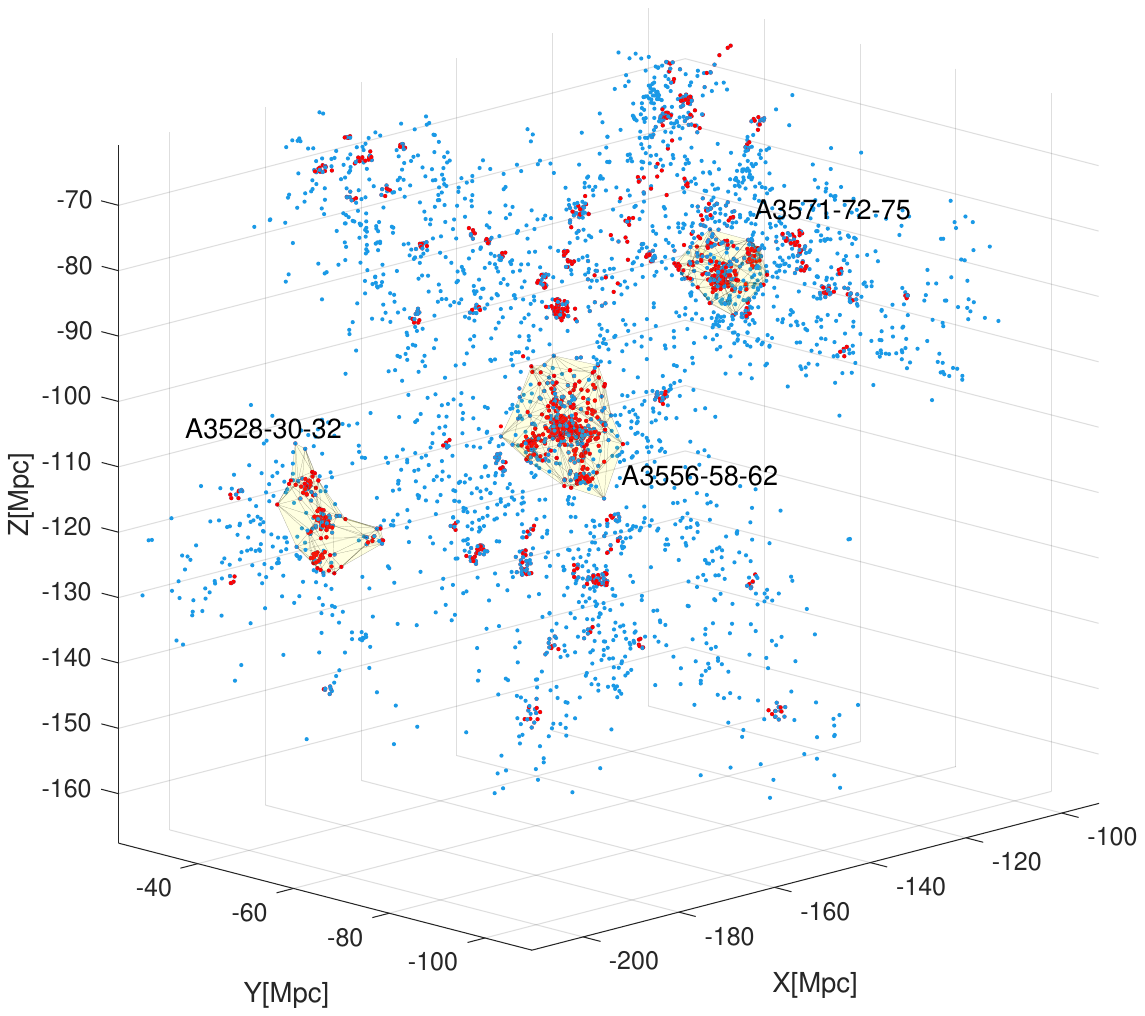} \\ 
 \includegraphics[trim={1.9cm 3.4cm 2.2cm 5.5cm},clip,scale=0.35]{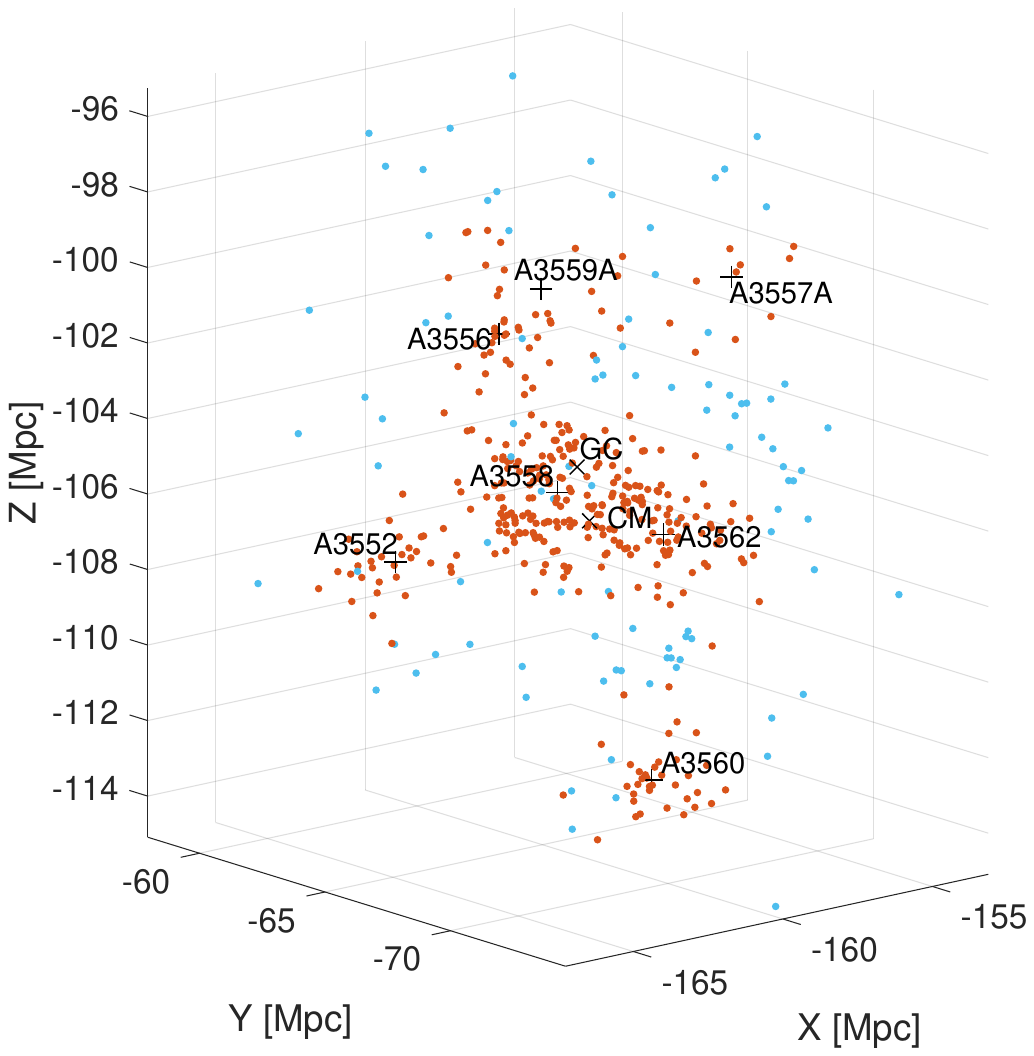} 
\caption{\textit{Top}: The three \emph{cores} identified for the \textit{Shapley Supercluster} (MSCC 389 and MSCC 401). The red points enclosed by yellow polyhedral surfaces correspond to the member galaxies of the systems of the respective \emph{cores}. The labels in each \emph{core} inform about the main Abell/ACO-clusters that comprise them. \textit{{Bottom:}} the main \emph{core} of the \textit{Shapley Supercluster} with 448 member 6dF-galaxies. The red dots represent the member galaxies of systems (Abell/ACO-clusters in this case), while the blue dots represent galaxies in bridges and disperse component; CM and GC are respectively the center of mass and the geometric center (centroid) of the \emph{core}; the black `+' symbols represent the centroids of the member clusters from the Abell/ACO catalogue.} 
 \label{f:cor}
 \end{figure}

\begin{figure}[] 
\centering
  \includegraphics[trim={0.5cm 4.5cm 0cm 5cm},clip,width=\columnwidth]{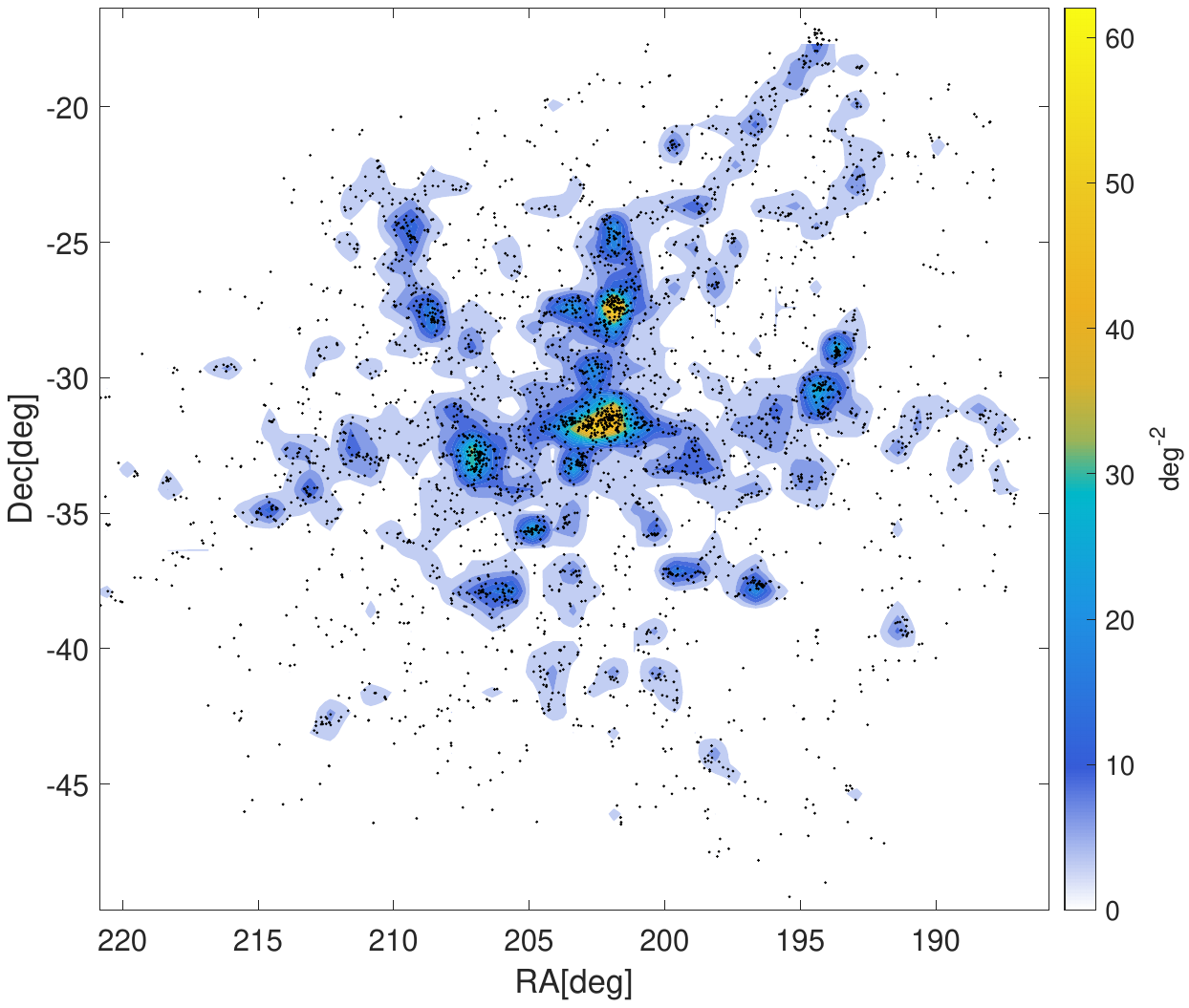}
\caption[]{2-D surface density map of the RA-Dec distribution of galaxies in the \textit{Shapley Supercluster} (MSCC 389 and MSCC 401). Three of the regions with the highest density of galaxies correspond to the \emph{cores} identified for the \textit{Shapley Supercluster}: DCC 066 ($\mathrm{RA}=194.211$, $\mathrm{Dec}=-30.698$), DCC 067 ($\mathrm{RA}=202.604$, $\mathrm{Dec}=-31.075$) and DCC 068 ($\mathrm{RA}=207.434$, $\mathrm{Dec}=-32.631$). The density peaks correspond to the most massive clusters of each \emph{core}.} 
 \label{f:density}
 \end{figure}

The top panel of Figure \ref{f:cor} shows an example of the \emph{cores} identified in the \textit{Shapley Supercluster} labeled with the respective main Abell/ACO-clusters that constitute them. The polyhedral surfaces fitted (with $s_\text{f}=0.5$) to the member galaxies of the \emph{core}, and used to determine the volumes of these, are also shown. The {bottom panel} of Figure \ref{f:cor} shows in greater detail the distribution of member galaxies in the so called main \emph{core} (\textit{i.e.}, A3556-A3558-A3562 and other clusters) of \textit{Shapley Supercluster} and the Abell/ACO-clusters that match its member systems. The three \emph{cores} identified for the \textit{Shapley Supercluster} also correspond to the regions with high surface density of galaxies in the RA-Dec plane, as can be seen in Figure \ref{f:density}. Note that there is a forth surface overdensity in this figure, north of the main \emph{core}, corresponding to the region of A1736 cluster: this region do not correspond to a \emph{core} because the overdensity is caused by a projection effect \citep[A1736 is composed of two rich systems aligned towards the LOS, \textit{e.g.},][]{ca2023}. These cases can be identified in all superclusters that have \emph{cores} in our sample.\\

\begin{figure}[t] 
\centering
  \includegraphics[trim={2.8cm 0cm 3.5cm 0.7cm},clip,width=\columnwidth]{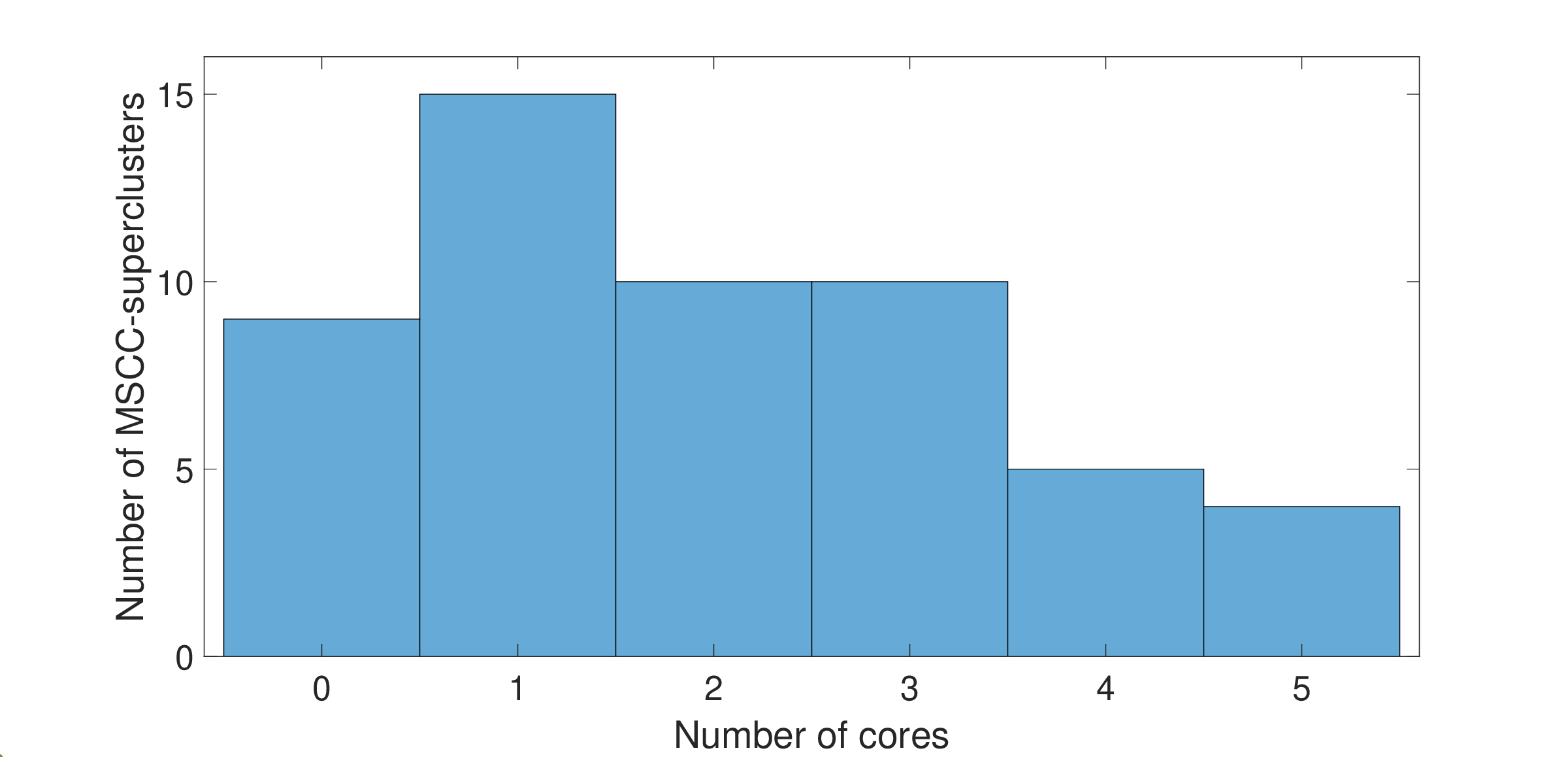}
  \includegraphics[trim={2.8cm 0.2cm 3.5cm 0cm},clip,width=\columnwidth]{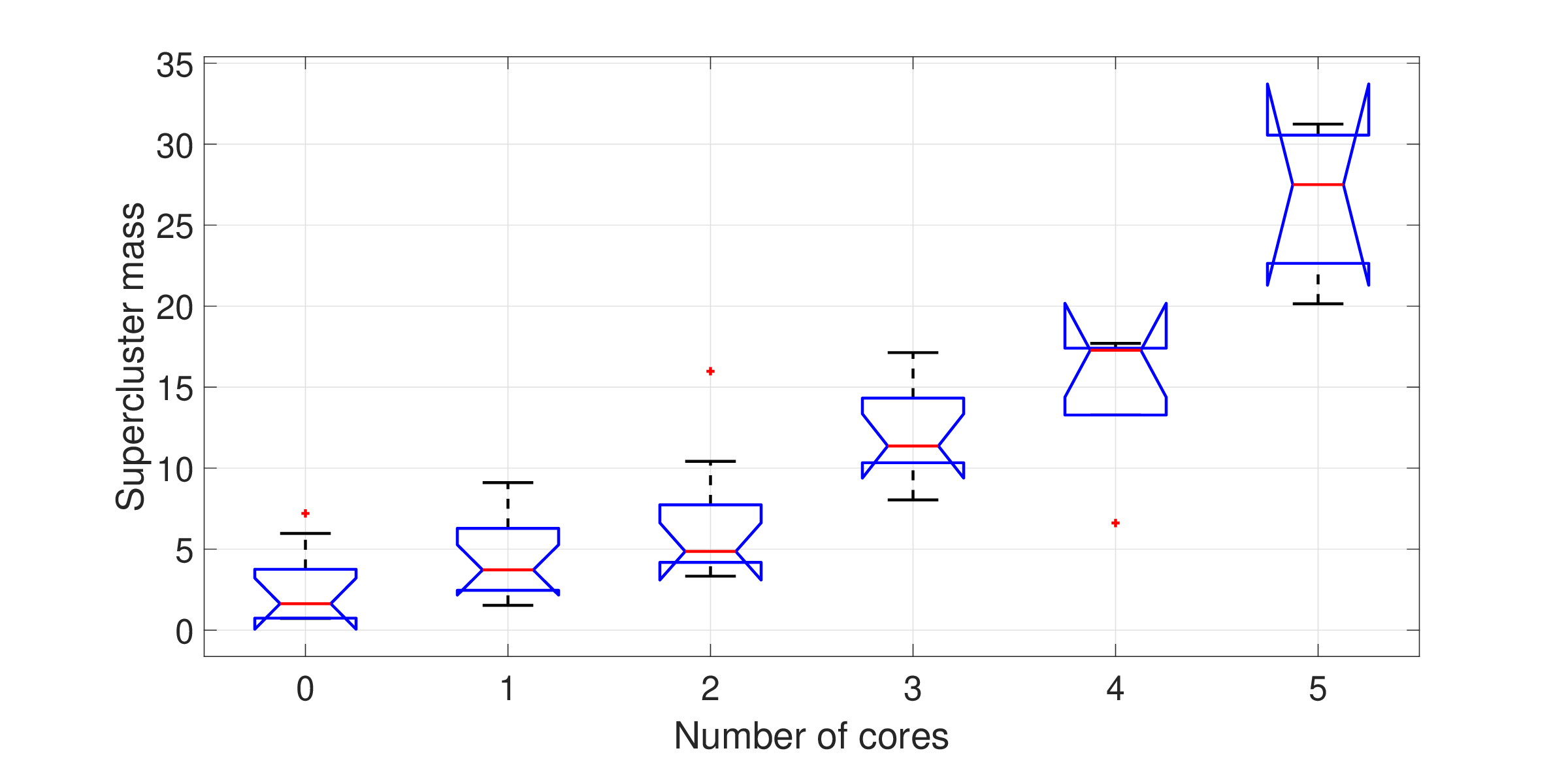}
\caption[]{\textit{Top}: The histogram shows the number of superclusters {as function of} the number of \emph{cores} they contain. \textit{Bottom}: Box plot showing the relationship between the extensive mass of superclusters ($\mathcal{M}_\text{ext}^\text{sc}$ in units of $10^{14}h^{-1}_{70}\mathcal{M}_\odot$) and the number of \emph{cores} they contain; the boxes represent the {interquartile} range with the red line 
indicating the median mass, the whiskers add 1.5 times the interquartile range, and notches display the confidence interval {(with a significance level of 0.05)} around the median.} 
 \label{f:stat}
 \end{figure}

The number of \emph{cores} ($N_\text{crs}$) found within each supercluster box, following the steps and criteria described above, is shown in column 11 of Table \ref{tab:SCprop}. From {the} histogram in {the} top panel of Figure \ref{f:stat} it can be noted that about 83\% of the MSCC rich superclusters in the sample have at least one \emph{core}, about 19\% have two \emph{cores}, and about 36\% have three or more \emph{cores}. Additionally, the boxplot in {the} bottom panel of Figure \ref{f:stat} shows a clear correlation between the number of \emph{cores} and the extensive mass (\textit{e.g}, $\mathcal{M}_{\text{ext}}^\text{sc}$ in Table \ref{tab:SCprop}) of their host superclusters. This correlation is expected given that {the more massive a supercluster is}, the greater the probability of forming regions (internal structures) with sufficient density to bind gravitationally and break away from the cosmic expansion. The sampled MSCC-superclusters with masses below $\sim 10^{15}h_{70}^{-1} \mathcal{M}_\odot$ tend to have fewer than three \emph{cores}, while those with masses above that value tend to have three or more.\\

\section{The DCC \emph{core}-catalogue}\label{flucc}
All the galaxy structures accepted as \emph{cores} were included in what we have called the Density-based \textit{Core} Catalogue (DCC). This catalogue contains a total of 105 massive nucleation regions inside nearby rich superclusters and with a high probability of future collapse and virialization, that are expected to survive the cosmic expansion to become ``island universes'' \citep[\textit{e.g.},][]{du2006}. To complement our previous definition, \emph{cores} are considered here as compact superclusters, \textit{i.e.}, dense ($\mathcal{R}\geq 7.86$ and $\Delta_\text{cr} \geq 1.36$) and probably gravitationally bound structures, formed by two or more galaxy systems of which at least one must be a rich and massive galaxy cluster (the ``seed'' cluster) of the Local Universe.\\ 

In Table \ref{tab:flucx} we list information on the 105 structures identified as \emph{cores} in the sample of 53 rich superclusters with redshift between 0 and $\sim$0.15: column 1 lists the number of the DCC and column 2 the number of the MSCC-supercluster it belongs to; the triplets of columns 3 to 5, 6 to 8 and 9 to 11 show the RA, Dec and $z$ values for the \emph{core} geometric centers (centroids), the centers of mass of the \emph{cores} and the centroid of the MMC associated with each \emph{core}, respectively; columns 12 and 13 present, respectively, the number of galaxies $N_\text{g}$ and the number $m_c$ (multiplicity) of galaxy systems identified for the \emph{cores}; finally, column 14 shows some matches between the \emph{core} member systems and the Abell/ACO clusters {\citep[including the S-clusters, \textit{e.g.},][]{ab1989}} as well as with clusters and groups from other catalogues mentioned in the table notes. This was performed in NED\footnote{The NASA/IPAC Extragalactic Database (NED) is funded by the National Aeronautics and Space Administration and operated by the California Institute of Technology. Website: \url{https://ned.ipac.caltech.edu/}.}, taking matches within a radius of 5 arcmin around the position of each \emph{core} system and with a radial velocity tolerance of $\pm 200$ km/s in LOS (when available). Column 14 is separated into two subcolumns which show the best matches (if any) for the MMC (left) and the other member systems (right) of each \emph{core}. \\

With the exception of a few structures for which a warning will be issued later, most DCC-\emph{cores} meet both the gravitational binding and density criteria required for their selection in the considered redshift range. Furthermore, some structures with extensive masses below $5\times 10^{14}$ $h_{70}^{-1} \mathcal{M}_\odot$ were included in the DCC catalogue because, despite their {relatively} low mass, they meet all other criteria established for selecting \emph{cores}. In Table \ref{tab:fluccP} we compile some additional properties estimated for the detected DCCs: column 1 shows the ID of the \emph{core} in the DCC catalogue; column 2 presents the line-of-sight velocity dispersion $\sigma_{\upsilon_\text{sys}}$ of member galaxy systems inside the \emph{core}; column 3 shows the extensive mass $\mathcal{M}^c_\text{ext}$ of the \emph{core}; column 4 contains the gravitational radius of the \emph{core}; column 5 presents the volume of the \emph{core} estimated by the polyhedral fit method, with $s_\text{f}=0.5$, from its FoG-corrected member galaxies; columns 6 and 7 show, respectively, the mass density and number density of galaxies of the \emph{core}; column 8 presents the galaxy number density contrast of the \emph{core} with respect to its respective host supercluster, $\delta_\text{g}^c=n_c/n_\text{sc}-1$; columns 9 and 10 show, respectively, the density ratio $\mathcal{R}$ (with respect to the local mass density) and the density contrast $\Delta_\text{cr}$ (with respect to the critical density at redshift $z$) of the \emph{core}, {with their respective uncertainties propagated through the calculations}; column 11 shows the gravitational binding state of the \emph{core} as a whole (\textit{i.e.}, using criterion \eqref{lig2}); and, finally, column 12 shows the 3-D distance $d_\mathrm{CMs}$ between the center of mass of the \emph{core} and the center of mass of its respective host supercluster. \\

\begin{table*}
\resizebox{17.5cm}{!}{
\begin{threeparttable}
\caption{The Density-based \textit{Core} Catalogue (DCC)}  
\label{tab:flucx}
\begin{tabular}{lrrrrrrrrrrrrcl}
\toprule \toprule
  \multicolumn{1}{c}{Name} &
  \multicolumn{1}{c}{ID} &
  \multicolumn{1}{c}{RA} &
  \multicolumn{1}{c}{Dec} &
  \multicolumn{1}{c}{$\bar{z}$} &
  \multicolumn{1}{c}{RA$_\text{CM}$} &
  \multicolumn{1}{c}{Dec$_\text{CM}$} &
  \multicolumn{1}{c}{$z_\text{CM}$} &
  \multicolumn{1}{c}{RA$_\text{MMC}$} &
  \multicolumn{1}{c}{Dec$_\text{MMC}$} &
  \multicolumn{1}{c}{$\bar{z}_\text{MMC}$} &
  \multicolumn{1}{c}{$N_\text{g}$} &
  \multicolumn{1}{c}{$m_c$} &
  \multicolumn{2}{c}{Cluster/group matches$^{2,3}$}\\
  
  \multicolumn{1}{c}{\emph{core}} &
  \multicolumn{1}{c}{MSCC$^1$} &
  \multicolumn{1}{c}{[deg]} &
  \multicolumn{1}{c}{[deg]} &
  \multicolumn{1}{c}{} &
  \multicolumn{1}{c}{[deg]} &
  \multicolumn{1}{c}{[deg]} &
  \multicolumn{1}{c}{} &
  \multicolumn{1}{c}{[deg]} &
  \multicolumn{1}{c}{[deg]} &
  \multicolumn{1}{c}{} &
  \multicolumn{1}{c}{} &
  \multicolumn{1}{c}{} &
  \multicolumn{1}{c}{MMC} &
  \multicolumn{1}{c}{Other \emph{core} members}\\
  
  \multicolumn{1}{c}{(1)} &
  \multicolumn{1}{c}{(2)} &
  \multicolumn{1}{c}{(3)} &
  \multicolumn{1}{c}{(4)} &
  \multicolumn{1}{c}{(5)} &
  \multicolumn{1}{c}{(6)} &
  \multicolumn{1}{c}{(7)} &
  \multicolumn{1}{c}{(8)} &
  \multicolumn{1}{c}{(9)} &
  \multicolumn{1}{c}{(10)}&
  \multicolumn{1}{c}{(11)}&
  \multicolumn{1}{c}{(12)}&
  \multicolumn{1}{c}{(13)}&
  \multicolumn{2}{c}{(14)}\\
\midrule \midrule

DCC 001 &   1 &   0.748 & -27.384 & 0.0646 &   0.748 & -27.353  &  0.0646  &   0.754 &  -26.874 & 0.0654 & 236  & 2  & {A2716} & S0003  \\
DCC 002 &   1 &   2.491 & -28.378 & 0.0612 &   2.847 & -28.742  &  0.0612  &   3.066 &  -28.890 & 0.0610 & 251  & 5  & {A2734} & A2726A \\
DCC 003 &   1 & 359.439 & -31.850 & 0.0588 & 359.261 & -32.186  &  0.0598  & 358.819 &  -32.902 & 0.0590 & 219  & 9  & - & - \\
DCC 004 &  27 &   9.059 & -25.705 & 0.0634 &   9.288 & -25.649  &  0.0633  &   9.527 &  -25.367 & 0.0629 & 189  & 8  & {A2800} & A0080A, EDCC 457, SWXCS J003759-2504.4\\
DCC 005 &  16/33 &   4.185 & -31.376 & 0.1068 &   4.185 & -31.376  &  0.1066  &   4.163 &  -31.351 & 0.1064 & 174  & 7  & {A2751B} & A2759A \\
DCC 006 &  33 &   5.610 & -33.756 & 0.1086 &   5.755 & -33.835  &  0.1086  &   5.717 &  -33.338 & 0.1099 & 96   & 5  & - & - \\
DCC 007 &  33 &   9.562 & -28.739 & 0.1124 &   9.484 & -28.759  &  0.1124  &   9.953 &  -28.903 & 0.1131 & 194  & 6  & {A2801} & A2798B, A2804 \\
DCC 008 &  33 &  10.257 & -26.428 & 0.1121 &  10.146 & -26.327  &  0.1123  &   9.962 &  -26.181 & 0.1124 & 89   & 4  & {EDCC 470} & A0088 \\
DCC 009 &  33 &  10.740 & -28.532 & 0.1082 &  10.647 & -28.528  &  0.1078  &  10.626 &  -28.547 & 0.1076 & 177  & 6  & {A2811B} & A2814B \\
DCC 010 &  39 &  10.431 &  -9.470 & 0.0549 &  10.420 &  -9.520  &  0.0547  &  10.467 &   -9.303 & 0.0555 & 65   & 2  & {A0085A} & SDSSCGA 00037 \\
DCC 011 &  39 &  15.234 & -14.992 & 0.0543 &  15.571 & -15.081  &  0.0542  &  17.408 &  -15.437 & 0.0538 & 86   & 5  & {A151B} & A126A \\
DCC 012 &  55 &  18.255 &  14.411 & 0.0594 &  18.050 &  14.182  &  0.0588  &  17.715 &   14.054 & 0.0579 & 192  & 7  & {A0150} & A0152A, A160B, RXC J0110.0+1358 \\
DCC 013 &  72 &  20.475 &   0.054 & 0.0777 &  20.421 &  -0.007  &  0.0777  &  20.272 &    0.045 & 0.0775 & 154  & 8  & {A0181A} & SDSSCGB 07949 \\
DCC 014 &  72 &  23.519 & -0.078 & 0.0799 &  23.572 &  -0.101  &  0.0801  &  22.841 &    0.519 & 0.0794 & 248  & 6  & {A0208A} & - \\
DCC 015 & 117 &  53.481 & -53.352 & 0.0604 &  53.712 & -53.160  &  0.0599  &  52.608 &  -52.599 & 0.0598 & 146  & 4  & {A3128B} & A3125A, A3158, S0366, RXC J0334.9-5342 \\
DCC 016 & 175 & 126.026 &  18.849 & 0.0979 & 126.069 &  19.291  &  0.0975  & 126.070 &   19.322 & 0.0975 & 106  & 3  & {A0659} & A650B, MSPM 04607, SDSSCGB 68634 \\
DCC 017 & 184 & 132.464 &  29.706 & 0.1043 & 132.285 &  29.651  &  0.1043  & 131.852 &   29.882 & 0.1049 & 93   & 3  & {A0705A} & SDSSCGA 00018 \\
DCC 018 & 219 & 152.862 &  19.169 & 0.1133 & 152.639 &  18.887  &  0.1134  & 152.408 &   18.560 & 0.1132 & 202  & 3  & {A0938B} & A0942A, A0952A  \\
DCC 019 & 222 & 156.006 &  49.460 & 0.1403 & 155.922 &  49.178  &  0.1420  & 155.893 &   49.078 & 0.1426 & 82   & 2  & {A0990} & SDSSCGB 03458 \\
DCC 020 & iso/222 & 158.128 &  53.074 & 0.1369 & 158.363 &  52.824  &  0.1371  & 158.405 &   52.779 & 0.1371 & 57   & 3  & {WHL J103344.3+524855} & A1027B, SDSSCGB 58169 \\
DCC 021 & 236 & 151.972 &  15.114 & 0.0292 & 150.397 &  15.228  &  0.0295  & 149.939 &   15.473 & 0.0295 & 147  & 4  & - & - \\
DCC 022 & 236 & 157.472 &  12.999 & 0.0325 & 157.725 &  13.661  &  0.0325  & 155.751 &   12.938 & 0.0320 & 230  & 8  & {A1016A} & A0999A, A1020A, SDSSCGB 08731, SDSSCGB 11341  \\
DCC 023 & 236 & 164.876 &   9.819 & 0.0350 & 165.012 &   9.978  &  0.0352  & 165.197 &   10.397 & 0.0354 & 153  & 4  & {A1142A} & SDSSCGB 27956, USGC U339  \\
DCC 024 & 278/236 & 173.329 &  22.675 & 0.0338 & 173.765 &  22.423  &  0.0326  & 172.261 &   22.900 & 0.0323 & 176  & 7  & {A1179B} & A1177B, SDSSCGB 66609, MSPM 00173 \\
DCC 025 & 238 & 155.446 &  41.292 & 0.0936 & 155.478 &  41.118  &  0.0927  & 155.040 &   41.011 & 0.0924 & 93   & 4  & {A0971A} & SDSSCGB 40601, SDSSCGB 58543 \\
DCC 026 & 238 & 156.534 &  37.597 & 0.1062 & 157.053 &  37.855  &  0.1050  & 157.184 &   37.919 & 0.1046 & 116  & 2  & {A1021B} & A0995A, A0997B \\
DCC 027 & 238 & 157.984 &  35.713 & 0.1236 & 157.995 &  35.183  &  0.1233  & 157.917 &   35.022 & 0.1233 & 125  & 5  & {A1033} & A1055, RXC J1031.7+3502, SDSSCGB 66740 \\
DCC 028 & iso/238 & 162.261 &  31.726 & 0.1145 & 162.349 &  31.776  &  0.1143  & 162.558 &   31.891 & 0.1138 & 81   & 2  & {MSPM 07607} & A1097, SDSSCGB 07211 \\
DCC 029 & 266 & 166.798 &  12.581 & 0.1261 & 166.786 &  12.632  &  0.1261  & 166.209 &   12.545 & 0.1277 & 99   & 3  & {A1159} & A1152, A1183A, SDSSCGA 01118 \\
DCC 030 & 272 & 167.795 &  40.530 & 0.0753 & 167.886 &  40.519  &  0.0753  & 167.993 &   40.717 & 0.0751 & 401  & 5  & {A1190} & A1173, A1187, A1203, SDSSCGB 30685\\
DCC 031 & 277 & 169.329 &  53.424 & 0.1060 & 169.200 &  53.431  &  0.1063  & 168.962 &   52.838 & 0.1077 & 159  & 3  & {A1218B} & A1225, SDSSCGB 02502, SDSSCGB 24283 \\
DCC 032 & 277 & 169.470 &  47.387 & 0.1119 & 169.768 &  47.573  &  0.1119  & 170.338 &   47.937 & 0.1121 & 181  & 4  & {A1227A} & A1202B, A1222, SDSSCGB 03830 \\
DCC 033 & 277 & 170.846 &  49.265 & 0.1108 & 170.808 &  49.410  &  0.1106  & 170.566 &   49.801 & 0.1111 & 115  & 3  & {A1231A} & WHL J112427.3+485246 \\
DCC 034 & 278 & 167.072 &  28.433 & 0.0335 & 167.676 &  28.085  &  0.0337  & 167.651 &   28.585 & 0.0329 & 360  & 4  & {A1185A} & SDSSCGB 11612, SDSSCGB 15832, SDSSCGB 41057 \\
DCC 035 & 278 & 173.385 &  22.446 & 0.0335 & 173.627 &  22.273  &  0.0335  & 173.633 &   22.266 & 0.0335 & 233  & 3  & {A1267A} & SDSSCGB 66609 \\
DCC 036 & 278 & 178.128 &  20.472 & 0.0241 & 177.170 &  20.583  &  0.0243  & 177.716 &   21.068 & 0.0246 & 399  & 5  & - & RXC J1204.1+2020, SWXCS J120409+2020.9 \\
DCC 037 & 283 & 172.040 &  17.496 & 0.1294 & 171.897 &  17.270  &  0.1292  & 171.797 &   17.119 & 0.1289 & 65   & 3  & {A1264B} & - \\
DCC 038 & 283 & 172.526 &  20.157 & 0.1332 & 172.530 &  20.192  &  0.1333  & 172.553 &   20.427 & 0.1335 & 90   & 2  & {A1278} & A1274 \\
DCC 039 & 283 & 172.700 &  24.146 & 0.1366 & 172.406 &  23.938  &  0.1369  & 172.369 &   23.856 & 0.1369 & 75   & 3  & {A1272} & A1268, MSPM 08938 \\
DCC 040 & 295 & 177.533 &  21.403 & 0.0229 & 177.133 &  20.889  &  0.0227  & 176.130 &   19.926 & 0.0216 & 727  & 7  & {A1367} & - \\
DCC 041 & 295 & 191.864 &  27.017 & 0.0235 & 192.734 &  27.521  &  0.0234  & 194.777 &   27.862 & 0.0233 & 1193 & 6  & {A1656} & [EKL2017] ECO0162, [EKL2017] ECO0207 \\
DCC 042 & 310 & 152.062 &  53.943 & 0.0460 & 151.735 &  54.015  &  0.0462  & 152.414 &   54.420 & 0.0460 & 192  & 6  & - & - \\
DCC 043 & 310 & 171.460 &  54.518 & 0.0696 & 171.062 &  54.467  &  0.0695  & 168.905 &   54.509 & 0.0699 & 258  & 5  & {RXC J1115.5+5426} & A1270 \\
DCC 044 & 310 & 178.083 &  54.588 & 0.0611 & 177.909 &  54.603  &  0.0611  & 177.058 &   54.643 & 0.0603 & 209  & 5  & {A1383} & A1396A, A1400B \\
DCC 045 & 310 & 179.455 &  54.959 & 0.0509 & 178.163 &  55.343  &  0.0514  & 176.806 &   55.693 & 0.0518 & 270  & 7 & {A1377} & A1324A, A1400A \\
DCC 046 & 310 & 181.666 &  56.943 & 0.0636 & 180.979 &  56.666  &  0.0642  & 180.070 &   56.204 & 0.0649 & 221  & 6 & {A1436} & - \\
DCC 047 & 311 & 174.837 &  13.478 & 0.0813 & 174.117 &  13.845  &  0.0810  & 173.270 &   14.457 & 0.0812 & 333  & 6 & {A1307} & - \\
DCC 048 & 311 & 175.230 &  11.494 & 0.0815 & 175.093 &  11.209  &  0.0812  & 175.078 &   10.903 & 0.0813 & 134  & 3 & {A1342A} & A1337A \\
DCC 049 & 311 & 177.739 &  12.261 & 0.0836 & 177.853 &  12.220  &  0.0833  & 177.493 &   12.305 & 0.0836 & 268  & 6 & {A1390} & A1372A, A1385A, MSPM 04978 \\
DCC 050 & 314 & 176.670 &  -1.306 & 0.0788 & 176.420 &  -1.707  &  0.0784  & 176.256 &   -1.992 & 0.0776 & 83   & 3 & {A1364A} & A1376A, SDSS-C4 1293 \\
DCC 051 & 317 & 176.126 &  -2.138 & 0.1209 & 176.186 &  -2.166  &  0.1209  & 176.425 &   -2.279 & 0.1206 & 57   & 2  & {A1373A} & - \\
DCC 052 & 317 & 176.675 &  -1.007 & 0.1179 & 176.761 &  -0.929  &  0.1181  & 176.483 &   -1.184 & 0.1176 & 36   & 2  & {A1376C} & - \\
DCC 053 & 323 & 179.073 &  24.399 & 0.1403 & 179.003 &  23.879  &  0.1405  & 178.821 &   23.496 & 0.1403 & 94   & 4  & {A1413B} & NSC J115657+241425, MaxBCG J179.49080+24.93397 \\
DCC 054 & 323 & 179.499 &  26.216 & 0.1386 & 179.596 &  26.243  &  0.1383  & 179.794 &   26.296 & 0.1379 & 44   & 2  & {A1425B} & A1420C \\
DCC 055 & 323 & 180.793 &  27.960 & 0.1371 & 180.739 &  28.071  &  0.1366  & 180.780 &   28.534 & 0.1348 & 102  & 4  & {A1449B} & A1455C, SDSSCGB 06034 \\
DCC 056 & 333 & 179.076 &  32.151 & 0.0794 & 178.891 &  32.740  &  0.0792  & 179.466 &   33.683 & 0.0798 & 94   & 6  & {A1423A} & A1427\\
DCC 057 & 333 & 180.356 &  29.677 & 0.0803 & 180.450 &  29.923  &  0.0805  & 180.672 &   29.048 & 0.0810 & 81   & 7  & {A1444B} & A1431A, A1449A \\
DCC 058 & 335 & 182.989 &  30.298 & 0.0743 & 182.748 &  30.726  &  0.0746  & 182.714 &   30.887 & 0.0742 & 71   & 3  & {A1480B} & A1478A, A1486A \\
DCC 059 & 311/343 & 177.234 &  10.623 & 0.0845 & 176.174 &   8.036 & 0.0841  & 177.021 &   7.983 & 0.0835 & 97   & 3  & {A1379} & A1358A, Mr20:[BFW2006] 25481 \\
DCC 060 & 343 & 182.284 &  15.300 & 0.0810 & 182.078 &  15.014  &  0.0806  & 182.033 &   14.951 & 0.0805 & 86   & 2  & {A1474} & A1481A \\
DCC 061 & 343 & 184.937 &  14.081 & 0.0820 & 185.341 &  13.882  &  0.0818  & 185.470 &   13.779 & 0.0816 & 82   & 4  & {A1526C} & A1499A \\
DCC 062 & 352/343 & 187.067 &  12.002 & 0.0851 & 187.429 &  11.720  &  0.0854  & 187.445 &   11.708 & 0.0853 & 143  & 2  & {A1552} & - \\
DCC 063 & 360 & 185.255 &  63.344 & 0.1061 & 184.568 &  63.617  &  0.1060  & 184.719 &   63.503 & 0.1062 & 88   & 4  & {A1518A} & A1539A, A1544A, SDSSCGB 16981 \\
DCC 064 & 386 & 196.775 &  39.710 & 0.0712 & 197.138 &  39.605  &  0.0713  & 197.770 &   39.273 & 0.0725 & 296  & 8  & {A1691} & A1680A \\
DCC 065 & 391/386 & 203.630 &  36.420 & 0.0595 & 203.733 &  36.351  &  0.0595  & 204.111 &  35.936 & 0.0597 & 139  & 6  & - & A1749A \\

\bottomrule
\end{tabular}
\end{threeparttable}}
\end{table*}

\begin{table*}
\resizebox{17.5cm}{!}{
\begin{threeparttable}
\textbf{Table \ref{tab:flucx}}. \textit{Continued}
\begin{tabular}{lrrrrrrrrrrrrcl}
\toprule \toprule

  \multicolumn{1}{c}{Name} &
  \multicolumn{1}{c}{ID} &
  \multicolumn{1}{c}{RA} &
  \multicolumn{1}{c}{Dec} &
  \multicolumn{1}{c}{$\bar{z}$} &
  \multicolumn{1}{c}{RA$_\text{CM}$} &
  \multicolumn{1}{c}{Dec$_\text{CM}$} &
  \multicolumn{1}{c}{$z_\text{CM}$} &
  \multicolumn{1}{c}{RA$_\text{MMC}$} &
  \multicolumn{1}{c}{Dec$_\text{MMC}$} &
  \multicolumn{1}{c}{$\bar{z}_\text{MMC}$} &
  \multicolumn{1}{c}{$N_\text{g}$} &
  \multicolumn{1}{c}{$m_c$} &
  \multicolumn{2}{c}{Cluster/group matches$^{2,3}$}\\
  
  \multicolumn{1}{c}{\emph{core}} &
  \multicolumn{1}{c}{MSCC$^1$} &
  \multicolumn{1}{c}{[deg]} &
  \multicolumn{1}{c}{[deg]} &
  \multicolumn{1}{c}{} &
  \multicolumn{1}{c}{[deg]} &
  \multicolumn{1}{c}{[deg]} &
  \multicolumn{1}{c}{} &
  \multicolumn{1}{c}{[deg]} &
  \multicolumn{1}{c}{[deg]} &
  \multicolumn{1}{c}{} &
  \multicolumn{1}{c}{} &
  \multicolumn{1}{c}{} &
  \multicolumn{1}{c}{MMC} &
  \multicolumn{1}{c}{Other \emph{cores} members}\\
  
  \multicolumn{1}{c}{(1)} &
  \multicolumn{1}{c}{(2)} &
  \multicolumn{1}{c}{(3)} &
  \multicolumn{1}{c}{(4)} &
  \multicolumn{1}{c}{(5)} &
  \multicolumn{1}{c}{(6)} &
  \multicolumn{1}{c}{(7)} &
  \multicolumn{1}{c}{(8)} &
  \multicolumn{1}{c}{(9)} &
  \multicolumn{1}{c}{(10)}&
  \multicolumn{1}{c}{(11)}&
  \multicolumn{1}{c}{(12)}&
  \multicolumn{1}{c}{(13)}&
  \multicolumn{2}{c}{(14)}\\

\midrule \midrule
DCC 066 & 389 & 194.211 & -30.698 & 0.0545 & 193.997 & -30.135  &  0.0546  & 193.641 &  -29.063 & 0.0542 & 141  & 5  & {A3528} & A3530, A3532   \\
DCC 067 & 389 & 202.604 & -31.075 & 0.0479 & 202.569 & -31.614  &  0.0479  & 202.182 &  -31.563 & 0.0477 & 451  & 7  & {A3558} & A3552, A3556, A3557A, A3559A, A3560, A3562 \\
DCC 068 & 401/389 & 207.434 & -32.631 & 0.0388 & 207.068 & -32.904  &  0.0388  & 206.863 &  -32.962 & 0.0389 & 178  & 3  & {A3571} & A3572, A3575 \\
DCC 069 & 414 & 204.778 &  26.019 & 0.0752 & 204.982 &  26.247  &  0.0753  & 205.461 &   26.403 & 0.0751 & 225  & 6  & {A1775B} & - \\
DCC 070 & 414 & 208.223 &  28.559 & 0.0760 & 208.827 &  28.130  &  0.0754  & 209.778 &   27.997 & 0.0751 & 349  & 7  & {A1831B} & A1800, A1817A  \\
DCC 071 & 414 & 208.723 &  27.350 & 0.0626 & 208.449 &  27.150  &  0.0625  & 207.242 &   26.693 & 0.0625 & 425  & 8  & {A1795} & A1818A, A1831A, WHL J135524.9+264738  \\
DCC 072 & 414 & 213.558 &  26.407 & 0.0705 & 212.826 &  26.197  &  0.0701  & 213.282 &   26.610 & 0.0694 & 150  & 4  & {A1886A} & SDSSCGB 13608 \\
DCC 073 & 414 & 213.864 &  30.358 & 0.0667 & 214.212 &  30.072  &  0.0670  & 215.058 &   30.397 & 0.0672 & 170  & 7  & - & A1869A \\
DCC 074 & 419 & 212.108 &   6.732 & 0.1128 & 212.510 &   6.623  &  0.1132  & 212.832 &    6.594 & 0.1134 & 166  & 5  & {A1870} & A1862, A1866A, WHL J141226.4+072020 \\
DCC 075 & 419 & 213.883 &   6.526 & 0.1096 & 213.577 &   6.781  &  0.1096  & 213.485 &    6.856 & 0.1095 & 81   & 2  & {A1881} & - \\
DCC 076 & 419 & 215.600 &   6.329 & 0.1132 & 215.720 &   6.177  &  0.1129  & 216.099 &    5.875 & 0.1124 & 83   & 3  & {WHL J142448.1+054809} & - \\
DCC 077 & 430 & 218.071 &  25.557 & 0.0958 & 217.797 &  25.570  &  0.0956  & 217.734 &   25.658 & 0.0953 & 109  & 3  & {A1927} & A1926A \\
DCC 078 & 430 & 219.279 &  24.530 & 0.0892 & 219.258 &  24.425  &  0.0894  & 219.230 &   24.283 & 0.0896 & 86   & 2  & {WHL J143758.3+240906} & A1939A \\
DCC 079 & 440 & 223.173 &  21.962 & 0.1171 & 223.156 &  21.888  &  0.1168  & 223.539 &   21.978 & 0.1166 & 194  & 4  & {A1986} & A1976, A1980, WHL J145513.4+222111 \\
DCC 080 & 441 & 223.827 &  27.906 & 0.1253 & 224.154 &  27.903  &  0.1252  & 224.693 &   27.892 & 0.1249 & 66   & 2  & {A2005B} & A1984, A1990A \\
DCC 081 & 454 & 221.279 &  10.496 & 0.0518 & 221.906 &  10.789  &  0.0521  & 222.171 &   11.264 & 0.0521 & 222  & 8  & - & MSPM 01932 \\
DCC 082 & 454 & 229.032 &   7.599 & 0.0450 & 229.228 &   7.536  &  0.0449  & 230.377 &    7.737 & 0.0445 & 337  & 8  & {A2063B} & A2028A, A2040,B A2055A \\
DCC 083 & 454 & 230.508 &   4.306 & 0.0379 & 229.773 &   4.421  &  0.0381  & 228.211 &    4.488 & 0.0385 & 533  & 9  & {SDSS-C4 1053} & - \\
DCC 084 & 454 & 230.527 &   3.536 & 0.0506 & 229.666 &   4.577  &  0.0492  & 229.927 &    4.076 & 0.0494 & 208  & 5  & {WHL J151903.5+042001} & WHL J151704.0+051522 \\
DCC 085 & 457 & 227.456 &   8.246 & 0.0791 & 227.277 &   8.265  &  0.0784  & 227.318 &    7.727 & 0.0772 & 305  & 6  & {A2028B} & A2040C, MSPM 05893 \\
DCC 086 & 457 & 228.045 &   5.240 & 0.0785 & 227.813 &   5.495  &  0.0785  & 227.743 &    5.857 & 0.0785 & 524  & 10 & {A2029} & A2033C, WHL J151052.2+045204 \\
DCC 087 & 457 & 231.024 &   7.498 & 0.0762 & 231.615 &   7.425  &  0.0760  & 232.621 &    7.968 & 0.0753 & 233  & 14 & - & - \\
DCC 088 & 457 & 231.580 &   3.251 & 0.0848 & 231.141 &   2.933  &  0.0845  & 230.498 &    2.360 & 0.0847 & 213  & 7  & {WHL J152203.5+022455} & A2082A, MSPM 05682, SDSS-C4 1070 \\
DCC 089 & 460 & 228.638 &  33.971 & 0.1132 & 228.308 &  33.538  &  0.1137  & 227.524 &   33.520 & 0.1135 & 213  & 4  & {A2034A} & WHL J151710.7+332153 \\
DCC 090 & 460 & 230.412 &  31.983 & 0.1133 & 230.490 &  31.815  &  0.1131  & 230.703 &   31.163 & 0.1137 & 165  & 3  & {A2067B} & A2062 \\
DCC 091 & 460 & 230.927 &  29.656 & 0.1142 & 231.015 &  29.721  &  0.1137  & 231.084 &   29.892 & 0.1136 & 283  & 4  & {A2069} & A2059B \\
DCC 092 & 463 & 230.572 &  31.115 & 0.0773 & 230.479 &  30.971  &  0.0771  & 230.357 &   30.675 & 0.0772 & 343  & 7  & {A2061A} & A2067A, MSPM 03940\\
DCC 093 & 463 & 232.261 &  27.782 & 0.0721 & 231.322 &  27.765  &  0.0718  & 230.668 &   27.739 & 0.0719 & 431  & 5  & {A2065} & A2089 \\
DCC 094 & 463 & 232.793 &  30.003 & 0.0659 & 232.676 &  29.323  &  0.0654  & 231.894 &   28.621 & 0.0673 & 343  & 7  & {A2073A} & A2079A, A2092A, WHL J152912.3+293823, MSPM 03789 \\
DCC 095 & 463 & 236.133 &  35.188 & 0.0683 & 235.768 &  35.132  &  0.0680  & 236.205 &   36.131 & 0.0662 & 247  & 10 & {A2122A} & A2106A, A2124, Mr20:[BFW2006] 33265 \\
DCC 096 & 463 & 238.869 &  28.553 & 0.0781 & 238.996 &  28.469  &  0.0781  & 237.922 &   27.862 & 0.0784 & 282  & 7  & {WHL J155203.6+275200} & MSPM 05829 \\
DCC 097 &458/474 & 230.182 &   7.473 & 0.0341 & 230.106 &   7.829  &  0.0343  & 230.746 &    8.610 & 0.0344 & 419  & 5  & {A2063A} & A2033A, A2040A, A2052 \\
DCC 098 & 474 & 237.025 &   9.212 & 0.0398 & 236.247 &   8.501  &  0.0401  & 235.814 &    8.157 & 0.0401 & 185  & 4  & {MSPM 00299} & MSPM 00295 \\
DCC 099 & 474 & 240.378 &  14.865 & 0.0361 & 240.839 &  16.056  &  0.0370  & 240.840 &   16.107 & 0.0374 & 1172 & 12 & {A2147} & A2151, A2152A, A2153A, MSPM 00504 \\
DCC 100 & 484 & 244.738 &  42.521 & 0.1362 & 244.812 &  42.508  &  0.1361  & 245.063 &   42.458 & 0.1353 & 43   & 4  & {A2179} & A2172, A2183 \\
DCC 101 & 509 & 297.799 & -30.414 & 0.0199 & 297.177 & -29.827  &  0.0200  & 297.886 &  -30.856 & 0.0205 & 77   & 3  & - & - \\
DCC 102 & 509 & 302.530 & -41.756 & 0.0188 & 302.928 & -40.533  &  0.0186  & 304.737 &  -41.159 & 0.0182 & 98   & 4  & {RXCJ2018.4-4102} & A3656 \\
DCC 103 & 574 & 351.646 & -24.333 & 0.1114 & 351.682 & -24.281  &  0.1116  & 351.659 &  -24.056 & 0.1119 & 67   & 4  & {A2599B} & A2601A, APMCC 894, EDCC 300 \\
DCC 104 & 574 & 352.502 & -29.706 & 0.1060 & 352.684 & -29.877  &  0.1062  & 353.071 &  -30.161 & 0.1064 & 101  & 5  & - & A4009, [RTL2009] 245 \\
DCC 105 & 579 & 351.451 &  14.630 & 0.0420 & 351.171 &  14.745  &  0.0419  & 351.100 &   14.520 & 0.0416 & 280  & 6  & {A2593A} & A2589 \\

\bottomrule
\end{tabular}
\begin{tablenotes}[hang]
\item[]Table note
\item[1]The separator `/' is used to indicate that the \emph{core} belongs to the MSCC supercluster with ID number on the left side, although it was detected in the supercluster box with ID number on the right side. The above is due to the overlap of supercluster boxes in the region. The special cases with the abbreviation `iso' on the left side correspond to structures that do not belong to any MSCC supercluster, they are isolated \emph{cores}.
\item[2]Used catalogues: Abell/ACO catalogues \citep{ab1958,ab1989}, Edinburgh-Durham Cluster Catalogue \citep[EDCC, \textit{e.g.},][]{lum1992}, Wen, Han \& Liu Cluster Catalogue \citep[WHL, \textit{e.g.},][]{wen2012}, Multiscale Probability Mapping Group and Cluster Catalogue \citep[MSPM, \textit{e.g.},][]{smi2012}, SDSS Compact Group Catalogue \citep[SDSSCGA and SDSSCGB, \textit{e.g.},][]{mcc2009}, UZC-SSRS2 Group Catalog \citep[USGC, \textit{e.g.},][]{ram2002}, C4 Cluster Catalog \citep[SDSS-C4, \textit{e.g.},][]{mil2005}, Mr20 Group and Cluster Catalog \citep[Mr20:BFW2006, \textit{e.g.},][]{ber2006}, Northern Sky Optical Cluster Survey \citep[NSC, \textit{e.g.},][]{gal2003}, MaxBCG Cluster Catalog \citep[MaxBCG, \textit{e.g.},][]{koe2007}, Meta-Catalogue of X-ray detected Clusters of Galaxies \citep[MCXC, \textit{e.g.},][]{pf2011}, and \textit{Swift} X-ray Cluster Survey catalogue \citep[SWXCS, \textit{e.g.},][]{liu2015}.
\item[3]The capital letters (A, B, C...) at the end of the Abell/ACO cluster names are used to identify the different components of a cluster along the line-of-sight \citep[that reflect the redshift information available in late 2012, see, for example,][]{and2005,chow2014}. 
\end{tablenotes}
\end{threeparttable}
}
\end{table*}
 
\begin{table*}
\resizebox{16cm}{!}{
\begin{threeparttable}
\caption{Basic properties estimated for DCCs}  
\label{tab:fluccP}
\begin{tabular}{cr |r@{\hspace{6pt}}c@{\hspace{6pt}}r| rr |rc@{\hspace{6pt}}r| cr |r@{\hspace{6pt}}c@{\hspace{6pt}}r| r@{\hspace{6pt}}c@{\hspace{6pt}}r| lr}
\toprule \toprule

  \multicolumn{1}{c}{ID} &
  \multicolumn{1}{c}{$\sigma_{\upsilon_\text{sys}}$} &
  \multicolumn{3}{c}{$\mathcal{M}^c_\text{ext}$} &
  \multicolumn{1}{c}{$r_G$} &
  \multicolumn{1}{c}{$V_c$} &
  \multicolumn{3}{c}{$\rho_c=\mathcal{M}^c_\text{ext}/V_c$} &
  \multicolumn{1}{c}{$n_c=N_\text{g}/V_c$} &
  \multicolumn{1}{c}{$\delta_\text{g}^c$} &
  \multicolumn{3}{c}{$\mathcal{R}$} &
  \multicolumn{3}{c}{$\Delta_\text{cr}$} &
  \multicolumn{1}{c}{Bind.} &
  \multicolumn{1}{c}{$d_\mathrm{CMs}$}\\
  
  \multicolumn{1}{c}{DCC} &
  \multicolumn{1}{c}{[km s$^{-1}$]} &
  \multicolumn{3}{c}{[$10^{14}h_{70}^{-1}\mathcal{M}_\odot$]} &
  \multicolumn{1}{c}{[$h_{70}^{-1}$ Mpc]} &
  \multicolumn{1}{c}{[$h_{70}^{-3}$ Mpc$^3$]} &
  \multicolumn{3}{c}{[$10^{10}h_{70}^{2}\mathcal{M}_\odot$Mpc$^{-3}$]} &
  \multicolumn{1}{c}{[$h_{70}^{3}$ Mpc$^{-3}$]} &
  \multicolumn{1}{c}{} &
  \multicolumn{3}{c}{} &
  \multicolumn{3}{c}{} &
  \multicolumn{1}{c}{state$^{1}$} &
  \multicolumn{1}{c}{[$h_{70}^{-1}$ Mpc]}\\
  
  \multicolumn{1}{c}{(1)} &
  \multicolumn{1}{c}{(2)} &
  \multicolumn{3}{c}{(3)} &
  \multicolumn{1}{c}{(4)} &
  \multicolumn{1}{c}{(5)} &
  \multicolumn{3}{c}{(6)} &
  \multicolumn{1}{c}{(7)} &
  \multicolumn{1}{c}{(8)} &
  \multicolumn{3}{c}{(9)} &
  \multicolumn{3}{c}{(10)} &
  \multicolumn{1}{c}{(11)} &
  \multicolumn{1}{c}{(12)}\\

\midrule \midrule

001 &  255  &  30.80  & $\pm$ & 4.51 &  33.25  &  1868.8  &  164.8 & $\pm$ &  24.1 &  0.1263  &   2.22  &   17.31  & $\pm$ &   2.68 &  11.12  & $\pm$ &  1.77 &  B  & 17.48 \\
002 &  125  &  22.53  & $\pm$ & 3.40 &  14.97  &   784.5  &  287.2 & $\pm$ &  43.4 &  0.3200  &   7.17  &   30.16  & $\pm$ &   4.81 &  20.12  & $\pm$ &  3.19 &  B  & 10.19 \\
003 &  248  &  10.42  & $\pm$ & 1.22 &  15.92  &   304.3  &  342.5 & $\pm$ &  40.1 &  0.7197  &  17.38  &   35.97  & $\pm$ &   4.60 &  24.19  & $\pm$ &  2.95 &  B  & 14.59 \\
004 &  217  &   7.10  & $\pm$ & 1.15 &  12.87  &   309.0  &  229.7 & $\pm$ &  37.3 &  0.6117  &  18.57  &   86.48  & $\pm$ &  15.90 &  15.89  & $\pm$ &  2.74 &  B  &  7.02 \\
005 &  192  &  13.71  & $\pm$ & 2.77 &  14.11  &   580.1  &  236.4 & $\pm$ &  47.7 &  0.3000  &  11.68  &   31.49  & $\pm$ &   6.52 &  16.38  & $\pm$ &  3.51 &  B  & 48.23 \\
006 &  269  &   9.09  & $\pm$ & 2.61 &  16.88  &   346.2  &  262.6 & $\pm$ &  75.6 &  0.2773  &  10.72  &   34.98  & $\pm$ &  10.19 &  18.31  & $\pm$ &  5.55 &  B  & 46.54 \\
007 &  150  &  19.30  & $\pm$ & 3.45 &  13.87  &   514.3  &  375.3 & $\pm$ &  67.2 &  0.3772  &  14.95  &   50.00  & $\pm$ &   9.23 &  26.60  & $\pm$ &  4.94 &  B  & 15.08 \\
008 &  65   &   6.41  & $\pm$ & 1.41 &  18.11  &   168.1  &  381.2 & $\pm$ &  84.4 &  0.5295  &  21.39  &   50.78  & $\pm$ &  11.48 &  27.03  & $\pm$ &  6.20 &  B  & 29.98 \\
009 &  296  &  19.58  & $\pm$ & 2.98 &  38.47  &   566.7  &  345.5 & $\pm$ &  52.6 &  0.3123  &  12.21  &   46.02  & $\pm$ &   7.32 &  24.40  & $\pm$ &  3.87 &  B  & 14.71 \\
010 &  197  &  18.48  & $\pm$ & 4.70 &  23.93  &   287.5  &  642.8 & $\pm$ & 163.6 &  0.2260  &  16.51  &   87.57  & $\pm$ &  25.04 &  46.27  & $\pm$ & 12.03 &  B  & 12.93 \\
011 &  179  &  12.38  & $\pm$ & 2.18 &  29.11  &   432.1  &  286.4 & $\pm$ &  50.6 &  0.1990  &  14.41  &   39.02  & $\pm$ &   8.56 &  20.06  & $\pm$ &  3.72 &  B  & 18.33 \\
012 &  261  &  17.49  & $\pm$ & 2.74 &  16.44  &   536.9  &  325.8 & $\pm$ &  51.0 &  0.3576  &  15.99  &   24.70  & $\pm$ &   5.95 &  22.96  & $\pm$ &  3.75 &  B  & 10.51 \\
013 &  169  &   3.83  & $\pm$ & 0.61 &  14.02  &   141.9  &  269.6 & $\pm$ &  43.4 &  1.0851  &  38.31  &   24.62  & $\pm$ &   4.57 &  18.83  & $\pm$ &  3.19 &  B  & 28.80 \\
014 &  312  &  14.25  & $\pm$ & 1.76 &  21.14  &   768.4  &  185.5 & $\pm$ &  23.0 &  0.3228  &  10.69  &   16.94  & $\pm$ &   2.62 &  12.64  & $\pm$ &  1.69 &  B  &  9.21 \\
015 &  270  &  29.32  & $\pm$ & 5.06 &  27.95  &  1414.1  &  207.4 & $\pm$ &  35.8 &  0.1032  &   8.22  &   29.43  & $\pm$ &   6.23 &  14.25  & $\pm$ &  2.63 &  B  & 26.33 \\
016 &  301  &  17.56  & $\pm$ & 3.36 &  64.03  &   347.6  &  505.3 & $\pm$ &  96.7 &  0.3050  &  18.84  &   94.95  & $\pm$ &  19.93 &  36.16  & $\pm$ &  7.11 &  U  & 14.48 \\
017 &  161  &  10.29  & $\pm$ & 2.18 &  27.93  &   181.8  &  566.0 & $\pm$ & 120.0 &  0.5117  &  70.68  &  310.64  & $\pm$ &  75.46 &  40.62  & $\pm$ &  8.82 &  B  & 15.64 \\
018 &  350  &  51.99  & $\pm$ & 7.83 &  38.47  &  1605.5  &  323.8 & $\pm$ &  48.8 &  0.1258  &  18.83  &  107.15  & $\pm$ &  19.90 &  22.81  & $\pm$ &  3.58 &  B  &  7.99 \\
019 &  747  &  17.50  & $\pm$ & 4.64 &  10.38  &   852.6  &  205.3 & $\pm$ &  54.4 &  0.0962  &   8.20  &   46.63  & $\pm$ &  14.15 &  14.09  & $\pm$ &  4.00 &  U  & 18.64 \\
020 &  19   &  15.19  & $\pm$ & 4.50 &  54.30  &   191.0  &  795.5 & $\pm$ & 235.7 &  0.2984  &  27.55  &  180.71  & $\pm$ &  59.80 &  57.49  & $\pm$ & 17.33 &  B  & 29.88 \\
021 &  58   &   7.21  & $\pm$ & 1.76 &  26.85  &   375.2  &  192.0 & $\pm$ &  46.9 &  0.3918  &   7.52  &   19.66  & $\pm$ &   4.99 &  13.12  & $\pm$ &  3.45 &  B  & 27.41 \\
022 &  167  &   5.07  & $\pm$ & 0.98 &  22.72  &   219.7  &  230.7 & $\pm$ &  44.7 &  1.0470  &  21.76  &   23.62  & $\pm$ &   4.86 &  15.96  & $\pm$ &  3.28 &  B  &  9.68 \\
023 &  142  &   6.07  & $\pm$ & 1.33 &  27.33  &   161.5  &  376.0 & $\pm$ &  82.6 &  0.9476  &  19.60  &   38.49  & $\pm$ &   8.87 &  26.65  & $\pm$ &  6.07 &  B  & 20.81 \\
024 &  593  &  28.49  & $\pm$ & 7.82 &  26.79  &  1614.9  &  176.4 & $\pm$ &  48.4 &  0.1090  &   1.37  &   18.06  & $\pm$ &   5.11 &  11.97  & $\pm$ &  3.56 &  B* & 33.21 \\
025 &  359  &  16.16  & $\pm$ & 3.70 &  19.61  &   267.4  &  604.3 & $\pm$ & 138.5 &  0.3477  &  71.06  &  482.08  & $\pm$ & 117.18 &  43.43  & $\pm$ & 10.18 &  B  & 58.86 \\
026 &  481  &  41.88  & $\pm$ & 9.17 &  59.79  &   813.3  &  514.9 & $\pm$ & 112.8 &  0.1426  &  28.56  &  410.76  & $\pm$ &  95.92 &  36.86  & $\pm$ &  8.29 &  B* & 10.80 \\
027 &  228  &  18.93  & $\pm$ & 4.09 &  72.16  &  1097.8  &  172.5 & $\pm$ &  37.3 &  0.1139  &  22.60  &  137.60  & $\pm$ &  31.77 &  11.68  & $\pm$ &  2.74 &  B  & 76.75 \\
028 &  212  &  11.27  & $\pm$ & 3.75 &  32.62  &   408.5  &  275.8 & $\pm$ &  91.8 &  0.1983  &  40.09  &  220.05  & $\pm$ &  75.35 &  19.28  & $\pm$ &  6.75 &  B  & 68.16 \\
029 &  8    &  36.13  & $\pm$ & 7.62 &  40.19  &   794.2  &  455.0 & $\pm$ &  95.9 &  0.1247  &   6.35  &   31.09  & $\pm$ &   8.64 &  32.45  & $\pm$ &  7.05 &  B  &  5.60 \\
030 &  94   &  18.12  & $\pm$ & 1.73 &  15.15  &   582.5  &  311.2 & $\pm$ &  29.7 &  0.6884  &  21.21  &   47.22  & $\pm$ &   6.28 &  21.88  & $\pm$ &  2.18 &  B  &  2.84 \\
031 &  552  &  26.12  & $\pm$ & 5.12 &  28.41  &  1091.3  &  239.4 & $\pm$ &  46.9 &  0.1457  &   7.62  &   24.06  & $\pm$ &   5.28 &  16.60  & $\pm$ &  3.45 &  B* & 29.81 \\
032 &  66   &  14.88  & $\pm$ & 2.37 &  24.21  &   525.3  &  283.3 & $\pm$ &  45.2 &  0.3446  &  19.39  &   28.48  & $\pm$ &   5.34 &  19.83  & $\pm$ &  3.32 &  B  & 22.29 \\
033 &  336  &  32.86  & $\pm$ & 6.34 &  34.31  &  1249.7  &  262.9 & $\pm$ &  50.7 &  0.0920  &   4.44  &   26.43  & $\pm$ &   5.72 &  18.33  & $\pm$ &  3.73 &  B  &  9.54 \\
034 &  277  &  13.11  & $\pm$ & 1.22 &  22.94  &   417.3  &  314.2 & $\pm$ &  29.3 &  0.8627  &  30.38  &   78.86  & $\pm$ &   8.11 &  22.11  & $\pm$ &  2.15 &  B  & 14.05 \\
035 &  33   &  28.63  & $\pm$ & 3.10 & 394.71  &  1037.8  &  275.8 & $\pm$ &  29.9 &  0.2245  &   7.17  &   69.23  & $\pm$ &   8.09 &  19.28  & $\pm$ &  2.20 &  B  & 12.21 \\
036 &  120  &   9.37  & $\pm$ & 1.23 &  12.60  &   335.9  &  278.9 & $\pm$ &  36.7 &  1.1879  &  42.22  &   70.00  & $\pm$ &   9.70 &  19.51  & $\pm$ &  2.70 &  B  & 33.65 \\
037 &  134  &  12.94  & $\pm$ & 3.12 &  38.81  &   174.3  &  742.3 & $\pm$ & 178.9 &  0.3729  &  71.65  &  267.95  & $\pm$ &  71.35 &  53.59  & $\pm$ & 13.16 &  B  & 40.86 \\
038 &  115  &  24.93  & $\pm$ & 5.69 &  24.41  &   882.0  &  282.7 & $\pm$ &  64.5 &  0.1020  &  18.88  &  102.04  & $\pm$ &  25.98 &  19.79  & $\pm$ &  4.74 &  B  & 11.14 \\
039 &  49   &  22.61  & $\pm$ & 5.16 & 112.77  &   471.7  &  479.2 & $\pm$ & 109.5 &  0.1590  &  29.98  &  172.98  & $\pm$ &  44.10 &  34.24  & $\pm$ &  8.05 &  B  & 32.23 \\
040 &  198  &  20.80  & $\pm$ & 1.79 &  18.84  &  1077.9  &  193.0 & $\pm$ &  16.6 &  0.6744  &   7.52  &   18.58  & $\pm$ &   1.79 &  13.19  & $\pm$ &  1.22 &  B  & 10.24 \\
041 &  103  &  25.91  & $\pm$ & 2.58 &  16.30  &  1094.5  &  236.7 & $\pm$ &  23.6 &  1.0900  &  12.77  &   22.79  & $\pm$ &   2.47 &  16.41  & $\pm$ &  1.73 &  B  & 17.60 \\
042 &  297  &   7.57  & $\pm$ & 1.27 &  13.62  &   253.6  &  298.7 & $\pm$ &  50.4 &  0.7571  &  25.02  &   41.76  & $\pm$ &   7.23 &  20.96  & $\pm$ &  3.71 &  B  & 77.45 \\
043 &  282  &  13.02  & $\pm$ & 1.79 &  27.16  &   601.9  &  216.3 & $\pm$ &  29.7 &  0.4286  &  13.73  &   30.25  & $\pm$ &   4.31 &  14.91  & $\pm$ &  2.18 &  B  & 37.74 \\
044 &  40   &  15.94  & $\pm$ & 2.85 &  22.57  &   455.5  &  350.0 & $\pm$ &  62.6 &  0.4588  &  14.77  &   48.93  & $\pm$ &   8.94 &  24.73  & $\pm$ &  4.60 &  B  & 12.55 \\
045 &  204  &  11.62  & $\pm$ & 1.66 &  17.12  &   621.1  &  187.1 & $\pm$ &  26.7 &  0.4347  &  13.94  &   26.16  & $\pm$ &   3.86 &  12.76  & $\pm$ &  1.96 &  B  & 40.37 \\
046 &  351  &  13.88  & $\pm$ & 2.09 &  29.71  &   695.2  &  199.6 & $\pm$ &  30.1 &  0.3179  &   9.93  &   27.91  & $\pm$ &   4.33 &  13.68  & $\pm$ &  2.21 &  B  & 27.40 \\
047 &  248  &  19.36  & $\pm$ & 2.11 &  26.02  &  1177.9  &  164.4 & $\pm$ &  17.9 &  0.2827  &  17.29  &   34.03  & $\pm$ &   4.16 &  11.09  & $\pm$ &  1.32 &  B  & 14.84 \\
048 &  279  &  11.83  & $\pm$ & 2.15 &  25.75  &   304.1  &  388.9 & $\pm$ &  70.7 &  0.4406  &  27.51  &   80.51  & $\pm$ &  15.29 &  27.60  & $\pm$ &  5.20 &  B  &  9.55 \\
049 &  240  &  19.65  & $\pm$ & 2.42 &  23.93  &   700.6  &  280.5 & $\pm$ &  34.5 &  0.3825  &  23.75  &   58.08  & $\pm$ &   7.82 &  19.63  & $\pm$ &  2.54 &  B  & 20.00 \\
050 &  298  &  13.97  & $\pm$ & 3.62 &  46.08  &   374.2  &  373.4 & $\pm$ &  96.7 &  0.2218  &   0.87  &    4.75  & $\pm$ &   1.45 &  26.46  & $\pm$ &  7.11 &  B  &  7.84 \\
051 &  82   &  20.25  & $\pm$ & 5.12 &  25.74  &   335.5  &  603.7 & $\pm$ & 152.6 &  0.1699  &  18.95  &   95.39  & $\pm$ &  27.86 &  43.39  & $\pm$ & 11.22 &  B  & 17.09 \\
052 &  112  &   9.72  & $\pm$ & 3.21 &  26.52  &   106.8  &  910.5 & $\pm$ & 301.2 &  0.3372  &  38.59  &  143.87  & $\pm$ &  52.03 &  65.95  & $\pm$ & 22.15 &  B  & 26.58 \\
053 &  453  &  38.06  & $\pm$ & 8.52 &  37.98  &  1293.7  &  294.2 & $\pm$ &  65.8 &  0.0727  &   6.34  &   48.07  & $\pm$ &  11.99 &  20.63  & $\pm$ &  4.84 &  B  & 25.04 \\
054 &  198  &   5.81  & $\pm$ & 2.04 &  33.67  &   161.6  &  359.6 & $\pm$ & 126.4 &  0.2723  &  26.52  &   58.76  & $\pm$ &  21.65 &  25.44  & $\pm$ &  9.29 &  B  &  2.73 \\
055 &  201  &  33.93  & $\pm$ & 6.74 &  30.36  &   609.6  &  556.5 & $\pm$ & 110.6 &  0.1673  &  15.91  &   90.95  & $\pm$ &  20.67 &  39.92  & $\pm$ &  8.13 &  B  & 22.57 \\
056 &  88   &   5.59  & $\pm$ & 1.43 &  40.94  &   358.5  &  156.0 & $\pm$ &  39.9 &  0.2622  &   2.16  &    6.02  & $\pm$ &   1.79 &  10.47  & $\pm$ &  2.94 &  B  & 19.49 \\
057 &  172  &   9.08  & $\pm$ & 4.21 &  27.35  &   290.9  &  312.2 & $\pm$ & 145.0 &  0.2784  &   2.36  &   12.04  & $\pm$ &   5.88 &  21.95  & $\pm$ & 10.66 &  B  &  7.94 \\
058 &  269  &   8.16  & $\pm$ & 5.70 &  22.67  &   429.5  &  190.0 & $\pm$ & 132.7 &  0.1653  &   3.48  &   19.73  & $\pm$ &  14.06 &  12.97  & $\pm$ &  9.76 &  B  &  9.01 \\
059 &  192  &  11.44  & $\pm$ & 2.92 &  53.24  &   252.0  &  454.1 & $\pm$ & 116.2 &  0.3850  &   3.92  &   14.97  & $\pm$ &   4.00 &  32.39  & $\pm$ &  8.54 &  B  & 34.34 \\
060 &  138  &   8.89  & $\pm$ & 1.87 &  75.95  &   204.7  &  434.2 & $\pm$ &  91.6 &  0.4200  &   4.37  &   14.32  & $\pm$ &   3.22 &  30.93  & $\pm$ &  6.73 &  B  & 15.21 \\

\bottomrule
\end{tabular}
\end{threeparttable}}
\end{table*}

\begin{table*}
\resizebox{16cm}{!}{
\begin{threeparttable}
\textbf{Table \ref{tab:fluccP}}. \textit{Continued}
\begin{tabular}{cr |r@{\hspace{6pt}}c@{\hspace{6pt}}r| rr |rc@{\hspace{6pt}}r| cr |r@{\hspace{6pt}}c@{\hspace{6pt}}r| r@{\hspace{6pt}}c@{\hspace{6pt}}r| lr}
\toprule \toprule

  \multicolumn{1}{c}{ID} &
  \multicolumn{1}{c}{$\sigma_{\upsilon_\text{sys}}$} &
  \multicolumn{3}{c}{$\mathcal{M}^c_\text{ext}$} &
  \multicolumn{1}{c}{$r_G$} &
  \multicolumn{1}{c}{$V_c$} &
  \multicolumn{3}{c}{$\rho_c=\mathcal{M}^c_\text{ext}/V_c$} &
  \multicolumn{1}{c}{$n_c=N_\text{g}/V_c$} &
  \multicolumn{1}{c}{$\delta_\text{g}^c$} &
  \multicolumn{3}{c}{$\mathcal{R}$} &
  \multicolumn{3}{c}{$\Delta_\text{cr}$} &
  \multicolumn{1}{c}{Bind.} &
  \multicolumn{1}{c}{$d_\mathrm{CMs}$}\\
  
  \multicolumn{1}{c}{DCC} &
  \multicolumn{1}{c}{[km s$^{-1}$]} &
  \multicolumn{3}{c}{[$10^{14}h_{70}^{-1}\mathcal{M}_\odot$]} &
  \multicolumn{1}{c}{[$h_{70}^{-1}$ Mpc]} &
  \multicolumn{1}{c}{[$h_{70}^{-3}$ Mpc$^3$]} &
  \multicolumn{3}{c}{[$10^{10}h_{70}^{2}\mathcal{M}_\odot$Mpc$^{-3}$]} &
  \multicolumn{1}{c}{[$h_{70}^{3}$ Mpc$^{-3}$]} &
  \multicolumn{1}{c}{} &
  \multicolumn{3}{c}{} &
  \multicolumn{3}{c}{} &
  \multicolumn{1}{c}{state$^{1}$} &
  \multicolumn{1}{c}{[$h_{70}^{-1}$ Mpc]}\\
  
  \multicolumn{1}{c}{(1)} &
  \multicolumn{1}{c}{(2)} &
  \multicolumn{3}{c}{(3)} &
  \multicolumn{1}{c}{(4)} &
  \multicolumn{1}{c}{(5)} &
  \multicolumn{3}{c}{(6)} &
  \multicolumn{1}{c}{(7)} &
  \multicolumn{1}{c}{(8)} &
  \multicolumn{3}{c}{(9)} &
  \multicolumn{3}{c}{(10)} &
  \multicolumn{1}{c}{(11)} &
  \multicolumn{1}{c}{(12)}\\

\midrule \midrule

061 &  98   &   4.79  & $\pm$ & 1.07 &  28.40  &   186.3  &  257.1 & $\pm$ &  57.9 &  0.4401  &   4.63  &    8.48  & $\pm$ &   2.02 &  17.90  & $\pm$ &  4.26 &  B  & 20.58 \\
062 &  78   &   9.24  & $\pm$ & 1.65 & 297.03  &   320.2  &  288.5 & $\pm$ &  51.8 &  0.4466  &   4.71  &    9.51  & $\pm$ &   1.86 &  20.21  & $\pm$ &  3.81 &  B  & 35.49 \\
063 &  66   &  14.26  & $\pm$ & 3.56 &  22.98  &   253.8  &  562.0 & $\pm$ & 140.6 &  0.3468  &  52.40  &  421.72  & $\pm$ & 116.46 &  40.32  & $\pm$ & 10.34 &  B  & 20.37 \\
064 &  286  &  31.93  & $\pm$ & 3.88 &  22.41  &   726.9  &  439.3 & $\pm$ &  53.4 &  0.4072  &  21.41  &   83.42  & $\pm$ &  11.80 &  31.30  & $\pm$ &  3.93 &  B  & 11.08 \\
065 &  341  &   6.12  & $\pm$ & 1.28 &  16.88  &   233.1  &  262.5 & $\pm$ &  55.0 &  0.5964  &  31.82  &   49.85  & $\pm$ &  11.04 &  18.30  & $\pm$ &  4.04 &  B  & 46.65 \\
066 &  188  &  18.36  & $\pm$ & 3.19 &  20.84  &   670.9  &  273.6 & $\pm$ &  47.6 &  0.2102  &   9.60  &   30.52  & $\pm$ &   5.55 &  19.12  & $\pm$ &  3.50 &  B  & 42.56 \\
067 &  175  &  50.67  & $\pm$ & 4.63 &  13.74  &  1532.8  &  330.6 & $\pm$ &  30.2 &  0.2942  &  13.84  &   36.87  & $\pm$ &   3.89 &  23.31  & $\pm$ &  2.22 &  B  &  7.52 \\
068 &  271  &  17.15  & $\pm$ & 2.89 &  35.32  &   539.7  &  317.8 & $\pm$ &  53.5 &  0.3298  &  15.63  &   35.44  & $\pm$ &   6.25 &  22.37  & $\pm$ &  3.93 &  B  & 34.90 \\
069 &  112  &  13.29  & $\pm$ & 2.26 &  14.02  &   335.0  &  396.7 & $\pm$ &  67.5 &  0.6717  &  17.04  &   32.14  & $\pm$ &   5.60 &  28.17  & $\pm$ &  4.96 &  B  & 47.63 \\
070 &  300  &  35.55  & $\pm$ & 4.18 &  32.55  &  1394.9  &  254.9 & $\pm$ &  30.0 &  0.2502  &   5.72  &   20.65  & $\pm$ &   2.55 &  17.74  & $\pm$ &  2.20 &  B  & 37.87 \\
071 &  138  &  25.63  & $\pm$ & 2.71 &  24.90  &  1110.5  &  230.8 & $\pm$ &  24.4 &  0.3827  &   9.28  &   18.70  & $\pm$ &   2.10 &  15.97  & $\pm$ &  1.79 &  B  & 25.92 \\
072 &  243  &  20.09  & $\pm$ & 3.67 &  27.89  &   454.9  &  441.7 & $\pm$ &  80.8 &  0.3298  &   7.86  &   35.79  & $\pm$ &   6.69 &  31.48  & $\pm$ &  5.94 &  B  & 12.35 \\
073 &  282  &  13.01  & $\pm$ & 2.12 &  20.74  &   507.9  &  256.2 & $\pm$ &  41.7 &  0.3347  &   7.99  &   20.76  & $\pm$ &   3.47 &  17.84  & $\pm$ &  3.07 &  B  & 18.40 \\
074 &  159  &  24.37  & $\pm$ & 4.40 &  29.26  &   800.9  &  304.3 & $\pm$ &  55.0 &  0.2073  &   9.92  &   22.36  & $\pm$ &   4.59 &  21.38  & $\pm$ &  4.04 &  B  &  7.75 \\
075 &  9    &  12.35  & $\pm$ & 3.04 &  80.82  &   178.7  &  691.1 & $\pm$ & 170.6 &  0.4533  &  22.88  &   50.78  & $\pm$ &  13.48 &  49.82  & $\pm$ & 12.54 &  B  & 11.42 \\
076 &  196  &  15.61  & $\pm$ & 4.05 &  36.29  &   416.3  &  375.0 & $\pm$ &  97.2 &  0.1994  &   9.50  &   27.56  & $\pm$ &   7.64 &  26.58  & $\pm$ &  7.15 &  B  & 24.03 \\
077 &  339  &   7.06  & $\pm$ & 1.43 &  20.37  &   319.0  &  221.2 & $\pm$ &  45.1 &  0.3416  &  24.01  &   43.65  & $\pm$ &   9.85 &  15.26  & $\pm$ &  3.31 &  U  & 10.03 \\
078 &  146  &  10.32  & $\pm$ & 2.27 &  24.35  &   149.4  &  691.0 & $\pm$ & 152.0 &  0.5757  &  41.15  &  136.35  & $\pm$ &  32.76 &  49.81  & $\pm$ & 11.17 &  B  & 27.00 \\
079 &  187  &  26.61  & $\pm$ & 3.51 &  29.16  &   867.8  &  306.6 & $\pm$ &  40.5 &  0.2236  &  19.87  &   89.65  & $\pm$ &  14.15 &  21.55  & $\pm$ &  2.98 &  B  &  7.14 \\
080 &  108  &  14.10  & $\pm$ & 3.69 &  66.33  &   426.1  &  330.9 & $\pm$ &  86.7 &  0.1549  &  30.36  &  281.88  & $\pm$ & 100.81 &  23.33  & $\pm$ &  6.38 &  B  &  1.97 \\
081 &  237  &   9.43  & $\pm$ & 1.53 &  14.96  &   555.9  &  169.6 & $\pm$ &  27.5 &  0.3994  &  16.82  &   24.48  & $\pm$ &   4.13 &  11.47  & $\pm$ &  2.02 &  B  & 43.97 \\
082 &  242  &  15.33  & $\pm$ & 1.64 &  18.12  &   551.3  &  278.2 & $\pm$ &  29.7 &  0.6113  &  26.27  &   40.14  & $\pm$ &   4.68 &  19.45  & $\pm$ &  2.19 &  B  &  4.73 \\
083 &  189  &  23.47  & $\pm$ & 2.49 &  15.68  &   764.0  &  307.2 & $\pm$ &  32.6 &  0.6976  &  30.12  &   44.33  & $\pm$ &   5.14 &  21.59  & $\pm$ &  2.40 &  B  & 25.98 \\
084 &  187  &  30.54  & $\pm$ & 4.41 &  22.32  &   674.2  &  453.0 & $\pm$ &  65.4 &  0.3085  &  12.76  &   65.37  & $\pm$ &   9.92 &  32.31  & $\pm$ &  4.81 &  B  & 23.54 \\
085 &  316  &  24.37  & $\pm$ & 2.82 &  18.91  &   785.7  &  310.2 & $\pm$ &  35.9 &  0.3882  &  17.01  &   31.76  & $\pm$ &   4.07 &  21.81  & $\pm$ &  2.64 &  B  & 20.37 \\
086 &  226  &  40.15  & $\pm$ & 3.81 &  21.97  &  1155.3  &  347.6 & $\pm$ &  33.0 &  0.4536  &  20.04  &   35.58  & $\pm$ &   3.91 &  24.56  & $\pm$ &  2.43 &  B  & 10.65 \\
087 &  269  &  13.49  & $\pm$ & 2.04 &  17.35  &   936.5  &  144.1 & $\pm$ &  21.8 &  0.2488  &  10.54  &   14.75  & $\pm$ &   2.37 &   9.60  & $\pm$ &  1.60 &  B  & 20.44 \\
088 &  216  &  33.02  & $\pm$ & 4.74 &  27.95  &   838.6  &  393.7 & $\pm$ &  56.5 &  0.2540  &  10.78  &   40.30  & $\pm$ &   6.20 &  27.95  & $\pm$ &  4.15 &  B  & 28.77 \\
089 &  259  &  43.32  & $\pm$ & 6.34 &  44.57  &  1758.6  &  246.3 & $\pm$ &  36.1 &  0.1211  &   6.64  &   23.77  & $\pm$ &   3.81 &  17.11  & $\pm$ &  2.65 &  B  & 23.34 \\
090 &  298  &  19.71  & $\pm$ & 3.29 &  34.21  &   807.3  &  244.2 & $\pm$ &  40.7 &  0.2044  &  11.90  &   23.57  & $\pm$ &   4.23 &  16.95  & $\pm$ &  2.99 &  B  &  8.37 \\
091 &  305  &  40.68  & $\pm$ & 5.10 &  25.16  &  1149.5  &  353.9 & $\pm$ &  44.3 &  0.2462  &  14.53  &   34.16  & $\pm$ &   4.83 &  25.02  & $\pm$ &  3.26 &  B  & 13.79 \\
092 &  496  &  32.56  & $\pm$ & 4.14 &  29.71  &   151.0  &  282.9 & $\pm$ &  35.9 &  0.2980  &  12.64  &   29.54  & $\pm$ &   3.95 &  19.80  & $\pm$ &  2.64 &  B  & 21.17 \\
093 &  287  &  44.27  & $\pm$ & 5.53 &  26.98  &  1504.6  &  294.2 & $\pm$ &  36.8 &  0.2865  &  12.11  &   30.73  & $\pm$ &   4.05 &  20.64  & $\pm$ &  2.70 &  B  & 12.55 \\
094 &  350  &  29.06  & $\pm$ & 4.13 &  25.66  &  1159.0  &  250.7 & $\pm$ &  35.6 &  0.2960  &  12.55  &   26.18  & $\pm$ &   3.88 &  17.44  & $\pm$ &  2.62 &  B  & 32.49 \\
095 &  462  &  26.23  & $\pm$ & 4.33 &  27.58  &  1076.0  &  243.7 & $\pm$ &  40.2 &  0.2296  &   9.51  &   25.45  & $\pm$ &   4.34 &  16.92  & $\pm$ &  2.96 &  B  & 39.19 \\
096 &  314  &  28.55  & $\pm$ & 4.22 &  20.03  &   504.8  &  565.5 & $\pm$ &  83.6 &  0.5586  &  24.57  &   59.05  & $\pm$ &   9.08 &  40.58  & $\pm$ &  6.15 &  B  & 36.61 \\
097 &  68   &  11.91  & $\pm$ & 1.25 &  16.16  &   511.9  &  232.7 & $\pm$ &  24.5 &  0.8184  &   6.94  &   12.43  & $\pm$ &   1.41 &  16.11  & $\pm$ &  1.80 &  B  & 32.61 \\
098 &  228  &  14.19  & $\pm$ & 2.40 &  24.49  &   566.7  &  250.4 & $\pm$ &  42.4 &  0.3265  &   2.17  &   13.37  & $\pm$ &   2.33 &  17.41  & $\pm$ &  3.11 &  B  & 27.40 \\
099 &  514  &  38.34  & $\pm$ & 2.15 &  23.31  &  1149.6  &  333.5 & $\pm$ &  18.7 &  1.0195  &   8.89  &   17.81  & $\pm$ &   1.24 &  23.53  & $\pm$ &  1.38 &  B  &  4.70 \\
100 &  250  &   7.63  & $\pm$ & 2.00 &  20.66  &   245.9  &  310.2 & $\pm$ &  81.4 &  0.1748  &   2.17  &    5.51  & $\pm$ &   1.73 &  21.81  & $\pm$ &  5.98 &  B  & 12.36 \\
101 &  144  &   3.98  & $\pm$ & 1.12 &  21.22  &    83.7  &  475.9 & $\pm$ & 134.1 &  0.9199  &  61.98  &  215.58  & $\pm$ &  63.32 &  33.99  & $\pm$ &  9.86 &  B  & 21.94 \\
102 &  99   &   4.64  & $\pm$ & 1.11 &  23.38  &   206.4  &  224.6 & $\pm$ &  53.7 &  0.4748  &  31.51  &  101.77  & $\pm$ &  25.78 &  15.52  & $\pm$ &  3.95 &  B  & 10.39 \\
103 &  116  &  15.70  & $\pm$ & 4.11 &  19.74  &   179.2  &  876.0 & $\pm$ & 229.7 &  0.3738  &  37.38  &  346.99  & $\pm$ &  94.79 &  63.41  & $\pm$ & 16.89 &  B  & 44.06 \\
104 &  58   &  13.45  & $\pm$ & 2.82 &  18.04  &   359.6  &  374.0 & $\pm$ &  78.6 &  0.2809  &  27.83  &  148.14  & $\pm$ &  33.14 &  26.50  & $\pm$ &  5.78 &  B  & 66.31 \\
105 &  273  &  10.46  & $\pm$ & 1.25 &  13.92  &   256.0  &  408.9 & $\pm$ &  49.0 &  1.0939  & 110.41  &  202.00  & $\pm$ &  33.32 &  29.06  & $\pm$ &  3.60 &  B  & 18.09 \\

\bottomrule
\end{tabular}
\begin{tablenotes}[hang]
\item[]Table note
\item[1]Gravitational binding state. B: structure globally bounded, B*: only one galaxy system is not bounded to the complete structure, U: structure not bounded as a whole.
\end{tablenotes}
\end{threeparttable}
}
\end{table*}


Note that the structures DCC 013, DCC 061, DCC 101 and DCC 102 have extensive masses less than $5\times 10^{14} h_{70}^{-1} \mathcal{M}_\odot$, the threshold mass established in the \emph{core} definition criteria of Section \ref{def}, however they are gravitationally bounded structures and their density contrasts $\mathcal{R}$ and $\Delta_\text{cr}$ are higher than the thresholds necessary to guarantee future collapse and virialization, which is why they were included in the DCC catalogue. On the other hand, the structures DCC 016, DCC 019 and DCC 077 were included in the catalogue even though they are not gravitationally bound structures yet, but they have enough mass and density contrast --with respect to the local environment and the critical density of the Universe-- to have already detached from cosmic expansion and it is very likely {that} they will begin a process of collapse in the future.\\ 

\begin{figure}[t] 
\centering
\includegraphics[trim={3cm 2.5cm 2.5cm 3cm},clip,width=\columnwidth]{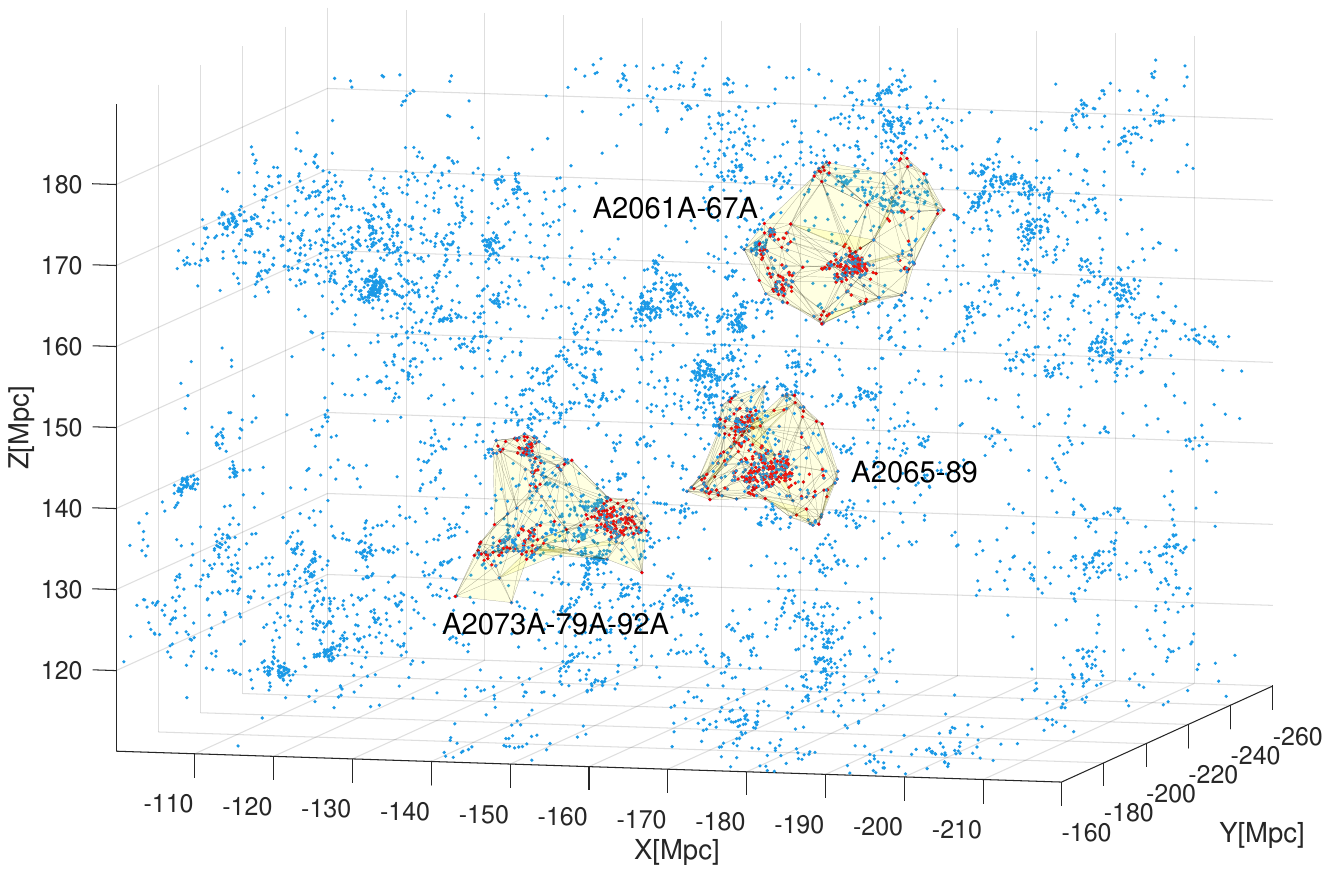} \\ 
\includegraphics[trim={1.5cm 1cm 1cm 1cm},clip,width=\columnwidth]{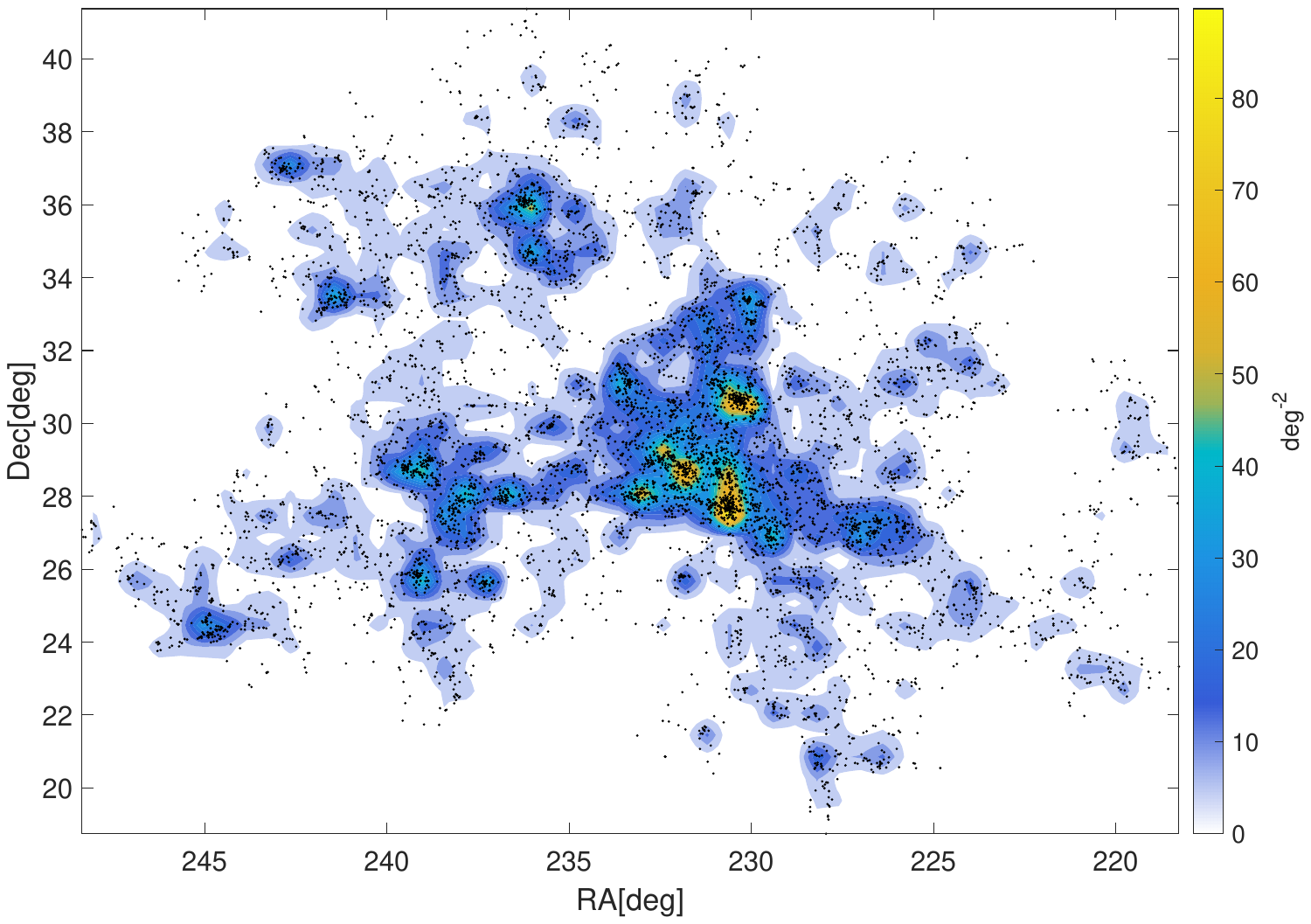} \\
 \caption[]{\textit{Top:} The \textit{Corona-Borealis Supercluster} (MSCC 463) and its three main \emph{cores}: DCC 092 (A2061A-A2067A), DCC 093 (A2065-A2089) and DCC 094 (A2073A-A2079A-A2092A). These three \emph{cores} were identified here as independent structures, but they could be a single central structure of the supercluster. \textit{Bottom:} 2-D surface density map of the RA-Dec distribution of galaxies in the \textit{Corona-Borealis Supercluster}. The three regions with the highest density of galaxies correspond to the DCC 092, DCC 093 and DCC 094 \emph{cores}. The other two \emph{cores} (DCC 095 and DCC 096) can be seen to the left (east) of the main concentrations. The density peaks correspond to the most massive clusters of each \emph{core}.} 
 \label{f:cor_bor}
 \end{figure}

Several of the structures compiled in the DCC catalogue correspond to identified \emph{cores} of known superclusters already reported in the literature. 
For example, for the \textit{Shapley Supercluster} (MSCC 389 and MSCC 401), one of the densest superclusters in the Local Universe, we have identified three \emph{cores}: DCC 066, DCC 067 and DCC 068. 
The second one corresponds to the so-called `central region' of \textit{Shapley} \citep[\textit{e.g.},][]{bre1994,bar2000,qui2000}, a well-studied concentration consisting mainly of the Abell clusters A3556, A3558 and A3562 (see Figure \ref{f:cor}). 
The other two structures identified here also correspond to important galaxy concentrations in \textit{Shapley} reported in the literature \citep[\textit{e.g.},][]{qui2000}, consisting mainly of clusters A3528, A3530 and A3532, as well as A3571, A3572 and A3575, respectively \citep[identified as MSCC 401 in][]{chow2014}. {The DCC 066 and 067 \emph{cores} correspond to the cluster complexes dominated by A3528 and A3558 as reported in \citet{bar2000}, with masses of $5.14 \times 10^{15} h_{70}^{-1}\mathcal{M}_\odot$ and $7.71\times 10^{15} h_{70}^{-1}\mathcal{M}_\odot$, respectively, which is consistent with our estimates (see Table \ref{tab:fluccP})}. The structures DCC 066, 067 and 068, constituted by well-known rich clusters, are clearly very massive and dense \emph{cores} ($\delta_\text{g}^c>9$, $\mathcal{R}>30$ and $\Delta_\text{cr}>18$) in process of collapse, in agreement with \citet{ch2015}. \\

For the \textit{Corona-Borealis Supercluster} (MSCC 463), the most distant of the best-known superclusters, there is good agreement between the detections made here and those reported in the literature: the so-called compact \emph{core} of \textit{Corona-Borealis} \citep[\textit{e.g.},][]{sm1997,sm1998,ko1998,mar2004}, {comprised by} the A2061, A2065 (the richest dominant cluster), A2067, A2089 and A2092 clusters, appears here fragmented by DBSCAN into the DCC 092, DCC 093 and DCC 094 structures (see Figure \ref{f:cor_bor}). In fact, gravitational binding analysis \eqref{lig2} shows that these three structures are globally bound and density contrast analysis shows that the composite structure (DCC 092-093-094), with $\mathcal{R}\sim29.22$ and $\Delta_\text{cr}\sim19.41$, could constitute a single nucleation region that is already in a stage of collapse, in agreement with \citet{sm1998}, \citet{ko1998}, \citet{pea2014} and \citet{ei2021}, who suggest that clusters A2061, A2065, A2067, A2089 and A2092, at the center of the supercluster, are in a stage of ``rapid gravitational collapse'' and will eventually form a large cluster. {This cluster complex will likely become one of the most massive virialized systems in the nearby Universe \citep{ei2021}}.  \\

\begin{figure}[t] 
\centering
\includegraphics[trim={3.5cm 2.5cm 2.5cm 3cm},clip,width=\columnwidth]{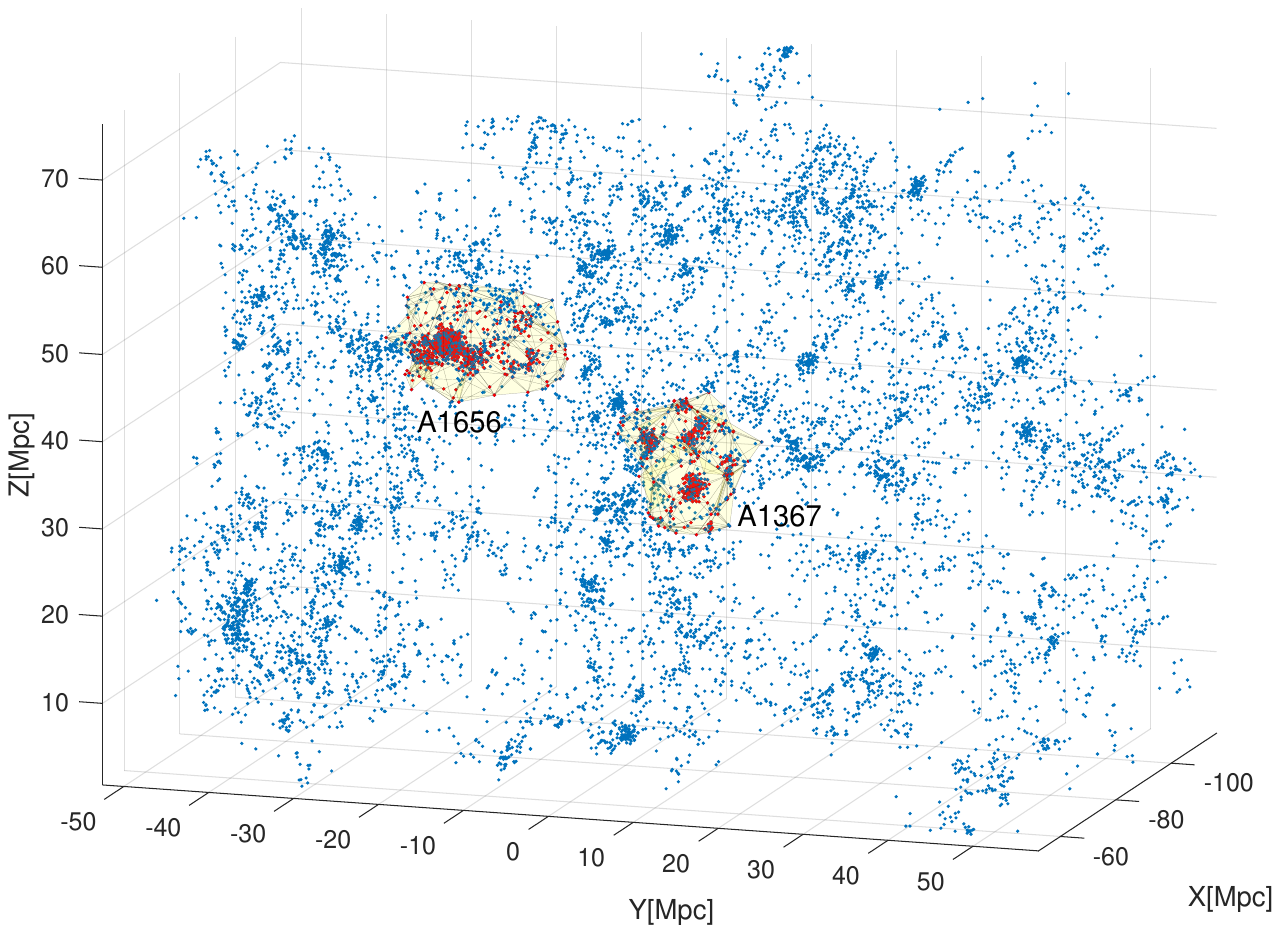} \\  
\includegraphics[trim={3cm 1cm 2.1cm 1cm},clip,width=\columnwidth]{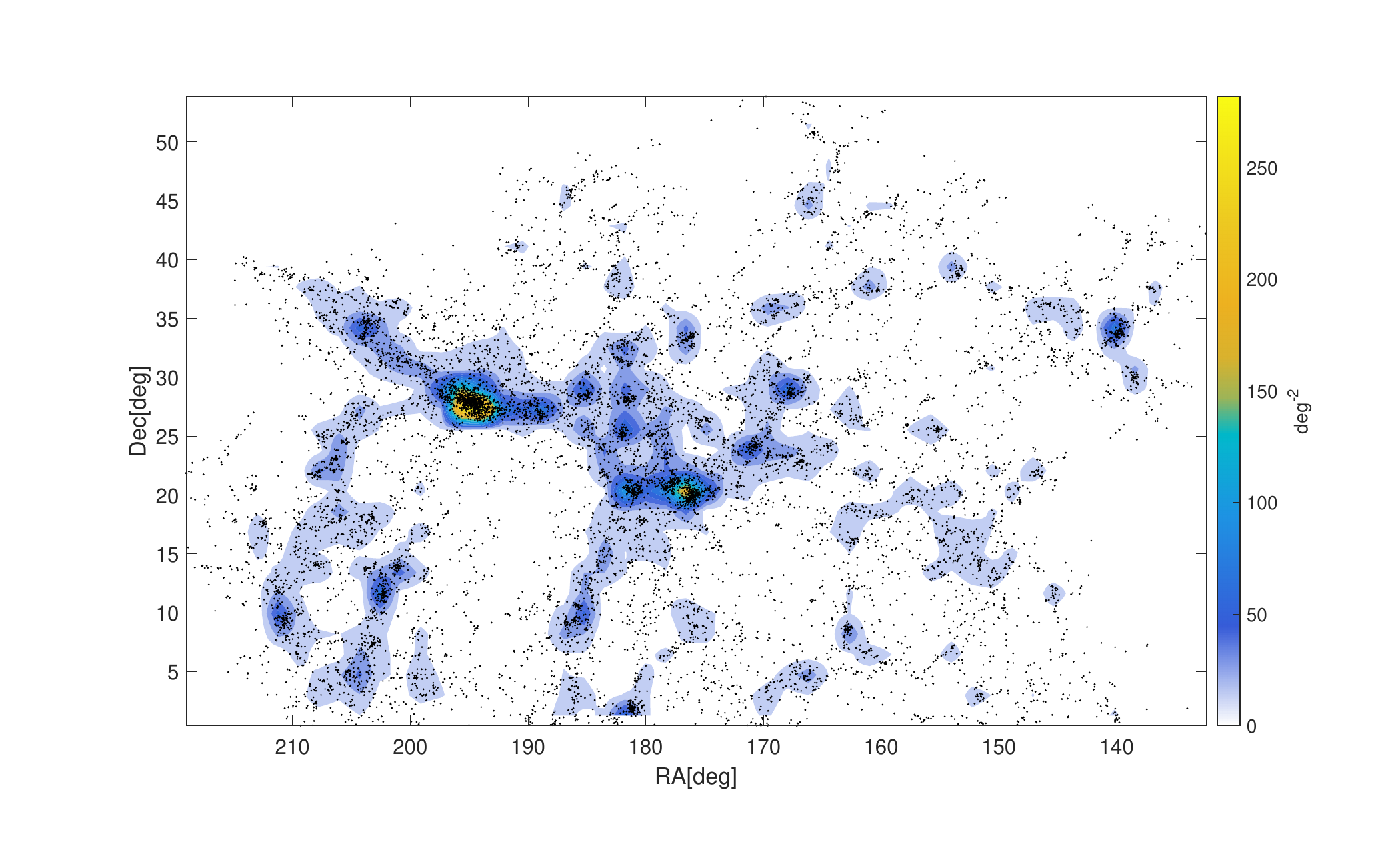} \\
 \caption{\textit{Top:} The \textit{Coma-Leo Supercluster} (MSCC 295) and its two \emph{cores}: DCC 040 (A1367, Leo Cluster) and DCC 041 (A1656, Coma Cluster). These two \emph{cores} were identified here as independent structures, but together they form a large filamentary structure that could be a single supercluster \emph{core}. \textit{Bottom:} 2-D surface density map of the RA-Dec distribution of galaxies in the \textit{Coma-Leo Supercluster}. The two regions with the highest density of galaxies correspond to the DCC 040 and DCC 041 \emph{cores}. The density peaks correspond to the most massive clusters of each \emph{core} (as expected A1367 and A1656).} 
 \label{f:com_leo}
 \end{figure}

Two other \emph{cores} worth mentioning are the structures DCC 040 and DCC 041 identified in the \textit{Coma-Leo Supercluster} (MSCC 295) and containing, respectively, the very rich clusters A1367 and A1656 (see Figure \ref{f:com_leo}). DCC 041 is the most prominent \emph{core} of the supercluster and contains the Coma Cluster, A1656, one of the most studied in the Local Universe, while DCC 040 contains the Leo Cluster, A1367. As with the three main \emph{cores} of the \textit{Corona-Borealis Supercluster}, structures DCC 040 and DCC 041 can be connected by DBSCAN by expanding the neighborhood radius from 6 $h_{70}^{-1}$ Mpc to as little as 6.5 $h_{70}^{-1}$ Mpc. The two structures are connected by bridges of galaxy systems located between them, but are not gravitationally bound to any separate \emph{core}. However, the composite structure, a large filament with densities $\mathcal{R}\sim 20.87$ and $\Delta_\text{cr}\sim 14.81$ seems to be globally bound and could be a single \emph{core} for the \textit{Coma-Leo supercluster} in a state of collapse, although {no pairwise} gravitational binding is verified between the Coma and Leo clusters. The three-dimensional distribution of galaxies in the \textit{Coma-Leo Supercluster} allows us to appreciate that both DCC 040 and DCC 041 are located at the intersection of filamentary substructures, close to the estimated center of mass for the supercluster, {so we may say} that \textit{Coma-Leo} presents a central binary \emph{core}.\\

{It is very interesting that the definitions of ``quasi-spherical superclusters'' (QSs) proposed by \citet{he2022} goes in the same direction as our definition of \emph{cores}. Of course our methods are different, but they give similar results, for example, for the masses of two objects we have in common: our DCC 046 (in the \textit{Ursa Major Supercluster}, MSCC 310), with $N_\mathrm{g}=221$ and $\mathcal{M}_\mathrm{ext}^c=(1.39 \pm 0.21)\times 10^{15}h_{70}^{-1}\mathcal{M}_\odot$, corresponds to its QS 524, with $N_\mathrm{g}=283$ and $\mathcal{M}=(1.82 \pm 0.64)\times 10^{15}h_{70}^{-1}\mathcal{M}_\odot$ centered on the A1436 cluster; and our DCC 097 (in the \textit{Hercules Supercluster}, MSCC 474), with $N_\mathrm{g}=419$ and $\mathcal{M}_\mathrm{ext}^c=(1.19 \pm 0.13)\times 10^{15}h_{70}^{-1}\mathcal{M}_\odot$, corresponds to its QS 550, with $N_\mathrm{g}=680$ and $\mathcal{M}=(2.58 \pm 0.73)\times 10^{15}h_{70}^{-1}\mathcal{M}_\odot$ around the A2052 cluster.}   

\section{Discussion and conclusions} \label{s:conc}
In this work we have identified \emph{cores} in nearby rich superclusters through two consecutive percolation processes, one for galaxies and the other for systems in a three-dimensional space, combining the complementary FoF and DBSCAN clustering algorithms. \textit{Cores} are understood here as gravitationally bound galaxy structures, comprised by two or more clusters and groups, with sufficient
matter density to survive cosmic expansion and virialize in the future. The \emph{cores} were selected from among several candidate structures --second-order galaxy `clusters'-- using density criteria \citep[\textit{e.g.},][]{du2006,ch2015} that define them as nucleation zones inside superclusters. In total, 105 structures were identified as \emph{cores} whose galaxy samples are {selected} in the SDSS (main sample), 2dF and 6dF (complementary samples) regions. These \emph{cores} and their estimated properties were compiled in the Density-based \textit{Core} Catalogue (DCC) presented here.\\

\begin{figure}[t] 
\centering
 \includegraphics[trim={3cm 0cm 3.5cm 0.5cm},clip, width=\columnwidth]{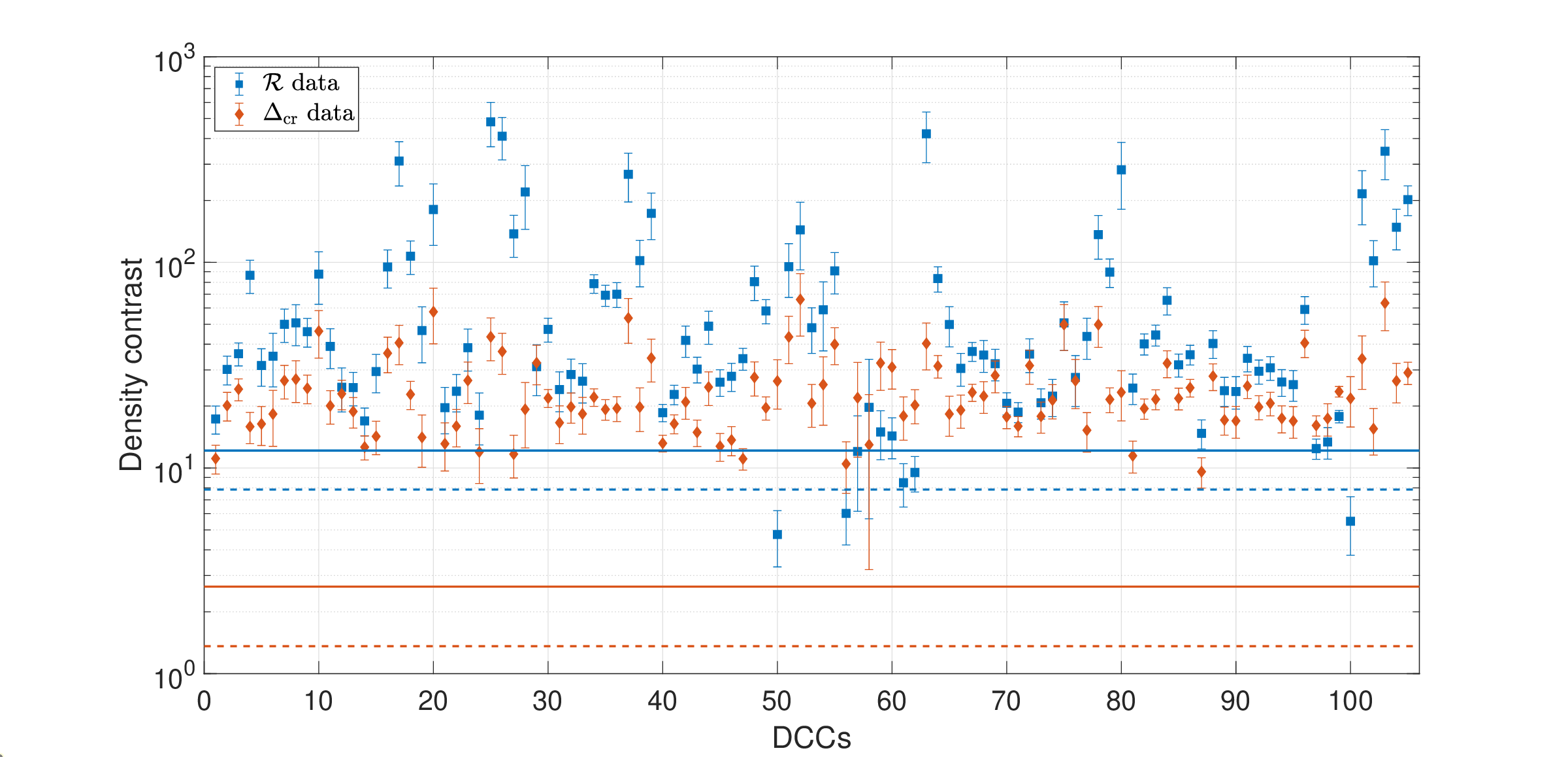} 
\caption[]{Density contrasts for DCCs. The horizontal axis shows the identifier of each core, while the vertical axis presents the two estimated values $\mathcal{R}$ and $\Delta_\text{cr}$. The bars represent the propagated uncertainties. The blue and red dashed lines represent, respectively, the $\mathcal{R}=7.86$ and $\Delta_\text{cr}=1.36$ values for structures prone to future collapse and the solid blue and red lines the respective $\mathcal{R}=12.15$ and $\Delta_\text{cr}=2.65$ values for structures at turn-around in the present epoch.}
\label{f:den_cont}
\end{figure}

Table \ref{tab:flucc_ps} presents a summary of the ranges of values that the \emph{core} properties compiled in Table \ref{tab:fluccP} take, as well as their mean and median values. {The mass and gravitational radius ranges of the DCCs are quite consistent with the mass and size ranges of the QSs of \citet{he2022} and the high-density cores found by \citet{ei2016} for the Sloan Great Wall supercluster complex}. In adittion, as can be seen from the mean/median values (see also Figure \ref{f:den_cont}), the DCC-\emph{cores} are structures that, in general, are already in the process of gravitational collapse since their density contrasts exceed the threshold values for structures at {turn-around} \citep[\textit{i.e.}, $\mathcal{R}\geq 12.15$ and $\Delta_\text{cr}\geq 2.65$, according to][]{ch2015} in the present epoch assuming a flat Universe with $\Omega_{m,0}=0.3$ and $\Omega_{\Lambda,0}=0.7$ (see also, Table \ref{tab:R_d}). These values are expected to guarantee that most of these structures will virialize in the future, reaching some state of dynamical equilibrium regardless of their current evolutionary degree. \\

{We have shown that the distinct luminosity limits of our three sub-samples do not affect the detection of systems and their membership in SysCat, as can be seen from the comparison of our richness, masses, and mass uncertainties to the ones of similar system catalogues in the literature, since we optimized the use of the data for each supercluster box. Also, the relative mass error distributions for the SysCat systems show no significant trend within the redshift range of our supercluster sample ($0.02 \leq z \leq 0.15$), indicating that selection effects do not significantly affect our calculations of density contrasts. Furthermore, the mass and richness distributions of our systems are also quite consistent when compared to other catalogues, so the systems in SysCat are reliable. Furthermore, as explained in Section \ref{core_sel}, the volumes of the structures were determined in such a way that they underestimated the density of the \emph{core} candidates and overestimated the densities of their host superclusters, this being the worst-case scenario for \emph{cores} selection. Even so, the selected structures presented high density contrasts with ranges of uncertainties that are within the established criteria as can be seen in Figure \ref{f:den_cont}.}\\

\begin{table*}
\caption[]{Outlier free range, mean (with standard deviation) and median (with the first-25\% and third-75\% quartiles) values for the basic properties estimated for the {DCCs} compiled in Table \ref{tab:fluccP}. {The outliers were obtained by establishing the lower ($\mathrm{Q1}-1.5\times\mathrm{IQR}$) and upper ($\mathrm{Q3}+1.5\times\mathrm{IQR}$) limits, where $\mathrm{IQR}=\mathrm{Q3}-\mathrm{Q1}$ and is the interquartile range and Q1 and Q3 are the first and third quartiles of the data in the total sample. Once the outlier data below and above these limits have been removed, the quartiles of the refined sample are recalculated, which are shown here.}}
\label{tab:flucc_ps}
\centering %
\resizebox{16cm}{!}{
\begin{tabular}{lccc}
\hline\hline
\textit{Core} property & Range  & Mean ($\pm$ std) & Median ($-\text{Q1}$/$+\text{Q3}$) \\ \hline
Velocity dispersion of galaxy systems, $\sigma_{\upsilon_\text{sys}}$  [km s$^{-1}$]  & $[8, 514]$         & 210 ($\pm 109$)       & 203 ($-77$/$+79$)          \\
Extensive mass, $\mathcal{M}_\text{ext}^c$ [$10^{14}h_{70}^{-1}\mathcal{M}_\odot$]                        & $[3.83, 44.27]$    & 18.34 ($\pm 10.10$)     & 15.62 ($-5.18$/$+9.80$)        \\
Gravitational radius, $r_G$ [$h_{70}^{-1}$ Mpc]                                       & $[12.60, 54.30]$  & 25.51 ($\pm 9.24$)      & 24.28 ($-6.16$/$+4.98$)        \\
Volume, $V_c$ [$h_{70}^{-3}$ Mpc$^3$]                                                     & $[83.7, 1614.9]$   & 610.2 ($\pm 389.3$)      & 514.3 ($-210.1$/$+318.0$)        \\
Mass density, $\rho_c$ [$10^{10}h_{70}^{2}\mathcal{M}_\odot$Mpc$^{-3}$]                                            & $[144.1, 604.3]$   & 314.5 ($\pm 107.0$)      & 291.3 ($-51.9$/$+82.6$)        \\
Galaxy density, $n_c$ [$h_{70}^{3}$ Mpc$^{-3}$]                                             & $[0.0727, 0.7571]$ & 0.3267 ($\pm 0.1597$)     & 0.3085 ($-0.1005$/$+0.0967$)       \\
Galaxy number density contrast, $\delta^c_\text{g}$                      & $[0.87, 42.22]$   & 16.02 ($\pm 10.04$)      & 14.42 ($-6.37$/$+7.25$)        \\
Mass density excess (with respect to background), $\mathcal{R}$              & $[4.75, 148.14]$   & 43.74 ($\pm32.26$)      & 32.14 ($-8.77$/$+18.64$)        \\
Mass density contrast (with respect to critical density), $\Delta_\text{cr}$ & $[9.60, 43.43]$    & 22.13 ($\pm 7.87$)      & 20.42 ($-3.82$/$+6.07$)        \\ 
\hline
\end{tabular}}
\end{table*}

A preliminary topological analysis based on percolation reveals that, within superclusters, the \emph{cores} are usually located at the intersections of large filaments, although \emph{cores} can also be found as part of the latter or, less frequently, as isolated regions (\textit{e.g.}, when they have cleaned/accreted {most the} matter around them) relatively close to the centers of mass of host superclusters. The mean and median values of the {--Euclidean-- 3D-distance}, $d_\mathrm{CMs}$, between the centers of mass of the \emph{cores} and those of their respective host superclusters are 23.4 $h_{70}^{-1}$ Mpc and 20.4 $h_{70}^{-1}$ Mpc, respectively. These centers of mass were estimated using the positions and virial masses of the member systems of each supercluster and each \emph{core} from the SysCat catalogues (see columns from 2 to 4 of Table \ref{tab:SCprop} and columns from 6 to 8 of Table \ref{tab:flucx}). The location of the \emph{cores} inside the superclusters depends mainly on the internal topology of the latter, which is closely related to their dynamical states \citep[\textit{e.g.},][]{ei2007b}.\\

The main conclusions of this work are the following: 
\begin{itemize}

\item Most of the rich superclusters (83\% in the case of our sample) have high-density regions that can be identified as \emph{cores}, structures characterized by having considerably larger concentrations of galaxies than other regions of the host superclusters and which can be considered as ``nucleation zones'' within the latter {in the sense that they are structures that still accrete matter from their surroundings (e.g., filaments and the dispersed component of superclusters), concentrating it in more compact regions within which their member galaxy systems undergo mergers and great dynamical activity.}

\item Rich superclusters can have more than one \emph{core}, and the number of these is proportional to the total mass and multiplicity of each host supercluster: superclusters with $\mathcal{M}_\mathrm{ext}^\mathrm{sc}\geq 10^{15}h^{-1}_{70}\mathcal{M}_\odot$ tend to have 3 or more \emph{cores}.

\item Within rich superclusters, \emph{cores} are usually located at intersections of large filaments or forming part of these, although they can also be found, less frequently, in isolated regions (usually near the centers of mass of the superclusters), all depending on the internal topology ---related to the dynamical state--- of the host superclusters.

\item \textit{Cores} are structures made up of rich and poor clusters, small galaxy groups and dispersed galaxies frequently located in bridges that connect systems. They present with significant contrasts, $\delta_\mathrm{g}^c$, $\mathcal{R}$ and $\Delta_{\mathrm{cr}}$, with respect to the local number and mass densities as well as to the critical density of the Universe, respectively. The \emph{cores} are themselves `compact superclusters' extending to scales up to $\sim 15.0 h_{70}^{-1}$ Mpc.

\item \textit{Cores} are the most massive and densest large-scale features that can be identified in the internal structure of rich superclusters. These are the largest gravitationally bound structures with a high probability of becoming virialized systems in the future which will survive the tear of cosmic expansion as `island universes'.
\end{itemize}
 
\begin{acknowledgement}
The authors appreciate the material provided through private communication with Dr. Iris Santiago-Bautista. The authors are grateful for the discussions and suggestions of Dr. Varun Sahni and Dr. Satadru Bag that helped improve the volume estimates of structures through polyhedral surface fitting. The authors are grateful for the valuable suggestions of the anonymous referee that helped to improve the final version of this paper.  \\

This research has made use of the VizieR catalogue access tool, CDS, Strasbourg, France (DOI: \url{10.26093/cds/vizier}). The original description  of the VizieR service was published by \citet{och2000}.\\ 

This research has made use of the NASA/IPAC Extragalactic Database, which is funded by the National Aeronautics and Space Administration and operated by the California Institute of Technology.
\end{acknowledgement}

\paragraph{Funding Statement}

This research was supported by CONAHCyT through a PhD grant and DAIP 0096/21 and 0162/22 projects grants.

\paragraph{Competing Interests}
`None'.

\paragraph{Data Availability Statement}

Data resulted from the present work are available in the manuscript, both printed and digitally, and codes can be requested to the corresponding author.

\printbibliography

@ARTICLE{ab1958,
  author = {Abell, G. O.},
  journal = {ApJS},
  year = {1958},
  volume = {3},
  pages = {211},
}

@ARTICLE{ab1961,
  author = {Abell, G. O.},
  year = {1961},
  journal = {AJ},
  volume = {66},
  pages = {607},
}

@ARTICLE{ab1989,
  author = {Abell, G. O. and Corwin, H. G. Jr. and Olowin, R. P.},
  year = {1989},
  journal = {ApJS},
  volume = {70},
  pages = {1},
}

@ARTICLE{al2017,
  author = {Albareti, F. D. and Allende Prieto, C. and Almeida, A. and et al.},
  year = {2017},
  journal = {ApJS},
  volume = {233},
  pages = {25},
}

@INPROCEEDINGS{and2005,
  author = {Andernach, H. and Tago, E. and Einasto, M. and Einasto, J. and Jaaniste, J.},
  year = {2005},
  BOOKtitle = {ASP Conf. Ser.},
  volume = {329},
  title = {Nearby Large-Scale Structures and the Zone of Avoidance},
  editor = {Fairall, A. P. and Woudt, P. A.},
  publisher = {Astron. Soc. Pac.},
  location = {San Francisco},
  pages = {283},
}

@ARTICLE{am2009,
  author = {Araya-Melo, P. A. and Reisenegger, A. and Meza, A. and van de Weygaert, R. and D\"unner, R. and Quintana, H.},
  year = {2009},
  journal = {MNRAS},
  volume = {399},
  pages = {97},
}

@ARTICLE{ba1984,
  author = {Bahcall, N. A. and Soneira, R. M.},
  year = {1984},
  journal = {ApJ},
  volume = {277},
  pages = {27},
}

@ARTICLE{ba1996,
  author = {Bahcall, N. A.},
  year = {1996},
  note = {preprint (arXiv:astro-ph/9611148v1)},
}

@ARTICLE{bar1994,
  author = {Bardelli, S. and Zucca, E. and Vettolani, G. and et al.},
  year = {1994},
  journal = {MNRAS},
  volume = {267},
  pages = {665},
}

@ARTICLE{bar2000,
  author = {Bardelli, S. and Zucca, E. and Zamorani, G. and Moscardini, L. and Scaramella, R.},
  year = {2000},
  journal = {MNRAS},
  volume = {312},
  pages = {540},
}

@ARTICLE{br1982,
  author = {Beers, T. C. and Geller, M. J. and Huchra, J. P.},
  year = {1982},
  journal = {ApJ},
  volume = {257},
  pages = {23},
}

@ARTICLE{br1990,
  author = {Beers, T. C. and Flynn, K. and Gebhardt, K.},
  year = {1990},
  journal = {AJ},
  volume = {100},
  pages = {32},
}

@ARTICLE{ber2006,
  author = {Berlind, A. A. and Frieman, J. and Weinberg, D. H. and et al.},
  year = {2006},
  journal = {ApJS},
  volume = {167},
  pages = {1},
}

@ARTICLE{bi2006,
  author = {Biviano, A. and Murante, G. and Borgani, S. and Diaferio, A. and Dolag, K. and Girardi, M.},
  year = {2006},
  journal = {A\&A},
  volume = {456},
  pages = {23},
}

@ARTICLE{bol2012,
  author = {Bolton, A. S. and Schlegel, D. J. and Aubourg, E. and et al.},
  year = {2012},
  journal = {AJ},
  volume = {144},
  pages = {144},
}

@ARTICLE{boh2021,
  author = {B\"ohringer, H. and Chon, G. and Tr\"umper, J.},
  year = {2021},
  journal = {A\&A},
  volume = {651},
  pages = {A16},
}

@ARTICLE{bre1994,
  author = {Breen, J. and Raychaudhury, S. and Forman, W. and Jones, C.},
  year = {1994},
  journal = {ApJ},
  volume = {424},
  pages = {59},
}

@ARTICLE{bry1998,
  author = {Bryan, G. L. and Norman, M. L.},
  year = {1998},
  journal = {ApJ},
  volume = {495},
  pages = {80},
}

@ARTICLE{ca2002,
  author = {Caretta, C. A. and Maia, M. A. and Kawasaki, W. and Willmer, C. N. A.},
  year = {2002},
  journal = {AJ},
  volume = {123},
  pages = {1200},
}

@ARTICLE{ca2023,
  author = {Caretta, C. A. and Andernach, H. and Chow-Mart\'inez, M. and Coziol, R. and De Anda-Su\'arez, J. and Hern\'andez-Aguayo C. and et al.},
  year = {2023},
  journal = {RMxAA},
  volume = {59},
  pages = {345},
}

@ARTICLE{ch2002,
  author = {Chiueh, T. and He, X.-G.},
  year = {2002},
  journal = {Phys. Rev. D},
  volume = {65},
  pages = {123518},
}

@ARTICLE{ch2013,
  author = {Chon, G. and B\"ohringer, H. and Nowak, N.},
  year = {2013},
  journal = {MNRAS},
  volume = {429},
  pages = {3272},
}

@ARTICLE{ch2015,
  author = {Chon, G. and B\"ohringer, H. and Zaroubi, S.},
  year = {2015},
  journal = {A\&A},
  volume = {575},
  pages = {L14},
}

@ARTICLE{chow2014,
  author = {Chow-Mart\'inez, M. and Andernach, H. and Caretta, C. A. and Trejo-Alonso, J. J.},
  year = {2014},
  journal = {MNRAS},
  volume = {445},
  pages = {4073},
}

@thesis{cw2019t,
  author = {Chow-Mart\'inez, M.},
  year = {2019},
  institution = {Universidad de Guanajuato, Mexico},
  type = {Doctoral thesis},
}

@ARTICLE{co2012,
  author = {Coil, A. L.},
  year = {2012},
  title = {Large Scale Structure of the Universe},
  note = {preprint (arXiv:1202.6633v1)},
}

@ARTICLE{col2001,
  author = {Colless, M. and Dalton, G. and Maddox, S. and et al. (the 2dFGRS team)},
  year = {2001},
  journal = {MNRAS},
  volume = {328},
  pages = {1039},
}

@ARTICLE{cd2011,
  author = {Costa-Duarte, M. V. and Sodr\'e Jr., L. and Durret, F.},
  year = {2011},
  journal = {MNRAS},
  volume = {411},
  pages = {1716},
}

@ARTICLE{da1985,
  author = {Davis, M. and Efstathiou, G. and Frenk, C. S. and White, S. D. M.},
  year = {1985},
  journal = {ApJ},
  volume = {292},
  pages = {371},
}

@ARTICLE{du2006,
  author = {D\"unner, R. and Araya, P. A. and Meza, A. and Reisenegger, A.},
  year = {2006},
  journal = {MNRAS},
  volume = {366},
  pages = {803},
}

@ARTICLE{em1994,
  author = {Edelsbrunner, H. and M\"ucke, E. P.},
  year = {1994},
  journal = {ACM Transactions on Graphics},
  volume = {13},
  number = {1},
  pages = {43},
}

@ARTICLE{ei1984,
  author = {Einasto, J. and Klypin, A. A. and Saar, E. and Shandarin, S. F.},
  year = {1984},
  journal = {MNRAS},
  volume = {206},
  pages = {529},
}

@ARTICLE{ei2001,
  author = {Einasto, M. and Einasto, J. and Tago, E. and M\"uller, V. and Andernach, H.},
  year = {2001},
  journal = {AJ},
  volume = {122},
  pages = {2222},
}

@ARTICLE{ei2007a,
  author = {Einasto, J. and Einasto, M. and Tago, E. and Saar, E. and H\"utsi, G. and et al.},
  year = {2007},
  journal = {A\&A},
  volume = {462},
  pages = {811},
}

@ARTICLE{ei2007b,
  author = {Einasto, M. and Saar, E. and Liivam\"agi, L. J. and Einasto, J. and et al.},
  year = {2007},
  journal = {A\&A},
  volume = {476},
  pages = {697},
}

@ARTICLE{ei2007c,
  author = {Einasto, M. and Einasto, J. and Tago, E. and Saar, E. and Liivam\"agi, L. J. and et al.},
  year = {2007},
  journal = {A\&A},
  volume = {464},
  pages = {815},
}

@ARTICLE{ei2008,
  author = {Einasto, M. and Saar, E. and Mart\'inez, V. J. and Einasto, J. and Liivam\"agi, L. J. and et al.},
  year = {2008},
  journal = {ApJ},
  volume = {685},
  pages = {83},
}

@ARTICLE{ei2010,
  author = {Einasto, J.},
  year = {2010},
  journal = {AIP Conference Proceedings},
  volume = {1205},
  pages = {72},
}

@ARTICLE{ei2015,
  author = {Einasto, M. and Gramann, M. and Saar, E. and Liivam\"agi, L. J. and Tempel, E. and et al.},
  year = {2015},
  journal = {A\&A},
  volume = {580},
  pages = {A69},
}

@ARTICLE{ei2016,
  author = {Einasto, M. and Lietzen, H. and Gramann, M. and Tempel, E. and Saar, E. and Liivam\"agi, L. J. and Hein\"am\"aki, P. and Nurmi, P. and Einasto, J.},
  year = {2016},
  journal = {A\&A},
  volume = {595},
  pages = {A70},
}

@ARTICLE{ei2018,
  author = {Einasto, J. and Suhhonenko, I. and Liivam\"agi, L. J. and Einasto, M.},
  year = {2018},
  journal = {A\&A},
  volume = {616},
  pages = {A141},
}

@ARTICLE{ei2019,
  author = {Einasto, J. and Suhhonenko, I. and Liivam\"agi, L. J. and Einasto, M.},
  year = {2019},
  journal = {A\&A},
  volume = {623},
  pages = {A97},
}

@ARTICLE{ei2021,
  author = {Einasto, M. and Kipper, R. and Tejens, P. and Lietzen, H. and Tempel, E. and et al.},
  year = {2021},
  journal = {A\&A},
  volume = {649},
  pages = {A51},
}

@ARTICLE{ei2024,
  author = {Einasto, M. and Einasto, J. and Tejens, P. and Korhonen, S. and Kipper, R. and Tempel, E. and et al.},
  year = {2024},
  journal = {A\&A},
  volume = {681},
  pages = {A91},
}

@INPROCEEDINGS{es1996,
  author = {Ester, M. and Kriegel, H. P. and Sander, J. and Xu, X.},
  year = {1996},
  BOOKtitle = {KDD-96 Proceedings},
  pages = {226},
}

@ARTICLE{gal2003,
  author = {Gal, R. R. and De Carvalho, R. R. and Lopes, P. A. A. and et al.},
  year = {2003},
  journal = {AJ},
  volume = {125},
  pages = {2064},
}

@ARTICLE{gra2002,
  author = {Gramann, M. and Suhhonenko, I.},
  year = {2002},
  journal = {MNRAS},
  volume = {337},
  pages = {1417},
}

@ARTICLE{hog2000,
  author = {Hogg, D. W.},
  year = {2000},
  note = {preprint (arXiv:astro-ph/9905116v4)},
}

@ARTICLE{hu1982,
  author = {Huchra, J. P. and Geller, M. J.},
  year = {1982},
  journal = {ApJ},
  volume = {257},
  pages = {423},
}

@ARTICLE{jo2009,
  author = {Jones, D. H. and Read, M. A. and Saunders, W. and et al.},
  year = {2009},
  journal = {MNRAS},
  volume = {399},
  pages = {683},
}

@ARTICLE{koe2007,
  author = {Koester, B. P. and McKay, T. A. and Annis, J. and et al.},
  year = {2007},
  journal = {ApJ},
  volume = {660},
  pages = {239},
}

@ARTICLE{ko1998,
  author = {Kopylova, F. G. and Kopylov, A. I.},
  year = {1998},
  journal = {AstL},
  volume = {24},
  pages = {491},
}

@ARTICLE{kr2011,
  author = {Kriegel, H. P. and Kr\"oger, P. and Sander, J. and Zimek, A.},
  year = {2011},
  journal = {Wiley Interdisciplinary Reviews: Data Mining and Knowledge Discovery},
  volume = {1},
  number = {3},
  pages = {231–240},
  doi = {10.1002/widm.30},
}

@ARTICLE{lib2018,
  author = {Libeskind, N. I. and van de Weygaert, R. and Cautun, M. and et al.},
  year = {2018},
  journal = {MNRAS},
  volume = {473},
  pages = {1195},
}

@ARTICLE{lii2012,
  author = {Liivam\"agi, L. J. and Tempel, E. and Saar, E.},
  year = {2012},
  journal = {A\&A},
  volume = {539},
  pages = {A80},
}

@ARTICLE{liu2015,
  author = {Liu, T. and Tozzi, P. and Tundo, E. and et al.},
  year = {2015},
  journal = {ApJS},
  volume = {216},
  pages = {28},
}

@ARTICLE{lh2002,
  author = {\L{}okas, E. L. and Hoffman, Y.},
  year = {2002},
  note = {preprint (astro/ph0108283)},
}

@ARTICLE{lum1992,
  author = {Lumsden, S. L. and Nichol, R. C. and Collins, C. A. and Guzzo, L.},
  year = {1992},
  journal = {MNRAS},
  volume = {258},
  pages = {1},
}

@ARTICLE{lu2011,
  author = {Luparello, H. and Lares, M. and Lambas, D. G. and Padilla, N.},
  year = {2011},
  journal = {MNRAS},
  volume = {415},
  pages = {964},
}

@ARTICLE{mar2004,
  author = {Marini, F. and Bardelli, S. and Zucca, E. and et al.},
  year = {2004},
  journal = {MNRAS},
  volume = {353},
  pages = {1219},
}

@manual{mw2022,
  author = {MATLAB},
  title = {MATLAB Documentation: Mathematics and Optimization},
  year = {2023},
  organization = {The MathWorks, Inc.},
  note = {Available in: https://www.mathworks.com/help/index.html},
}

@ARTICLE{mcc2009,
  author = {McConnachie, A. W. and Patton, D. R. and Ellison, S. L. and Simard, L.},
  year = {2009},
  journal = {MNRAS},
  volume = {395},
  pages = {255},
}

@ARTICLE{me1983,
  author = {Melott, A. L. and Einasto, J. and Saar, E. and et al.},
  year = {1983},
  journal = {Phys. Rev. Lett.},
  volume = {51},
  pages = {935},
}

@ARTICLE{mil2005,
  author = {Miller, C. J. and Nichol, R. C. and Reichart, D. and et al.},
  year = {2005},
  journal = {AJ},
  volume = {130},
  pages = {968},
}

@ARTICLE{oo1983,
  author = {Oort, J. H.},
  year = {1983},
  journal = {A\&A},
  volume = {21},
  pages = {373},
}

@ARTICLE{pea2014,
  author = {Pearson, D. W. and Batiste, M. and Batuski, D.},
  year = {2014},
  journal = {MNRAS},
  volume = {441},
  pages = {1601},
}

@BOOK{pe1980,
  author = {Peebles, P. J. E.},
  year = {1980},
  title = {The Large-Scale Structure of the Universe},
  publisher = {Princeton Univ. Press},
  address = {Princeton, NJ},
}

@ARTICLE{pr2021,
  author = {Pe\~naranda-Rivera, J. D. and Paipa-Le\'on, D. L. and Hern\'andez-Charpak, S. D. and Forero-Romero, J. E.},
  year = {2021},
  journal = {MNRAS},
  volume = {500},
  issue = {1},
  pages = {L32},
}

@ARTICLE{pf2011,
  author = {Piffaretti, R. and Arnaud, M. and Pratt, G. W. and Pointecouteau, E. and Melin, J.-B.},
  year = {2011},
  journal = {A\&A},
  volume = {534},
  pages = {A109},
}

@ARTICLE{qui2000,
  author = {Quintana, H. and Carrasco, E. R. and Reisenegger, A.},
  year = {2000},
  journal = {AJ},
  volume = {120},
  pages = {511},
}

@ARTICLE{ram2002,
  author = {Ramella, M. and Geller, M. and Pisani, A. and Da Costa, L.},
  year = {2002},
  journal = {AJ},
  volume = {123},
  pages = {2976},
}

@incollection{sa2011,
  author = {Sander, J.},
  year = {2011},
  title = {Density-Based Clustering},
  BOOKtitle = {Encyclopedia of Machine Learning},
  editor = {Sammut, C. and Webb, G. I.},
  publisher = {Springer},
  address = {Boston, MA},
  doi = {10.1007/978-0-387-30164-8_211},
}

@ARTICLE{ir2020,
  author = {Santiago-Bautista, I. and Caretta, C. A. and Bravo-Alfaro, H. and Pointecouteau, E. and Andernach, H.},
  year = {2020},
  journal = {A\&A},
  volume = {637},
  pages = {A31},
}

@ARTICLE{st1977,
  author = {Sargent, W. L. W. and Turner, E. L.},
  year = {1977},
  journal = {ApJ},
  volume = {212},
  pages = {L3},
}

@ARTICLE{se1996,
  author = {Serna, A. and Gerbal, D.},
  year = {1996},
  journal = {A\&A},
  volume = {309},
  pages = {65},
}

@ARTICLE{sh1983,
  author = {Shandarin, S. F.},
  year = {1983},
  journal = {SvAL},
  volume = {9},
  pages = {104},
}

@ARTICLE{sh2004,
  author = {Shandarin, S. F. and Sheth, J. V. and Sahni, V.},
  year = {2004},
  journal = {MNRAS},
  volume = {353},
  pages = {162},
}

@BOOK{sc2015,
  author = {Schneider, P.},
  year = {2015},
  title = {Extragalactic Astronomy and Cosmology: An Introduction},
  edition = {Second},
  publisher = {Springer},
}

@ARTICLE{sm1997,
  author = {Small, T. A. and Sargent, W. L. W. and Hamilton, D.},
  year = {1997},
  journal = {ApJS},
  volume = {111},
  pages = {1},
}

@ARTICLE{sm1998,
  author = {Small, T. A. and Ma, C. P. and Sargent, W. L. W. and Hamilton, D.},
  year = {1998},
  journal = {ApJ},
  volume = {492},
  pages = {45},
}

@ARTICLE{smi2012,
  author = {Smith, A. G. and Hopkins, A. M. and Hunstead, R. W. and Pimbblet, K. A.},
  year = {2012},
  journal = {MNRAS},
  volume = {422},
  pages = {25},
}

@ARTICLE{st1979,
  author = {Stauffer, D.},
  year = {1979},
  journal = {Phys. Rep.},
  volume = {54},
  issue = {1},
  pages = {1},
}

@incollection{the2009,
  author = {Theodoridis, S. and Koutroumbas, K.},
  year = {2009},
  title = {Pattern Recognition},
  BOOKtitle = {Pattern Recognition, 4th edn.},
  editor = {T. Sergios and K. Konstantinos},
  publisher = {Academic Press},
  address = {Boston},
  pages = {595},
}

@ARTICLE{tl2014,
  author = {Tully, R. B. and Courtois, H. and Hoffman, Y. and Pomar\`ede, D.},
  year = {2014},
  journal = {Nature},
  volume = {513},
  number = {7516},
  pages = {71},
}

@ARTICLE{tl2015,
  author = {Tully, R. B.},
  year = {2015},
  journal = {AJ},
  volume = {149},
  pages = {54},
}

@ARTICLE{wen2012,
  author = {Wen, Z. L. and Han, J. L. and Liu, F. S.},
  year = {2012},
  journal = {ApJS},
  volume = {199},
  pages = {34},
}

@ARTICLE{zel1982,
  author = {Zeldovich, Y. B. and Einasto, J. and Shandarin, S. F.},
  year = {1982},
  journal = {Nature},
  volume = {300},
  pages = {407},
}

@ARTICLE{och2000,
  author = {Ochsenbein, F. and Bauer, P. and Marcout, J.},
  year = {2000},
  journal = {A\&AS},
  volume = {143},
  pages = {23},
}

@article{cole2005,
    author = {Cole, S. and Percival, W. J. and Peacock, J. A. and Norberg, P. and Baugh, C. M. and Frenk, C. S. and Baldry, I. and Bland-Hawthorn, J. and et al. (2dFGRS team)},
    journal = {MNRAS},
    volume = {362},
    number = {2},
    pages = {505},
    year = {2005},
}

@article{ham2001,
    author = {Hambly, N.C. and MacGillivray, H.T. and Read, M.A. and Tritton, S.B. and Thomson, E.B. and Kelly, B.D. and Morgan, D.H. and Smith, R.E. and Driver, S.P. and Williamson, J. and Parker, Q.A. and Hawkins, M.R.S. and Williams, P.M. and Lawrence, A.},
    journal = {MNRAS},
    volume = {326},
    number = {4},
    pages = {1279},
    year = {2001},
}

@ARTICLE{ad2020,
  author = {Abdullah, M. H. and Wilson, G. and Klypin, A. and Old, L. and Praton, E. and Ali, G. B.},
  journal = {ApJS},
  year = {2020},
  volume = {246},
  pages = {2},
}

@ARTICLE{tm2014,
  author = {Tempel, E. and Tamm, A. and Gramann, M. and Tuvikene, T. and Liivam\"agi, L. J. and et al.},
  journal = {A\&A},
  year = {2014},
  volume = {566},
  pages = {A1},
}

@ARTICLE{san2023,
  author = {Sankhyayan, S. and Bagchi, J. and Tempel, E. and More, S. and Einasto, M. and et al. },
  journal = {ApJ},
  year = {2023},
  volume = {958},
  pages = {62},
}

@ARTICLE{he2022,
  author = {Hein\"am\"aki, P. and Teerikorpi, P. and Douspis, M. and Nurmi, P. and Einasto, M. ang Gramann, M. and Nevalainen, J. and Saar, E.},
  year = {2022},
  journal = {A\&A},
  volume = {668},
  pages = {A37},
}

@ARTICLE{sd2011,
  author = {Sheth, R. and Diaferio, A.},
  journal = {MNRAS},
  year = {2011},
  volume = {417},
  pages = {4},
}

@ARTICLE{ra2006,
  author = {Ragone, C. J. and Muriel, H. and Proust, D. and et al.},
  journal = {A\&A},
  year = {2006},
  volume = {445},
  pages = {819},
}

@ARTICLE{bc2021,
  author = {B\"ohringer, H. and Chon, G.},
  journal = {A\&A},
  year = {2021},
  volume = {656},
  pages = {A144},
}

@ARTICLE{na2003,
  author = {Nakamura, O. and Fukugita, M. and Yasuda, N. and Loveday, J. and Brinkmann, J. and et al.},
  journal = {AJ},
  year = {2003},
  volume = {125},
  pages = {1682},
}

@ARTICLE{zu2024,
  author = {Z\'u\~niga, J. M. and Caretta, C. A. and Gonz\'alez, A. P. and Garc\'ia-Manzan\'arez E.},
  journal = {RMxAA},
  year = {2024},
  volume = {60},
  pages = {141},
}
\end{document}